\documentclass[aps,
prd,
superscriptaddress,showpacs,showkeys,nofootinbib,eqsecnum,floats]{revtex4}
\usepackage{graphicx,color}
\usepackage{float}
\usepackage{amsmath}
\usepackage{amssymb}
\usepackage{graphicx}
\usepackage{bm}
\usepackage{color,colordvi}
\usepackage{epsfig}
\usepackage{natbib}
\usepackage{mciteplus}
\usepackage{mathcomp}
\usepackage{cancel}  
\usepackage{ulem}    
\usepackage[colorlinks=true,urlcolor=blue]{hyperref}
\usepackage{multirow}
\hypersetup{                
backref = true,             
pagebackref = true,                 
hyperindex = true,          
colorlinks = true,          
breaklinks = true,          
urlcolor = blue,            
linkcolor = blue,           
bookmarks = true,           
bookmarksopen = true,       
citecolor=red,
}
\begin{document}
\begin{flushright}
\end{flushright}
\newcommand  {\ba} {\begin{eqnarray}}
\newcommand  {\ea} {\end{eqnarray}}
\def\m{\text{m}}
\def\c{\text{c}}
\def\cM{{\cal M}}
\def\cO{{\cal O}}
\def\cK{{\cal K}}
\def\cS{{\cal S}}
\def\d{\mathrm{d}}
\newcommand{\mh}{m_{h^0}}
\newcommand{\mw}{m_W}
\newcommand{\mz}{m_Z}
\newcommand{\mt}{m_t}
\newcommand{\mb}{m_b}
\newcommand{\p}{\sim_\Phi}
\newcommand{\ps}{\sim_{\Phi,S}}
\newcommand{\be}{\beta}\newcommand{\al}{\alpha}
\newcommand{\lam}{\lambda}
\newcommand{\beq}{\begin{equation}}
\newcommand{\eeq}{\end{equation}}
\newcommand{\bee}{\end{equation}}
\newcommand{\beqa}{\begin{eqnarray}}
\newcommand{\eeqa}{\end{eqnarray}}
\newcommand{\noi}{\noindent}
\newcommand{\nn}{\nonumber}
\newcommand{\comment}[1]{\textcolor{red}{\bf{[\sl #1}]}}
\newcommand{\Mcomment}[1]{\textcolor{blue}{\bf{[\sl #1}]}}
\newcommand{\Dcomment}[1]{\textcolor{black}{\bf{[ #1}]}}
\newcommand{\bpm}{\begin{pmatrix}}
\newcommand{\epm}{\end{pmatrix}}

\def\ga{\mathrel{\raise.3ex\hbox{$>$\kern-.75em\lower1ex\hbox{$\sim$}}}}
\def\la{\mathrel{\raise.3ex\hbox{$<$\kern-.75em\lower1ex\hbox{$\sim$}}}}


\newtheorem{theorem}{Theorem}[section]
\newtheorem{acknowledgement}[theorem]{Acknowledgement}
\newtheorem{algorithm}[theorem]{Algorithm}
\newtheorem{axiom}[theorem]{Axiom}
\newtheorem{claim}[theorem]{Claim}
\newtheorem{conclusion}[theorem]{Conclusion}
\newtheorem{condition}[theorem]{Condition}
\newtheorem{conjecture}[theorem]{Conjecture}
\newtheorem{corollary}[theorem]{Corollary}
\newtheorem{criterion}[theorem]{Criterion}
\newtheorem{definition}[theorem]{Definition}
\newtheorem{example}[theorem]{Example}
\newtheorem{exercise}[theorem]{Exercise}
\newtheorem{lemma}[theorem]{Lemma}
\newtheorem{notation}[theorem]{Notation}
\newtheorem{problem}[theorem]{Problem}
\newtheorem{proposition}[theorem]{Proposition}
\newtheorem{remark}[theorem]{Remark}
\newtheorem{solution}[theorem]{Solution}
\newtheorem{summary}[theorem]{Summary}
\newcommand{\qeed}{\hfill\textrm{$\blacksquare$}\break\null}
\newenvironment{demo}{\noindent\textit{Proof.}~}{\qeed}
\newcommand{\gilbert}[1]{\textcolor{black}{ \bf{#1}}}
\newcommand{\gilbertzero}[1]{#1}
\newcommand{\gilbertred}[1]{\textcolor{red}{ \bf{#1}}}

\title{Low Energy Supergravity Revisited (I)}


\author{Gilbert Moultaka}
\email{gilbert.moultaka@umontpellier.fr}
 \thanks{corresponding author}
\affiliation{Laboratoire Charles Coulomb (L2C), UMR 5221 CNRS-Universit\'e de Montpellier, Montpellier, F-France}
\author{ Michel Rausch de Traubenberg}
\email{michel.rausch@iphc.cnrs.fr}
\author{Damien Tant}
\email{damien.tant@iphc.cnrs.fr}
\affiliation{Universit\'e de Strasbourg, CNRS, IPHC UMR 7178, F-6700 Strasbourg, France}

\date{\today}

\begin{abstract}
General forms of the K\"ahler and superpotenials that lead to 
consistent low energy broken Supersymmetry originating from $N=1$ 
Supergravity have been classified and used for model building 
since more than three decades. We point out the incompleteness of 
this classification \gilbertzero{when hidden sector vacuum expectation values are of the order of the Planck mass}. Focusing in this paper mainly on the case of
minimal K\"ahler potential, we adopt a rigorous approach that 
retrieves on the one hand the known forms, and demonstrate on the other hand 
the existence of a whole set of new forms for the superpotential of which we
give a complete classification. The latter forms involve a new type of 
chiral superfields having the unusual property of belonging neither to the hidden sector nor to 
the conventional observable sector. Comparing the obtained forms with the conventional ones, we argue how new possibilities for model building can arise, and
discuss the gravity mediation of soft as well as additional hard (but parametrically small) Supersymmetry breaking,  in the presence 
of the new type of chiral superfields. In the simplest case, we study the vacuum structure, characterize the masses and couplings
of the scalar components to 
the hidden and observable sectors and discuss briefly the physical 
role they could play. \gilbertzero{In the generic case, we estimate the magnitude and possible consequences of the hard breaking 
of \gilbertzero{Supersymmetry in terms of the interplay between hidden and visible sectors mass scales.}} 
\end{abstract}

\maketitle
\section{Introduction}
\noindent
The experimental discovery of the ${\cal H}$-boson at the Large Hadron Collider by the ATLAS \cite{Aad:2012tfa} and CMS 
\cite{Chatrchyan:2012xdj} collaborations has completed the  
building blocks of the standard model (SM) of particle physics. 
This great experimental success of the SM, the properties of the  ${\cal H}$-boson having so far proven to be very close to the those 
of the SM Higgs \cite{Englert:1964aa,Higgs:1964aa,Higgs:1964ab,Guralnik:1964aa}, 
has been accompanied by a forewarning regarding the search for new physics beyond the SM which remains for the time 
being elusive. In particular the absence of direct experimental signals of low energy broken Supersymmetry (SUSY)
may ultimately question, if not its existence, at least the way this symmetry is realized in Nature. Given the theoretical motivations
for SUSY, it is of paramount importance to constantly verify the degree of universality on which the low energy models are
based since the latter condition the interpretation of the data. Among the various ways to  
communicate SUSY breaking to the low energy sector, the mediation through gravitational 
interactions \cite{ca,bfs,iba,ohta,ent,apw,Polonyi:1977pj,Cremmer:1982vy,
Nilles:1982dy,Nilles:1982xx,SW,gm,Nilles:1983ge,Hall:1983iz} in the context of Supergravity, is probably one of the most appealing
scenarios as it provides a unique framework encompassing the four fundamental interactions.   

In this paper we reconsider the formal consequences of the general requirement any supersymmetric extension of the SM,
viewed as an effective low energy remnant of Supergravity, should satisfy \gilbertzero{{\sl in the class of models where hidden sector fields take
vacuum expectation values (VEVs) of the order of the Planck mass (${\cal O}(m_{p\ell})$)}}.
\gilbertzero{In this context} 
we revisit the approach of Soni and Weldon\cite{SW} (SW) 
\gilbertzero{who analyzed a long time ago the acceptable forms of the 
superpotential and 
K\"ahler potential in that class of models. As these forms 
encompassed the ones pioneered previously \cite
{ca,bfs,iba,ohta,ent,apw,Polonyi:1977pj,Cremmer:1982vy,
Nilles:1982dy,Nilles:1982xx,SW,gm,Nilles:1983ge,Hall:1983iz} and were more
general, they have been since relied upon when constructing
models where Supersymmetry breaking is mediated by gravitational
interactions (see e.g. \cite{Brignole:2010sax} for a review). 
It should be noted, though, that having hidden sector VEVs of ${\cal O}(m_{p\ell})$ became somewhat secondary as the
interest shifted towards the celebrated requirement of {\sl soft}
breaking of Supersymmetry that these forms ensure\cite{Chung:2003fi}; even more so, since consistent scenarios 
of gravity
mediation that did not assume such big values of the VEVs 
(see e.g. \cite{Banks:1993en}) were also examined along the line, 
somewhat lessening the pertinence of
the approach adopted initially by SW. However, it could be of some 
interest to revive such an approach. We see a few reasons
for that: {\sl (i)} the consistency requirement {\sl \`a la} SW is 
nothing else but a {\sl tree-level} protection of 
low-scale physics from very large Planck scale effects. 
In that sense this is just one instance of the tree-level 
prerequisites when dealing with issues such as the hierarchy 
problem and the stabilization of largely different scales against 
quantum effects. 
{\sl (ii)} In the presence of ${\cal O}(m_{p\ell})$ 
hidden sector VEVs, 
the general consistency requirements {\sl \`a la} SW} 
are found to lead to much richer structures compatible with low 
energy physics, than previously established. 
\gilbertzero{{\sl (iii)} These structures allow consistently not only 
soft but also hard breaking  contributions. {\sl (iv)} The approach can be in principle extended to cases where the hidden
sector VEVs are much smaller than ${\cal O}(m_{p\ell})$ but still much larger than other physical scales.}

\gilbertzero{The main goal of the present paper is to demonstrate {\sl (ii)} and {\sl (iii)}.}
We provide rigorous proofs that allow a complete classification of 
the acceptable  structures
of the superpotential 
in the case of canonical K\"ahler potential, and give a few 
illustrative examples in the case of the general K\"ahler 
potential as well. 
We then discuss on general grounds the possible physical relevance 
of the novel forms and their associated fields.  


The paper is organized as follows: in Section \ref{sec:gen}
we recall the general setting of $N=1$ Supergravity. 
A summary of the main ingredients of the approach of Ref.~\cite{SW} is given in Section \ref{sec:SW},
as well as a reminder of the general requirement for the consistency of any low energy visible sector 
of (broken) Supersymmetry \gilbertzero{in the presence of at least
one ${\cal O}(m_{p\ell})$ VEV.} Section \ref{sec:NSW-flat} is devoted to the main
results of the paper: we describe our approach and argue why the results of \cite{SW} cannot encompass 
all possible forms of the K\"ahler potential and superpotential that comply with this requirement.  
We then give, in the case of the flat K\"ahler metric, the complete classification of the general forms of the superpotential of $N=1$ 
Supergravity consistent with low energy physics, yielding not only the conventional ones of \cite{SW}, 
but also a whole class of new solutions. Complete details of the rigorous
proof of this classification are relegated to appendices  
\ref{app:prelimin} and \ref{app:proof-flat}. 
A comparison between the features of the new and conventional forms is addressed in section 
\ref{sec:comparison}. In Section \ref{sec:NSW-non-flat} and \ref{app:non-flat} we exhibit and discuss briefly 
examples of new solutions in the more involved general K\"ahler case.
A qualitative discussion of the possible
implications of the new forms for model building
is given in Section \ref{sec:model-pheno-1}.
\gilbertzero{ In Section \ref{sec:model-pheno-2} we carry out,
in the case of flat K\"ahler metric, 
a comparison between the conventional and new forms regarding 
the gravity mediation of SUSY breaking, and determine the structure of the low energy soft-SUSY breaking terms
\gilbertzero{as well as additional 
hard breaking terms}
in presence of
the fields that enter the new forms.}
Section \ref{sec:simple-NSWS} is devoted to an analysis of the simplest new form of the superpotential.
There we study the vacuum structure, characterize the masses and couplings of the
new fields to the hidden and observable sectors and discuss briefly the physical 
role they could play. \gilbertzero{In section \ref{sec:hardbreaking} we survey briefly the magnitude and possible impact on phenomenology of the hard SUSY breaking terms that are present in the new solution, ending with a few related comments.} 
We conclude in Section \ref{sec:conclusion}.

\section{$N=1$ Supergravity, a quick reminder}
To set the stage we recall first the main ingredients of $N=1$ Supergravity theory and the way it can
encompass consistently the low energy models with broken global Supersymmetry. This also serves to define the notations.

\subsection{The general setting \label{sec:gen}}
Consider an $N=1$ Supergravity theory with $k+\ell$ chiral superfields whose scalar field components
will be generically denoted by
$Z^I, I=1,\cdots, k+\ell$. 
[Throughout the paper we use the following convention: $Z$ denotes the full set of $Z^I$ fields, $Z^\dag$ denotes the full set of 
their complex conjugates $Z^{I^*}$, and complex conjugation of a function of these fields is denoted by a bar.]
Introducing  the K\"ahler
potential $K(Z,Z^\dag)$ and superpotential $W(Z)$,
the $F$-term contribution to the scalar potential takes the form
\beqa
\label{eq:V}
V_F=e^{\frac{K}{m_{p\ell}^2}}\Big({\cal D}_I W K^{I{} {J^\ast}}{\cal D}_{J^\ast}\overline W  -  \frac3{m_{p\ell}^2} |W|^2\Big) \ , 
\eeqa
with
\beqa
{\cal D}_I W  = W_I + \frac1{m_{p\ell}^2} K_I W  \label{eq:DI} \ ,
\eeqa
where we define
\beqa
W_I \equiv \frac{\partial W}{\partial Z^I}, \;\;  (\overline{W}_{I^*} \equiv \frac{\partial \overline{W}}{\partial Z^{I^*}}), 
\label{eq:WI}\\
K_I \equiv \frac{\partial K}{\partial Z^I}, \;\;  (K_{I^*} \equiv \frac{\partial K}{\partial Z^{I^*}}), 
\label{eq:KI}
\eeqa
and
\beqa
K_{I^*J}= \frac{\partial^2 K}{\partial  Z^{I^*} \partial Z^J } \label{eq:KIJ}
\eeqa
is the K\"ahler metric, 
$K^{I{} {J^\ast}} \equiv K^{-1}_{I{} {J^\ast}}$ its inverse, 
and $m_{p\ell}$ denotes the reduced Planck mass. 

We also recall for later reference the alternative form 
\beqa
V_F= m_{p\ell}^2e^{\frac{\cal G}{m_{p\ell}^2}} \Big({\cal G}_{I} {\cal G}^{I J^\ast} {\cal G}_{J^\ast}-  3 m_{p\ell}^2   \Big)  
\eeqa
of Eq.~(\ref{eq:V}), where the notational convention is as in Eqs.~(\ref{eq:KI}, \ref{eq:KIJ}) and ${\cal G}(Z,Z^\dag)$
denotes the  generalized
K\"ahler potential 
\beqa
{\cal G} = K + m_{p\ell}^2    \ln \Big|\frac{ W}{m_{p\ell}^3}\Big|^2 \ . \
\eeqa
The gauge sector contributions to the potential 
\beqa
V_D = \frac12 (\text{Re} f)^{\alpha \beta} D_\alpha D_\beta, \label{eq:VD}
\eeqa
where $D_\alpha$ denote all the $D-$terms corresponding to the gauge interactions, 
$f_{\alpha \beta}$ a general gauge kinetic function and $(\text{Re} f)^{\alpha \beta}$ the inverse
of its real part 
as well as all the 
remaining parts of the Supergravity Lagrangian, including the gauge sector, 
the superpotential and K\"ahler geometry dependent scalar-fermion 
interactions, (see for 
instance \cite{WessBagger199203}), are not directly relevant to 
our analysis. 

The general setting assumes two distinct
sectors: --an {\sl observable 
sector} describing the Minimal Supersymmetric Standard Model (MSSM) 
or any of its extensions including possibly a Grand Unified (GUT)
sector --a {\sl hidden sector} in which local 
SUSY is assumed to be broken.  The notions of {\sl visible} and {\sl hidden} sectors 
are used loosely here. Their physical justification will be made clear in the next section. 
The chiral superfields split accordingly into $k$ chiral superfields in the observable sector with scalar components
denoted by $\Phi^a=(\Phi^1,\cdots,\Phi^k)$, and
$\ell$ chiral superfields in the hidden sector with scalar components denoted by $\zeta^i=(\zeta^1,\cdots,\zeta^\ell)$.
The spontaneous breaking of local SUSY occurs when (a combination of) the $\zeta^i$ fields develop a vacuum expectation value
(VEV) \gilbertzero{such that 
\begin{equation}
\langle F^I \rangle  \neq 0 \label{eq:Fvev} , 
\end{equation}
where
\begin{equation}
F^I \equiv m_{p\ell} e^{\frac12\frac{\cal G}{m_{p\ell}^2}} {\cal G}^{I{} {J^\ast}} {\cal G}_{J^\ast} + \cdots
\label{eq:Fterm}, 
\end{equation}
and $I$ labels any of the scalar fields $Z^I$; (the dots stand for contributions from the chiral fermionic partners of the $Z^I$  
or from fermions of vector supermultiplets, whose VEVs will be assumed to vanish in the present study.)}
\gilbertzero{The hidden sector VEV is also assumed} to yield a nonvanishing  gravitino mass
\beqa
m_{3/2} = \frac1{m_{p\ell}^2}\Big< |W| e^{\frac12 \frac{K}{m_{p\ell}^2}} \Big> = m_{p\ell} \Big< e^{\frac12 \frac{\cal G}{m_{p\ell}^2}} \Big> \ ,
\label{eq:mgrav}
\eeqa
while keeping the graviton massless. \gilbertzero{Although the hidden sector dynamics could be such that $\langle W \rangle \neq 0$
whenever SUSY is broken, this is in general neither necessary nor sufficient unless one imposes  a vanishing 
vacuum energy as well leading to the tree-level relation $\sqrt{3} m_{3/2} m_{p\ell}= (\langle F^I F_I\rangle)^{1/2}\equiv M_S$ where
$M_S$ denotes the SUSY breaking scale.\footnote{\gilbertzero{This remains essentially true when allowing a tiny cosmological constant, albeit issues related
to (anti)deSitter spaces in which we do not need to enter in the present study. See for instance \cite{Knoops:2016cps}
and references therein.}}  }
The gravitational interactions communicate then  
SUSY breaking to the visible sector through the 
generation of soft-SUSY breaking terms triggered by $m_{3/2}$
\cite
{ca,bfs,iba,ohta,ent,apw,Polonyi:1977pj,Cremmer:1982vy,
Nilles:1982dy,Nilles:1982xx,SW,gm,Nilles:1983ge,Hall:1983iz}. 


\subsection{The Soni-Weldon approach \label{sec:SW}}
It is important to stress that there is in general no guarantee that the above scenario 
of mediation of SUSY breaking from the hidden 
sector to the visible sector through gravitational interactions,    
would not lead to inconsistencies at low energy.
Since Supergravity is non-renormalizable and
should thus be viewed as an effective description of yet another layer of a more fundamental theory,
the K\"ahler potential and superpotential can {\sl a priori} be arbitrary functions of the fields
and contain arbitrary powers of the Planck mass $m_{p\ell}$. 
Moreover, \gilbertzero{in the scenarios we are considering}, 
one expects generically some
of the VEVs developed by (some of) the $Z^I$ fields to be of order $m_{p\ell}$. But other $Z^I$ fields should develop much smaller 
VEVs so as to set off the electroweak scale and possibly other
intermediate scales corresponding to extensions of the SM. 
\gilbertzero{Following SW\cite{SW}, the
scalar fields that
acquire ${\cal O}(m_{p\ell})$ VEVs are called {\sl hidden
sector} fields $\zeta^i$ responsible for local SUSY breaking, while the {\sl visible sector} fields $\Phi^a$ have much smaller or vanishing VEVs. In order to retrieve 
consistently the low energy Lagrangian for the visible sector,
it is required that:}
\beqa
&\text{ {\sl all visible sector fields should not appear in the operators of the Lagrangian }} \nonumber \\
&\text{ {\sl that diverge in the flat limit  $m_{p\ell} \to \infty$.}}
\label{eq:the_condition}
\eeqa
From now on we focus on the scalar potential part of 
the Lagrangian. \gilbertzero{Near the minimum of the potential,
requirement (\ref{eq:the_condition}) implies automatically that 
the $\zeta$ fields have Planck suppressed couplings to the 
$\Phi$ fields.
Indeed, near the minimum  the scalar fields read
$\zeta = \langle \zeta \rangle + \delta \zeta$ and $\Phi = \langle \Phi \rangle + \delta \Phi$, with
$|\langle \zeta \rangle| \sim {\cal O}(m_{p\ell}) \gg |\langle \Phi \rangle| \sim |\delta \zeta| \sim
|\delta \Phi|$.
If, following (\ref{eq:the_condition}), $\Phi$ should be absent from an operator when $\zeta = \langle \zeta \rangle \sim {\cal O}(m_{p\ell})$, then it obviously remains so by continuity when $\zeta$ is allowed to vary slightly 
around  its VEV value. This is most easily seen when reasoning
on polynomial operators. In particular,  
operators coupling $\zeta$ and $\Phi$ are forbidden if their
dimension is $\leq 4$, and    
 should be suppressed by powers of $m_{p\ell}$ if their dimension
 is strictly larger than four.}
 It thus appears natural in retrospect to distinguish, as done in the previous section, \gilbertzero{the notion of {\sl hidden sector} as comprising $\zeta^i$ fields that develop 
 ${\cal O}(m_{p\ell})$ VEVs} from that of  
{\sl observable sector} fields $\Phi^a$ that have small or vanishing VEVs.

\gilbertzero{It is important at this point to stress that this definition of {\sl hidden} and {\sl visible} sectors is 
not at odds with the modern and more general use of these terms,
namely that their mutual interactions are Planck mass suppressed.
Our definition is a special case of the latter and is a {\sl sufficient} but not necessary condition. As such, \gilbertzero{while visible
sector scalar fields should have VEVs much smaller than $m_{p\ell}$,  there should be no ambiguities in the fact
that, in contrast, }a field having a VEV much smaller than $m_{p\ell}$ is not
necessarily in the visible sector. For instance a set of fields
$Z^I$ related by some unbroken symmetry and developing a VEV 
of order $m_{p\ell}$ in just one direction would, due to 
requirement (\ref{eq:the_condition}), have automatically the couplings of all its fields to another sector (that is neutral under this symmetry) suppressed by powers of $m_{p\ell}$.\footnote{\gilbertzero{Of course one could also contemplate
 scenarios 
where none of the hidden sector VEVs is of order $m_{p\ell}$, (see e.g. \cite{Banks:1993en} for a discussion). We have nothing to say
about such scenarios in the present paper, except that a similar
approach as the one considered hereafter can be adapted to include several different mass scales.}}}

We move now to the classification carried out by
 (SW) \cite{SW} of the most general forms of the 
K\"ahler potential and the superpotential compatible with the above consistency requirement.
In order to treat properly the limit $m_{p\ell} \to + \infty$   
they introduced dimensionless fields
in the hidden sector
\beqa
\zeta^i = m_{p\ell} z^i \ , \label{eq:reduced-hidden}
\eeqa
and started off with a K\"ahler potential $K$ and superpotential $W$ having the general expansions
\beqa
K(Z,Z^\dag) &=& \sum \limits_{n=0}^N m_{p\ell}^n K_n(z,z^\dag,\Phi,\Phi^\dag) \ ,  \label{eq:Kexpansion} \\ 
W(Z) &=& \sum \limits_{n=0}^M m_{p\ell}^n W_n(z,\Phi) \ . \label{eq:Wexpansion}
\eeqa
Since $W$ and $K$ have respectively mass dimension three and two, the $W_n$ and the $K_n$ should in general contain
other mass scales much smaller than $m_{p\ell}$ (\gilbertzero{corresponding for instance to some gauge sectors}), and/or non canonical powers of $\Phi^a$. Note that at low energy the dimensionless 
hidden sector fields $z$ are by definition ${\cal O}(1)$, cf. Eq.~(\ref{eq:reduced-hidden}).
Then, imposing (\ref{eq:the_condition}) they required the potential to split into two independent sectors: 
one sector depending exclusively on the hidden sector with possibly positive powers of the Planck mass, the other
sector remaining finite with possibly Planck suppressed hidden/observable sectors interactions, but reducing to the standard
flat Supersymmetry form ({\it i.e.} the usual $F-$ and $D-$terms) for the observable sector in the large $m_{p\ell}$ limit. 
Analyzing this strong constraint they arrived at the following special reduced forms of 
\eqref{eq:Kexpansion}, \eqref{eq:Wexpansion},   
\beqa
K(Z,Z^\dag) &=& m_{p\ell}^2 K_2(z,z^\dag) + m_{p\ell} K_1(z,z^\dag) + K_0(z,z^\dag,\Phi,\Phi^\dag) \ , \label{eq:SWKgen}\\
W(Z) &=& m_{p\ell}^2 W_2(z) + m_{p\ell} W_1(z) + W_0(z,\Phi), \label{eq:SWWgen}
\eeqa
where the $K_i$'s and $W_i$'s are arbitrary functions. The main message here is that the observable sector fields
can enter only the $m_{p\ell}$ to the power zero components of Eqs.~(\ref{eq:Kexpansion}, \ref{eq:Wexpansion}), the otherwise arbitrary 
$K_0, W_0$ functions, provided that the exclusively
hidden sector components do not have $m_{p\ell}$ powers greater than two.
Plugging Eqs.~(\ref{eq:SWKgen}, \ref{eq:SWWgen}) in Eq.~(\ref{eq:V}), one can easily check that these forms are indeed 
{\sl sufficient} solutions to satisfy the physical requirement 
(\ref{eq:the_condition}). 
However, the claim in \cite{SW} was that Eqs.~(\ref{eq:SWKgen}, \ref{eq:SWWgen}) are general 
enough to exhaust all possible forms consistent with (\ref{eq:the_condition}). In the next section we argue against
this claim then revisit the analysis of \cite{SW} and indeed exhibit new solutions. 

\section{The general solutions}
We consider here separately the cases of minimal and non-minimal K\"ahler potentials, but focus mainly on the former giving
a thorough discussion and a detailed proof of the classification of all possible solutions.
\subsection{The flat K\"ahler metric case \label{sec:NSW-flat}}
\noindent
Assuming only canonical kinetic terms for all the 
chiral superfields, the K\"ahler potential reads, 
\beqa
\label{eq:Kf}
K(z,z^\dag,\Phi,\Phi^\dag) = m_{p\ell}^2 z^{i *} z^i + \Phi^{a *} \Phi^a \ . 
\eeqa
This form is a special case of Eq.~(\ref{eq:SWKgen}).
Making use of \eqref{eq:Wexpansion}, the potential \eqref{eq:V} can be straightforwardly cast in the form
\beqa
V_F &=& \exp\left(\frac{Z^I Z^{I^*}}{m_{p\ell}^2}\right) \sum_{0 \leq m \leq M, \; 0 \leq n \leq M} \left\{
m_{p\ell}^{m+n} \frac{\partial W_m}{\partial \Phi^a} \frac{\partial \overline W_n}{\partial \Phi^{a *}} \; + \right.  \nonumber \\
 &&m_{p\ell}^{m+n-2}\left( \Big(\frac{\partial W_m}{\partial z^i} + z^{i *} W_m\Big) 
\Big(\frac{\partial \overline W_n}{\partial z^{i*}} + z^i \overline W_n\Big) \right. \nonumber \\
&& ~~~~~~~~~~~~~~\left.
+ \;\Phi^a \frac{\partial W_m}{\partial \Phi^a} \overline W_n  + 
\Phi^{a *} \frac{\partial \overline W_m}{\partial \Phi^{a *}} W_n - 3 W_m \overline W_n\right)  \nonumber \\
 &&~~~~~~~~~~~~~~~~~~~~~~~~~~~~~~~~~~~~~~~~~~\left. + \;m_{p\ell}^{m+n-4} W_m \overline W_n \Phi^{a *} \Phi^a \right\}  \label{eq:Vgen00}\\
   &=& \exp\left(\frac{Z^I Z^{I^*}}{m_{p\ell}^2}\right) \sum^{2 M}_{c = 0}  V_{M,c}[z, z^\dag, \Phi, \Phi^\dag] \; m_{p\ell}^c + {\cal O}(m_{p\ell}^{-1}),
\label{eq:Vgen0}
\eeqa
where in the last equation we collected over all powers of $m_{p\ell}$.  
[See Eq.~(\ref{eq:Vgen}) for the explicit expressions of the $V_{M,c}$.] 
The general requirement (\ref{eq:the_condition}) discussed in the previous section implies that
each $V_{M,c}$  with $c\geq1$ should be {\sl functionally independent} of $\Phi, \Phi^\dagger$.
That is:
\beqa
\frac{\delta }{\delta \Phi^a(x)} \, V_{M,c}[z, z^\dag, \Phi, \Phi^\dag]  = \frac{\delta}{\delta \Phi^{a *}(x)} 
\, V_{M,c}[z, z^\dag, \Phi, \Phi^\dag] = 0, \; \forall a, x, M, c\geq1. \label{eq:masterequation}
\eeqa 
As can be seen from  Eq.~(\ref{eq:Vgen}), this constraint turns into a tower of partial differential equations (PDEs)
in the unknown  superpotential components $W_n(z, \Phi)$ of the expansion Eq.~(\ref{eq:Wexpansion}) and their complex conjugates.
In order to determine all the superpotential forms that are consistent with Eq.~(\ref{eq:the_condition}) one needs to fully solve 
this tower of PDEs, taking moreover into account the non-trivial constraint of holomorphy of the superpotential. \\

We are now ready to state the reason for questioning the claim of generality made by the authors of Ref.~\cite{SW}:
rather than following the exhaustive approach we have just described, the focus in Ref.~\cite{SW} was put exclusively on two terms 
in (\ref{eq:Vgen00}), namely, the term containing a linear factor of  $\Phi^a$ (or $\Phi^{a *}$) present in the coefficient of 
$m_{p\ell}^{m+n-2}$, and the coefficient of $m_{p\ell}^{m+n-4}$ that contains a $\Phi^a \Phi^{a *}$ factor. Requiring these two terms to 
be separately  independent of the observable sector fields $\Phi^a, \Phi^{a *}$, the authors of \cite{SW} 
were lead to the superpotential given in Eq.~(\ref{eq:SWWgen}) as the only possibility.
While this approach is consistent with requirement (\ref{eq:the_condition}) individually for each one of these two terms, 
it so happens that the obtained form of the 
superpotential is {\sl sufficient}, together with Eq.~(\ref{eq:Kf}), to satisfy (\ref{eq:the_condition}). However,
 obviously this {\sl does not} guarantee obtaining all the sufficient {\sl and necessary} general forms!
For one thing, the considered terms can mix with other terms contributing to the {\sl same} $V_{M,c}[z, z^\dag, \Phi, \Phi^\dag]$. 
For another, the various superpotential functions $W_n$ are still arbitrary at this stage, and there is {\sl a priori} no mathematical 
reason that forbids cancellations from taking place among the various terms in each $V_{M,c}[z, z^\dag, \Phi, \Phi^\dag]$ 
that would lead to further possibilities consistent with (\ref{eq:the_condition}). 
This is precisely what the tower of equations (\ref{eq:masterequation}) encodes.\\

The main purpose of the present paper is to solve completely this tower of equations to clarifiy the seemingly dubious approach
of Ref.~\cite{SW}. This is presented in \ref{app:prelimin} and \ref{app:proof-flat} to which the reader is referred
for full details. Here we describe the procedure in a nutshell and summarize the main results of the analysis:  
starting from the expansion given in Eq.(\ref{eq:Vgen0}), each value of $c \geq 1$ implies a PDE, where the unknown function is $W_n$
 for a given $n$, and the $\Phi^a$'s the variables. 
To proceed we first solve the PDE corresponding to the highest power of $m_{p\ell}$, that is for $c=2 M$. This gives a general
solution for $W_M$. Using this solution we then solve the PDE corresponding to $c=2 M -1$ leading to a general solution for
$W_{M-1}$. These general solutions are not explicitly holomorphic in the $z^i$ and $\Phi^a$ fields. Moreover, the structure
of  $W_{M-1}$ requires distinguishing two types of $\Phi^a$ fields, dubbed $\widetilde{\Phi}^a$ and $S^p$. A detailed
and lengthy investigation of the holomorphy requirement leads then to the final general forms of $W_{M}$ and $W_{M-1}$, see 
Eqs.~(\ref{eq:finalformWM}, \ref{eq:finalformWM-1}). If the expansion in Eq.~(\ref{eq:Wexpansion})
is truncated at $M=1$, then there are no further PDEs to solve. If $M > 1$, examination of the PDEs corresponding 
to $c=2 M -2, 2 M-3, ...$ allows to determine the general forms of $W_{M-2}, W_{M-3}, ...$ but shows that the expansion in 
Eq.~(\ref{eq:Wexpansion}) should be truncated at $M=2$. Moreover, the general forms in the latter case
do not allow for fields of the $S^p$ type. We summarize the results as follows:

\begin{itemize}
\item if the series (\ref{eq:Wexpansion}) extends beyond $M=2$, with $W_n(z, \Phi)$ {\sl non vanishing} functions for $n \geq 3$, then 
there exist no solutions for Eq.~(\ref{eq:Wexpansion}) satisfying  (\ref{eq:the_condition}).
\item if the series (\ref{eq:Wexpansion}) is truncated at $M=2$, i.e. $W_n(z, \Phi)$ are possibly non vanishing only when $n < 3$, 
there exist two general solutions satisfying (\ref{eq:the_condition}),

(1) {\bf Soni-Weldon}: 
\begin{equation} 
W(\zeta, \Phi) = m_{p\ell}^2 W_2(z) + m_{p\ell} W_1(z) + W_0(z, \Phi), 
\label{eq:SW}
\end{equation}
where the $W_{1,2}$ are arbitrary holomorphic functions of the $z^i$ fields and $W_0$ an arbitrary holomorphic function of the $z^i$ and 
$\Phi^a$ fields, corresponding to the solution found in \cite{SW}. We will 
refer to this solution as SWS.

(2) {\bf Non-Soni-Weldon}: in this case $W_2 \equiv 0$ and the visible sector fields 
$\Phi$ should be split into two types denoted $\widetilde{\Phi}$ 
and $S$. Moreover, different solutions entail different partitions of the set of $S$ fields. 
The general superpotential reads, 
 \begin{equation} 
 W(\zeta, \Phi) = m_{p\ell} W_1(z,S) + W_0(z, S,  \widetilde{\Phi}),
\label{eq:Non-SWgen}
\end{equation}
with
\beqa
&W_1(z, S)= W_{1,0}(z) +  \sum_{p \geq 1}^P W_{1,p}(z) \sum_{s \geq 1}^{n_p}  \; \mu_{p_s}^* \,S^{p_s},~~~~~~~~~~~~~~~~& \label{eq:finalformWM0} \\
\text{and} \nonumber \\
&W_{0}(z, S, \widetilde{\Phi})= \sum_{q \geq 1}^{k_1} \! W_{0,q}(z) 
\, S^{q}  + \Xi(...,
{\cal U}^{p p_s}_S...;...,\widetilde{\Phi}^a,...; ...,z^i,...),~~~~& 
\label{eq:finalformWM-10}
\eeqa
where $W_{1,0}(z), W_{1,p}(z)$ and $W_{0,p}(z)$ are arbitrary holomorphic functions of the $z^i$ fields, 
$\Xi$ an arbitrary function of all its entries and holomorphic in the $z^i, \widetilde{\Phi}^a, S^q$ fields, 
and we define  
\beqa
{\cal U}^{p p_s}_{S} \equiv \xi_{p_s}(z) S^{p_s} - 
\xi^{p_s}(z) S^{p_1}, \; {\rm with}\; p=1,\dots,P, \; {\rm and} \; s=1, \dots, n_p, 
\label{eq:UdefFINAL0}
\eeqa
the two sets of functions $\xi^{p_s}(z)$ and $\xi_{p_s}(z)$ denoting arbitrary holomorphic functions of the $z^i$ fields satisfying
\beqa
\mu_{p_s}  \; \xi_{p_s}(z) = \mu_{p_1}  \; \xi^{p_s}(z) \ . \label{eq:xirelation0}
\eeqa
The $\mu_{p_s}$'s entering Eq.~(\ref{eq:xirelation0}), and their complex-conjugate entering Eq.~(\ref{eq:finalformWM0}), 
denote arbitrary {\sl nonvanishing} complex-valued $k_1$ constants,
$k_1$ being the total number
of $S$ fields, while $P$ and $n_p$ are arbitrary (positive) integers satisfying 
\beqa
k_1 = \sum_{p\geq 1}^P n_p \label{eq:partition0} \ .
\eeqa
The latter define a given partition of the set of $S$ fields into $P$ subsets labeled by $p=1,...,P$, where
$n_p$ denotes the number of elements in the $p^{th}$ set.
See also the discussion following Eq.~(\ref{eq:partition}) and Eq.~(\ref{eq:notation}). 
We will refer to this new  solution as non-Soni-Weldon (NSWS). It represents the central result of the paper.
\end{itemize}

\subsection{Comparing SWS and NSWS \label{sec:comparison}}
The two classes of solutions are different in various ways, and NSWS is clearly not a special case of SWS. 
We list here some of the salient features:
\begin{itemize}
\item[1.] The SWS confines the dependence on the observable sector fields $\Phi$ to the $m_{p\ell}$ independent component of the 
superpotential.   
In contrast, NSWS requires to make a distinction between two subsets of the observable sector, the $\widetilde{\Phi}$-type and 
the $S$-type fields.
\item[2.] Both $\widetilde{\Phi}$ and $S$ have the properties  of the observable sector in the sense that they satisfy (\ref{eq:the_condition}). 
\item[3.] The  $\widetilde{\Phi}$-type fields have a status similar to that of $\Phi$ in SWS, they enter
only the $m_{p\ell}$ independent component of the superpotential through the arbitrary function $\Xi$. All the conventional implementations
of the SUSY extensions of the SM in the observable sector (from \cite{ca,bfs}, \cite{Cremmer:1982vy}, onwards) that rely on
SWS, can thus be carried over unchanged to the $\widetilde{\Phi}$ sector of NSWS.
\item[4.] The presence of the $S$ fields in the NSWS case offers new possibilities.
A striking difference with SWS is that the $S$ fields, even though observable, appear in the term of the superpotential 
that is proportional to $m_{p\ell}$. This seemingly counter-intuitive result comes, however, at the price of
requiring linearity in the $S$-type in $W_1$, and in $W_0$, the arbitrary dependence {\sl exclusively through the combinations} 
${\cal U}^{p p_s}_{S}$ in $\Xi$, modulo an extra linear piece in the $S$-type. It should be clear that the set of all possible 
general functions of ${\cal U}^{p p_s}_{S}$ does not contain all the set of possible general functions of the $S$ fields. 
\item[5.] The above point is the key issue distinguishing the $\widetilde{\Phi}$-type from the $S$-type. It is easy to see that 
this difference is triggered by the nonvanishing $\mu_{p_s}$ constants; if a given $S^{p_s}$ field is functionally absent from $W_1$, 
that is if $\mu_{p_s}=0$, then the corresponding ${\cal U}^{p p_s}_{S}$ reduces to $\xi_{p_s}(z) S^{p_s}$ as a consequence of
Eq.~(\ref{eq:xirelation0}). The function $\Xi$ becomes arbitrary in $S^{p_s}$ so that the latter field has the same status
as the $\widetilde{\Phi}$-type fields and becomes part of the latter set. On the other hand, a given $S^{p_s}$ can
be present in $W_1$ and absent from $\Xi$ and still remain part of the $S$-type set. Indeed this would correspond to a choice of
the partition in $P$ subsets with $\{S^{p_s}\}$ being one element of this partition. 
\item[6.] An important peculiarity of the NSWS
is that direct couplings with arbitrary strength between the  $\widetilde{\Phi}$ and $S$ sectors (as well as within 
the $S$ sector itself) in the superpotential, 
are allowed only if there exist {\sl two or more} different $S$ fields in the theory. This is again a 
direct consequence of the fact that the $S$ dependence in $\Xi$ enters exclusively through the ${\cal U}^{p p_s}_{S}$ combinations.
\item[7.] With at least two $S$-type fields in the theory, the $\Xi$ part of the superpotential will always contribute F-flat 
directions to the potential, defined by  ${\cal U}^{p p_s}_{S}=0$. These directions are however generically lifted by the mandatory 
linear $S$ dependence in $W_1$, and possibly the linear dependence in $W_0$. \gilbertzero{We come back to this point at the end of Section \ref{sec:simple-NSWS}.}
\item[8.] If, in contrast, there is only one single $S$-type field in the theory, then it appears only linearly in the 
superpotential and, as we show in Section  \ref{sec:simple-NSWS}, has Planck mass suppressed couplings to both the hidden and 
the remaining part of the observable sectors.
\item[9.] Finally, note that there is only one configuration where
both SWS and NSWS reduce to the same form: if  $\mu_{p_s}=0$, $\forall p_s$, in
Eq.~(\ref{eq:finalformWM0}) then the relation (\ref{eq:xirelation0}) becomes trivial and $\Xi$ an arbitrary function of the $S$
fields. The fields $\widetilde{\Phi}$ and $S$ are then formally indistinguishable and $W_0(z, S, \widetilde{\Phi})$ corresponds
to $W_0(z, \Phi)$ of SWS. The two classes of solutions become then equivalent if 
$W_2$ in Eq.~(\ref{eq:SW}) is a vanishing function of $z$. 
\end{itemize}

We will see in Section \ref{sec:model-pheno} how the above formal properties
come into play in relation with model building. 

\subsection{The general K\"ahler case \label{sec:NSW-non-flat}}
This section is somewhat outside the scope of the present paper. It aims
at pointing out that the existence of new forms as found in Section \ref{sec:NSW-flat}
is not a peculiarity of the minimal K\"ahler assumption Eq.~(\ref{eq:Kf}). 


We exhibit here two non-Soni-Weldon solutions in the
case of non minimal K\"ahler. Starting from the general superpotential and
K\"ahler potential with $M=N=2$ in Eqs.~(\ref{eq:Kexpansion}, \ref{eq:Wexpansion}), 

\beqa
K(z,z^\dag,\Phi,\Phi^\dag) = m_{p\ell}^2 K_2(z,z^\dag,\Phi,\Phi^\dag) + m_{p\ell} K_1(z,z^\dag,\Phi,\Phi^\dag) + K_0(z,z^\dag,\Phi,\Phi^\dag), \ \\
W(z, \Phi) =m_{p\ell}^2  W_2(z,\Phi)+ m_{p\ell} W_1(z,\Phi) + W_0(z,\Phi),~~~~~~~~~~~~~~~~~~~~~~~~~~\, 
\eeqa
we find that the consistency requirement (\ref{eq:the_condition}) is satisfied if:
\begin{itemize}
\item $W_1,K_1$ are {\sl vanishing} functions,
\item $W_0,K_0$ are {\sl arbitrary} functions,
\item $W_2,K_2$ are arbitrary functions that should depend {\sl explicitly} on $\Phi, \Phi^\dag$, subject to a no-scale-like condition
\cite{cfkn,ekn,sm-no-scale,ln}
\beqa
\partial_I {\cal G}_2 \Bigg(\frac{\partial^2 {\cal G}_2}{\partial Z^I \partial Z^{J^*}}\Bigg)^{-1}
\partial_{J^*} {\cal G}_2 =3 \ , \ \ \text{with} \ \ 
{\cal G}_2 = K_2+ \ln \Big|\frac{W_2}{m_{p\ell}}\Big|^2 \ ,
\eeqa
\end{itemize}
or alternatively if:
\begin{itemize}
\item $W_2$ is a {\sl vanishing} function,
\item $W_0,W_1, K_0, K_1, K_2$ are {\sl arbitrary} functions that should depend {\sl explicitly} on $\Phi, \Phi^\dag$.
\end{itemize}
We stress the difference with the general form given in \cite{SW} where the functions $K_1, K_2, W_1, W_2$,
although arbitrary, had to be {\sl independent} of the observable fields, $\Phi, \Phi^\dag$, 
while for instance in the first solution given above $K_2$ and $W_2$ must depend on these fields provided that
$K_1$ and $W_1$ are vanishing. See \ref{app:non-flat} for more details.

\section{Towards model building and phenomenology from the NSWS
\label{sec:model-pheno}}
We return here to the NSWS in the minimal K\"ahler case to discuss possible realizations of
the role the $S$-type sector can play in model building. 

\subsection{General considerations}
\label{sec:model-pheno-1}
This section is devoted to a non-exhaustive broad-brush discussion of the various possibilities.
So far we have ignored features related to gauge or global symmetries.  As already stated in Section \ref{sec:comparison}, 
the $\widetilde{\Phi}^a$ fields of the NSWS
can play the same role as the $\Phi^a$ fields of the SWS for model building. The $\Xi$ function can thus contain the superpotential 
of the MSSM or any other SUSY extension of the SM by implementing the corresponding fields and their quantum numbers in the 
$\widetilde{\Phi}^a$ set. Can the $S$ fields play the same role? There is an obstruction because of the linear
terms in $S$ in $W_1$: since the hidden sector fields are assumed to be neutral under the SM gauge groups, all the $W_{1, p}$ are singlets,
thus all the $S$ fields must be singlets under these same groups so as to ensure the invariance of $W_1$. 
{\sl The $S$-type sector is thus particularly
suited to non-minimal SUSY extensions with SM singlet fields.} The existence of the NSWS can even be seen as a motivation for
such extensions. The next to minimal MSSM (NMSSM) \cite{Fayet:1974pd,Fayet:1977yc} is the first example that comes to mind. 
However, as stated in point 6. 
of Section (\ref{sec:comparison}), a realization in the $S$ sector should have at least two different gauge singlet fields, and thus 
corresponds to extensions of the NMSSM itself. Such extra singlets could be welcome for various reasons (see e.g. \cite{Ellwanger:2009dp}
for a review). But it should be kept in mind that their embedding as $S$-type fields is physically not equivalent to their 
conventional embedding as $\widetilde{\Phi}$-type fields. Among the usual terms of the NMSSM Higgs superpotential,
\beqa
\lambda S H_u \cdot H_d, \; \xi_F S, \; \frac12 \mu' S^2, \; \frac13 \kappa S^3,
\eeqa 
(we use the notation of \cite{Ellwanger:2009dp}), 
only the ``tadpole'' can preserve its form, while in all the others 
the field $S$ should be replaced by, at least, one combination
of the form ${\cal U}_S^{a b}$:
\beqa
\lambda {\cal U}_S^{a b} H_u \cdot H_d, \; \xi_F^a S^a + 
\xi_F^b S^b, \; \frac12 \mu' [{\cal U}_S^{a b}]^2, \; 
\frac13 \kappa [{\cal U}_S^{a b}]^3 \ ,  \label{eq:extNMSSM}
\eeqa
where $S^a$ and $S^b$ denote two independent $S$-type fields.
As can be seen from
Eqs.~(\ref{eq:Non-SWgen}, \ref{eq:finalformWM-10}) the tadpole 
parameters $\xi_F^a, \xi_F^b$ have now their origin
in the hidden sector of Supergravity. They are given respectively
by $\langle m_{p\ell} W_{1,a} ( z ) + W_{0,a} (z) \rangle$ and 
$\langle m_{p\ell} \mu_b^* W_{1,a} (z) + W_{0,b} (z) \rangle$ 
at low energies 
where the hidden sector degrees of freedom are frozen at their
VEVs and we assumed the normalisation $\mu_a=1$. Similarly,  
${\cal U}_S^{a b} =\xi_{b}(\langle z \rangle )  (S^{b} - 
\mu_b S^{a})$, cf.  
Eq.~(\ref{eq:UdefFINAL0}) where we have used Eq.~(\ref{eq:xirelation0}) and the same normalisation.   
More generally, perturbing around $\langle z \rangle$ one has a definite structure for the interaction between the hidden and the
observable sector through the singlet fields in this extended version of the NMSSM.
Another issue is related to the so-called `tadpole problem'. In the NSWS case the two parameters
$\xi_F^a, \xi_F^b$ have contributions proportionnal to the Planck mass. However by construction
in the NSWS such a term should vanish in the potential. \\

Another possibility for model building is to consider $S$-type fields that are non-singlets under symmetries involving
the hidden sector while remaining SM singlets. For instance a subset of these fields could be charged under 
hidden abelian or non-abelian gauge groups. If direct couplings of such fields to the observable sector
are forbidden through the $\Xi$ function then whatever physics occuring in the hidden sector would be communicated to the visible 
sector only by gravitionally suppressed effects. This would be necessarily the case if there is only one single $S$-type multiplet
charged under a given hidden gauge group $\mathfrak{G}$. Indeed in this case the full set of $S^q$ fields spans the components of a 
multiplet $\mathbf S$ of a given representation of $\mathfrak{G}$. In order for the term 
$\displaystyle \sum_{p \geq 1}^P W_{1,p}(z) \sum_{s \geq 1}^{n_p}  \; \mu_{p_s}^*\,S^{p_s}$
in Eq.~(\ref{eq:finalformWM-10}) to be gauge invariant, $W_{1,p}(z)$ should be the $p^{\text{th}}$ component of the conjugate
representation and simultaneously the sum $\displaystyle \sum_{s \geq 1}^{n_p}$ should reduce to only one term, i.e. $n_p=1$ for each $p$.
This corresponds to the singleton partition of the set of $S^q$ fields made of the $k_1$ subsets $\{S^q\}$ with $k_1=P$ being equal to the 
dimension of the considered group representation. It follows that all ${\cal U}^{p p_s}_{S}$ are vanishing and thus no direct
coupling can be constructed between the $S$-type and $\widetilde{\Phi}$-type fields in $\Xi$, (see also item five of Section (\ref
{sec:comparison})). Note that the same conclusion holds trivially when $\mathfrak{G}$ is abelian, since in this case to have 
a single multiplet means to have one single $S$ field. 

Direct coupling between the $S$-type and $\widetilde{\Phi}$-type fields requires at least two multiplets $\mathbf{S}_1$, $\mathbf{S}_2$
belonging to the {\sl same} group representation. The gauge invariance of the sum just discussed enforces a unique choice for the
 partition of the full set of $S$ fields. Each element of the partition should now be of the form $\{ S_1^r, S_2^r\}$, the label $r$
 spanning the components of the representation under consideration. With the correspondence $p_1 \to (r,1), p_{n_p} \to (r,2)$,
the sum reads 
\beqa
\displaystyle \sum_{r \geq 1}^{N} W_{1,r}(z) \; ( \mu_{(r,1)}^* \,S^{r}_1 + \mu_{(r,2)}^* \,S^{r}_2) \ , \label{eq:thesum}
\eeqa
where $N =k_1/2$
is the dimension of the representation. 
One can also construct the combinations ${\cal U}_S$, cf. Eqs.~(\ref{eq:UdefFINAL0}, \ref{eq:xirelation0}) to write, adapting
slightly the notation, 
\beqa
{\cal U}_{S_{1 2}}^{r} =\xi_{r}(z) \; (\mu_{(r,1)} \, S_2^{r} - \mu_{(r,2)} S_1^{r}) \ .
\eeqa
It is noteworthy that in general the constant factors $\mu_{(r,1)}, \mu_{(r,2)}$ cannot be absorbed in a redefinition of
the $S$ fields as they multiply them differently in the above expressions. Gauge invariance, however, requires them
to be equal. Thus choosing  $\mu_{(r,1)} = \mu_{(r,2)}$ gives now the freedom to reabsorb this common factor
in a redefinition of the functions $W_{1,r}(z)$ and $\xi_{r}(z)$.
Requiring furthermore $W_{1,r}(z)$ and $\xi_{r}(z)$ to be the components of
the conjugate representations, the term in Eq.~(\ref{eq:thesum}) becomes gauge invariant,
as well as the sum of the ${\cal U}_{S}$ combinations
\beqa
{\cal U}_{S_{1 2}} \equiv \sum_{r \geq 1}^{N} {\cal U}_{S_{1 2}}^{r} \ .
\eeqa
The latter constitutes an interesting building block for the construction of new extensions.
For instance all the terms listed in Eq.~(\ref{eq:extNMSSM}) can be simply generalized by
the replacements
\beqa
{\cal U}_S^{a b} &\to& {\cal U}_{S_{1 2}} \ , \nonumber \\
\xi_F^a S^a + \xi_F^b S^b &\to&  \sum_{r \geq 1}^{N} (\xi_{F_1}^r(z) S^r_1 + \xi_{F_2}^r(z) S^r_2) \ ,
\eeqa
where
\beqa
\xi_{F_1}^r(z) = m_{p\ell} W_{1,r} (z) + W^{(1)}_{0,r} (z) \ , \;
\xi_{F_2}^r(z) = m_{p\ell} W_{1,r} (z) + W^{(2)}_{0,r} (z) \ .
\eeqa 
Recall that the arbitrary functions $W^{(1)}_{0,r} (z), W^{(2)}_{0,r} (z)$, that correspond to 
the functions $W_{0,q} (z)$ of Eq.~(\ref{eq:finalformWM-10}), are not mandatory for the $S$-type
characterization, and can be chosen vanishing contrary to the $W_{1,r} (z)$ functions. The obtained
terms in the superpotential provide an unconventional general structure for the interaction between
the hidden and visible sectors in the realm of singlet extensions of the MSSM that complies with
the general consistency requirement for low energy Supergravity. 
Moreover, when the hidden sector fields acquire VEVs the replacement $z \to \langle z \rangle$
generates dynamically  the low energy superpotential terms of the MSSM singlet extensions that
are invariant under the SM gauge groups and break spontaneously the hidden gauge group $\mathfrak{G}$. 
From this perspective the NMSSM extension corresponding to the terms given in Eq.~(\ref{eq:extNMSSM}) can
be interpreted as a special case of a spontaneously broken $U(1)$ factor in $\mathfrak{G}$.

Another interesting possibility arises when considering secluded sectors such as in gauge mediation \cite{fa-gmsb,cans1,cans2,dersav,df,no,acw,dn1,dn2,dn3,gr}, or dark sectors
\cite{ArkaniHamed:2008qp}. These sectors are still considered as {\sl observable} with respect to the hidden sector
of Supergravity, but are not directly coupled to the SM observable sector. Some of the $S$-type fields can now be charged under the 
symmetry groups of these sectors, thus linking them naturally to the hidden sector at least through the mandatory linear terms
in the superpotential Eq.~(\ref{eq:finalformWM0}).  
Moreover, in contrast with the case of the observable sector, some of the $S$-type fields can now play the role of the matter superfields 
in these sectors. For instance, in a dark copy of the SM the NSWS structure of the superpotential $\Xi$ 
predicts the existence of at least two `quark' and `lepton' families. Also a dark or secluded Yukawa sector should necessarily be of the 
form
\beqa
({\cal U}_{S}^{a b})_L \cdot H \,({\cal U}_{S}^{a b})_R,
\eeqa
where $({\cal U}_{S}^{a b})_{L,R}$ denote doublets and singlets of the dark $SU(2)$, 
thus predicting naturally nonzero off-diagonal Yukawa couplings.

\subsection{Supergravity mediated SUSY breaking}
  \label{sec:model-pheno-2}
  In this section we compare explicitly the usual gravity mediation SUSY breaking assuming
  the superpotential to be a direct sum \cite{ca,bfs} of the hidden and visible sectors functions in the SWS case,
  to the gravity mediation in the NSWS case with a similar direct sum assumption. 
  The first important general difference
  lies in the fact that the conventional SWS allows a nonzero $W_2$ term in the superpotential, 
  while the NSWS {\sl forbids} it allowing at most a $W_1$ term. 
  This implies typically that after SUSY breaking the gravitino mass Eq.~(\ref{eq:mgrav}) is suppressed
  by a factor $M/m_{p\ell}$ in the NSWS case as compared to the SWS case, where $M (< m_{p\ell})$ denotes the largest
  mass scale other than the Planck scale present in the theory, e.g. a GUT scale.
  Another important difference comes from the structure of the ensuing SUSY breaking parameters. 

\subsubsection{The SWS case \label{subsec:SWS}}  
  Let us first recall how one arrives at the well-known structure of the potential when positing a direct 
  sum superpotential in the SWS case. We start from
  
\beqa
W(\zeta, \Phi) = h(\zeta) + g(\Phi),
\eeqa
corresponding to Eq.~(\ref{eq:SW}) with
\beqa
W_2(z) = m_{p\ell}^{-2} h(\zeta) \equiv M_2 \omega_2(z) , \; W_1(z)=0 , \; \text{and} \; W_0(z, \Phi) = g(\Phi) \ ,
\label{eq:W2non0}
\eeqa
where $M_2$ is a mass scale much smaller than $m_{p\ell}$. 
Assuming a minimal K\"ahler, the $F$-term contribution to the potential, Eq.~(\ref{eq:V}), now reads 
\beqa
V_F= e^{\frac{|\zeta^i|^2 + |\Phi^a|^2}{m_{p\ell}^2 }} \Big ( |h_i + \frac{\zeta^{i *}}{m_{p\ell}^2} W|^2 + |g_a + \frac{\Phi^{a *}}{m_{p\ell}^2} W|^2 -  \frac3{m_{p\ell}^2} |W|^2\Big) , \label{eq:VFDS}
\eeqa
with the notation as in Eq.~(\ref{eq:WI}). 
When the hidden sector fields $\zeta^i$ acquire a VEV the gravitino mass Eq.~(\ref{eq:mgrav}) is expressed as
\beqa
\label{eq:gravSWDS}
m_{3/2} = \frac1{m_{p\ell}^2}\Big \langle |h(\zeta)| e^{\frac12 \frac{|\zeta^i|^2}{m_{p\ell}^2 }} \Big \rangle 
=M e^{\frac12 \langle |z^i|^2 \rangle } \ ,  
\eeqa
where we define $M \equiv M_2 \langle \omega_2(z) \rangle$.
To obtain the low energy form of the potential $V_F$ in the flat SUSY limit 
we make the substitution $z \to \langle z \rangle$ and retain the leading terms in the
flat 
limit $m_{p\ell} \to \infty$ with $m_{3/2}$ fixed. 
Adopting slightly modified notations from the conventional ones \cite{Nilles:1983ge}, we write
\beqa
&&\langle \zeta^i \rangle = m_{p\ell} \langle z^i \rangle  \equiv m_{p\ell}  b_i \ , \label{eq:VEVzeta} \\
&&\langle h \rangle \equiv M m_{p\ell}^2 \ , \; 
\langle h_i \rangle = \langle \frac{\partial h}{\partial \zeta^i} \rangle  \equiv a_i^* M m_{p\ell} \ . \label{eq:VEVhSWS}
\eeqa
Plugging these back in Eq.~(\ref{eq:VFDS}) and keeping the leading operators, we retrieve the well-known low energy softly broken 
SUSY potential \cite{ca,bfs}, \cite{Nilles:1983ge}, with SUSY breaking mass terms for the $\Phi^a$ fields and couplings:
\beqa
V_{\text{LE}}^{\text{SWS}}=  \Big|\frac{\partial \widehat{g(\Phi)}}{\partial \Phi^{a}}\Big|^2 + m_{3/2}^2 |\Phi^{a}|^2 +  
m_{3/2} \Big( \, ( A-3 ) \; \widehat{g(\Phi)} 
 + \Phi^{a} \frac{\partial \widehat{g(\Phi)}}{\partial \Phi^{a}}  + \text{h.c.}\Big) \nonumber \\
 + {\cal O}(m_{p\ell}^{-2}) \ ,
 \label{eq:VDSSWS}
\eeqa
where
\beqa
A \equiv \sum_i b_i^* (a_i + b_i), \label{eq:Adef} 
\eeqa
 a factor $\exp \frac12 |b_i|^2$ has been absorbed in the hatted $g$, 
and we have \gilbertzero{omitted} an additive constant with positive powers of $m_{p\ell}$, originating from the VEVed purely hidden sector (\gilbertzero{see footnote \ref{footnote:x}}).
Recall that choosing a direct product potential \cite{Cremmer:1982vy} would have given a similar result albeit with the special
relation $A=3$.  

\subsubsection{The NSWS case \label{subsec:NSWS}}
To compare with the NSWS case, we assume for the latter 
\gilbertzero{a minimal K\"ahler potential,} 
\beqa
K(z,S,\widetilde{\Phi}) = m_{p\ell}^2 z^i z^{i *} + S^p S^{p *} +  \widetilde{\Phi}^a \widetilde{\Phi}^{a *} , \label{eq:K-Non-SW}
\eeqa
\gilbertzero{where we distinguish explicitly the $S^p$ fields, }
and the following direct sum superpotential:
\beqa
W(\zeta, S, \widetilde{\Phi}) = h(\zeta^i, S^q) + g(S^q, \widetilde{\Phi}^a), \label{eq:spotDS}
\eeqa
with
\beqa
h(\zeta^i, S^q) = m_{p\ell} \big(W_{1,0}(z^i) +  W_{1,1}(z^i) \sum_{q \geq 1}^{k_1}  \mu_{q}^* \,S^{q} \big) \label{eq:hDSNSWS} \ ,
\eeqa
and
\beqa
g(S^q, \widetilde{\Phi}^a) = \Xi(..., \mu_1 S^q - \mu_q S^1,..., \widetilde{\Phi}^a,...). \label{eq:gDSNSWS}
\eeqa
This corresponds to choosing the trivial partition in the $S$-type sector, i.e.
$W_{1,q}(z) := W_{1,1}(z), \, \forall q$,  and we take for simplicity $W_{0, q}(z) :=0$. 
Note that by construction the NSWS solutions do not allow the separation of the $S$-type fields from the hidden
sector fields in the form of a direct sum. By taking
$\Xi$ independent of the hidden sector our choice is the closest possible to such a separation. (However
a separation in the form of a direct product is possible and comes naturally with the present choice of singleton 
partition, provided that $W_{1,0}(z):=0$ and $\Xi$ is appropriately chosen as a direct product.)

Defining
\beqa
W_{1,0}(z) \equiv M_{1 0}^2 \omega_{1 0}(z), \;  W_{1,1}(z) \equiv M_{1 1} \omega_{1 1}(z),
\eeqa
where, as in the previous case, $M_{1 0}$ and  $M_{1 1}$ denote some physical mass scales much smaller than $m_{p\ell}$,
and
\beqa
M\equiv \big|M_{1 0}^2 \langle \omega_{1 0}(z) \rangle + M_{1 1} \langle \omega_{ 1 1} (z) \rangle \sum_q \mu_{q}^* \, 
\langle S^{q} \rangle \big|^\frac12 \ ,
\eeqa
the gravitino mass is given by
\beqa
\label{eq:gravNSWDS}
m_{3/2} = \frac1{m_{p\ell}^2}\Big \langle \left|h(\zeta, S)\right| e^{\frac12 \frac{|\zeta^i|^2}{m_{p\ell}^2 }} \Big \rangle 
= \frac{M^2}{m_{p\ell}} e^{\frac12 |b_i|^2 } \ ,  
\eeqa
where we neglected the VEV of the $S$ fields in the exponential, 
\gilbertzero{as well as possible contributions from the visible sector
Eq.~(\ref{eq:gDSNSWS}) since they are typically suppressed by 
one power of $m_{p\ell}$ as compared to the hidden sector
contribution.}\footnote{\label{foot:b} \gilbertzero{This 
approximation is however not essential here 
and depends on the actual magnitudes of the physical scales involved.
We come back to this point in section \ref{sec:hardbreaking}}.} 
(To keep the discussion at the generic level, we do not consider possible unabsorbable complex phases in the various VEVs.)
It is noteworthy that in the present case the mass scale $M$ 
can get contributions from the VEV of the $S$-type fields, which can even be the leading
effects, depending on the relative magnitudes of the mass scales $M_{1 0}, M_{1 1}$ and the dimensionless numbers 
$\langle \omega_{1 0}(z) \rangle, \langle \omega_{1 1}(z) \rangle$. One can thus consider model settings where the $S$ fields
trigger the gravity mediation of SUSY breaking to the visible sector, even though those fields are not part of the hidden sector.
More generally, when the mass scale $M \ll m_{p\ell}$ has a comparable magnitude to that of the SWS case, 
the gravitino mass is expected to be typically smaller due to the Planck scale suppression in 
Eq.~(\ref{eq:gravNSWDS}) as compared to Eq.~(\ref{eq:gravSWDS}). To proceed, let us define
\beqa
&&\langle h \rangle \equiv M^2 m_{p\ell}, \; 
\langle \frac{\partial h}{\partial \zeta^i} \rangle =\langle h_i \rangle \equiv a_i^* M^2 , \label{eq:VEVhNSWS}\\
&&\langle \frac{\partial h}{\partial S^q} \rangle =\langle h_q \rangle=m_{p\ell} M_{1 1} \langle \omega_{1 1}(z) \rangle \mu_q^* 
\equiv a_q^*  M m_{p\ell} . \label{eq:VEVhSNSWS}
\eeqa  
Note the different mass powers in $\langle h \rangle$ and $\langle h_i \rangle$ between Eq.~(\ref{eq:VEVhSWS}) and Eq.~(\ref{eq:VEVhNSWS}).
We can now consider the low energy potential $V_F$ in the SUSY flat limit by taking as previously the limit
$m_{p\ell} \to \infty$ with fixed $m_{3/2}$ and making the substitution $\zeta \to \langle \zeta \rangle$ to which we add the substitution
$S^q \to S^q + \langle S^q \rangle$ thus allowing for nonvanishing $S$-fields VEVs. 
The latter shift leads to the relation
\beqa
h_i = \langle h_i \rangle + M_{1 1} \langle \frac{ \partial\omega_{1 1}(z)}{\partial z^i} \rangle \sum_q \mu_q^* S^{q} \equiv
a_i^* M^2 + a'^*_{i} M \sum_q \mu_q^* S^q \label{eq:hSHIFTNSWS} ,
\eeqa
which also defines the parameters $a'^*_{i}$. 
Plugging back in Eq.~(\ref{eq:VFDS}) the various
pieces of the superpotential Eqs.~(\ref{eq:spotDS}, \ref{eq:hDSNSWS}, 
\ref{eq:gDSNSWS}) and their derivatives, adding the $S$-fields contribution 
$|\frac{\partial W}{\partial S^q} + \frac{S^{q *}}{m_{p\ell}^2} W|^2$, and taking into account  
Eqs.~(\ref{eq:VEVzeta}, \ref{eq:VEVhNSWS}, \ref{eq:VEVhSNSWS}, \ref{eq:hSHIFTNSWS}), we find (a sum over all indices being understood),
\beqa
V_{\text{LE}}^{\text{NSWS}}&=& \Big|\frac{\partial \widehat{\Xi}}{\partial \widetilde{\Phi}^a}\Big|^2 + 
\Big|\frac{\partial \widehat{\Xi}}{\partial S^q}\Big|^2
+ m_{3/2}^2 \Big( |\widetilde{\Phi}^a|^2 + |S^q + \langle S^q \rangle|^2 \Big) |1 + A^{(S)}|^2 \label{eq:1stline}\\
&& + \; m_{3/2} \Big( \, \big( A -3 + \langle A^{(S)} \rangle  + (|b_i|^2 -2) A^{(S)} + b^*_i A'^{(S)}_i \big)\; \widehat{\Xi} 
\label{eq:2ndline} \\
&&+   ( 1 + A^{(S)}) \, \widetilde{\Phi}^a \frac{\partial \widehat{\Xi}}{\partial \widetilde{\Phi}^a}  
+   ( 1 + A^{(S)}) \, (S^q + \langle S^{q} \rangle) \frac{\partial \widehat{\Xi}}{\partial S^q}    
+ \text{h.c.} \Big) \label{eq:3rdline}\\
&& + \; e^{|b_i|^2} M^2 {\cal A}_{q r} S^q S^{r *}  + e^{|b_i|^2} M^3 \Big(\big( (A + \langle A^{(S)} \rangle -2 ) a_q^* + A' \mu^*_q \big)S^{q} + \text{h.c.}  \Big)  \nonumber \\ 
&&+ {\cal O}(m_{p\ell}^{-2}) \ , \label{eq:4thline}
\label{eq:VDSNSWS}
\eeqa
where we have again omitted additive constants to the potential originating from the VEVed hidden sector or
from VEVs of the coupled hidden and $S$ sectors  \footnote{\gilbertzero{\label{footnote:x}Here we are only interested in the structure of the SUSY breaking terms. A thourough treatment 
would have required a detailed study of the vacuum structure and an assessment of an almost vanishing cosmological constant, as well as the consistency requirement $\langle S \rangle \ll
{\cal O}(m_{p\ell})$. This will be carried out in some detail in a simpler setting in Section \ref{sec:simple-NSWS}.}}, and defined $A$ as in Eq.~(\ref{eq:Adef}) and 
\beqa
A' &\equiv& \sum_i a'^*_{i} (a_i + b_i), \\
A^{(S)} &\equiv& \frac1M \sum_q a_q S^{q *} = \frac{M_{1 1}}{M^2} \langle \omega_{1 1}(z) \rangle^* \sum_q \mu_q S^{q *} \label{eq:ASdef},   \\
A'^{(S)}_i &\equiv& \frac1M a'_i \sum_q \mu_q S^{q *} = \frac{M_{1 1}}{M^2} \langle \frac{ \partial\omega_{1 1}(z)}{\partial z^i} \rangle^*
\sum_q \mu_q S^{q *}, \label{eq:ASprimedef} \\
{\cal A}_{q r} &\equiv& \sum_i (b_i a_q + a'_i \mu_q)^* (b_i a_r + a'_i \mu_r) - a^*_q a_r \ .
\eeqa
 A few comments are in order here. Note first that dangerous Planck enhanced contributions
involving the $S$ fields and originating from the crossed terms in $|\frac{\partial W}{\partial S^q} + \frac{S^{q *}}{m_{p\ell}^2} W|^2$ have, as 
expected, canceled out as a result of the interplay between the specific forms of the $S$-dependent terms in Eqs.~(\ref{eq:hDSNSWS}, 
\ref{eq:gDSNSWS}). This confirms that the $S$-type fields are part of the visible sector \gilbertzero{in the sense of (\ref{eq:the_condition})}.
 We also retrieve the usual SUSY 
preserving contributions, the first two terms of Eq.~(\ref{eq:1stline}). Second, the soft SUSY breaking contribution
to the masses of the $\widetilde{\Phi}$-type and $S$-type fields, proportional to $m_{3/2}^2$ in Eq.~(\ref{eq:1stline}), are the same
as in $V_{\text{LE}}^{\text{SWS}}$, Eq.~(\ref{eq:VDSSWS}), but there are new interaction terms with the $S$-type fields due to
the presence of $A^{(S)}$ \gilbertzero{that we will discuss separately at the end of this subsection}.  
Third, the SUSY breaking terms proportional to $m_{3/2}$ have now a much richer structure in comparison with 
$V_{\text{LE}}^{\text{SWS}}$:
apart from the presence of new interaction terms with the $S$ fields \gilbertzero{that we will also discuss separately at the end of this subsection}, 
one should note the new contribution $\langle A^{(S)} \rangle$ 
to the \gilbertzero{soft breaking}   
$(A-3)$ parameter. This leads to a novel aspect since the magnitude of the soft SUSY breaking parameter in the conventional visible
sector, that is involving the $\widetilde{\Phi}$ fields, can now depend on the dynamics
of a visible sector field! Moreover, since $V_{\text{LE}}^{\text{NSWS}}$, like $V_{\text{LE}}^{\text{SWS}}$, is supposed to be generated
at a high scale where SUSY is broken in the hidden sector, this scale does not need to match to the one where the $S$-fields
develop a VEV, the latter being presumably lower. It follows that the running of the conventional soft SUSY breaking parameters
can be modified by a different initial condition at the scale where the VEV of the $S$ fields sets in. This is not to be confused
with the usual threshold effects due to the decoupling of heavy states in the running quantities, and obviously not with gauge mediated 
SUSY breaking scenarios-like effects since the mediation is only gravitational. A more drastic scenario obtains if, as mentioned above, 
the gravitino mass itself is triggered by the VEVs of the $S$-fields. For instance if $|M_{1 0}^2 \langle \omega_{1 0}(z)
 \rangle| \ll  |M_{1 1} \langle \omega_{ 1 1} (z) \rangle \sum \mu_{q}^* \, 
\langle S^{q} \rangle|$ then the bulk of the gravity mediation SUSY breaking to the observable sector occurs not at the scale where SUSY 
is broken but at the scale where the $S$-fields acquire a VEV. Moreover, in this case $\langle A^{(S)} \rangle \approx {\cal O}(1)$ 
leading to a sensible modification of $(A-3)$.  

The bilinear and linear contributions in the $S$ fields of the last line, Eq.~(\ref{eq:4thline}), are unusual. 
They are formally not enhanced by powers of
$m_{p\ell}$ as expected for the $S$-fields sector. However they can be large since $M \sim {\cal O}( (m_{3/2} m_{p\ell})^{1/2})$ as seen
from Eq.~(\ref{eq:gravNSWDS}), so that on general grounds $M$ is related to the SUSY breaking scale 
$M_S= (\sqrt{3} \, m_{3/2} m_{p\ell})^{1/2}$ up to some numerical factors. They can also play an important role in triggering the
magnitudes of $\langle S^p \rangle$. These VEVs are supposed to remain consistently much below $m_{p\ell}$ which is not in 
general dynamically guaranteed. We come back to this last issue in Section \ref{sec:simple-NSWS} for the simplest NSWS case.
\gilbertzero{We note also in passing that in both SWS and NSWS examples we assumed  for simplicity $\langle h \rangle$ to be real-valued, 
cf. Eqs.~(\ref{eq:VEVhSWS}, \ref{eq:VEVhNSWS}). In the case where $\langle h \rangle$ develops a phase, there is in the SWS the possibility
to choose $\Xi$ in such a way that this phase is not physical and can be absorbed in a redefinition of the fields in the terms 
proportional to $m_{3/2}$ in Eq.~(\ref{eq:VDSSWS}). This contrasts with the NSWS case where the presence of terms linear in $S$ 
forbids such a redefinition.}

\vspace{.3cm}
\noindent
\gilbertzero{\underline{\sl Hard breaking terms}: The terms containing $A^{(S)}$ in Eqs.~(\ref{eq:1stline}, \ref{eq:2ndline}, \ref{eq:3rdline}) as well as the term containing $A'^{(S)}_i$ in
Eq.~(\ref{eq:2ndline}) are not of the soft breaking type 
due to their dependence on $S^{q *}$ \cite{Girardello:1981wz}. 
We will see that they generically lead to {\sl hard} breaking. 
The presence of such 
{\sl hard} SUSY breaking terms in 
$V_{\text{LE}}^{\text{NSWS}}$ is an uncommon feature. It is 
important to identify the underlying reason for the appearance 
of these terms as it is usually taken for granted that gravity 
mediated scenarios lead only to soft breaking at low energy.
In fact hard breaking terms are present in the standard scenarios and
can be traced back to the presence nonrenormalizable 
operators in the Supergravity potential
that are suppressed by negative powers of $m_{p\ell}$.\footnote{ 
\gilbertzero{Given that $N=1$ Supergravity breaking can be recast
in terms of a global supersymmetric nonrenormalizable theory
 (see for instance chapter 8 of \cite{Gates:1983nr}), it is in principle possible to classify the hard breaking terms using 
 supergraph techniques\cite{Grisaru:1979wc} as was done in \cite{Girardello:1981wz}, by allowing spurion couplings beyond power
 counting renormalizability criteria.}} One can easily see this in the simple setting of Section \ref{subsec:SWS}; take for example the second term
in Eq.~(\ref{eq:VFDS}),  $\exp\left({\frac{|\zeta^i|^2 + |\Phi^a|^2}{m_{p\ell}^2 }}\right) |g_a + \frac{\Phi^{a *}}{m_{p\ell}^2} W|^2$. It leads after SUSY breaking to the term 
$(m_{3/2}/m_{p\ell}^2) \Phi^{a *} \Phi^{a} \widehat{ g(\Phi)} + h.c.$
in the potential, which is clearly \gilbertzero{not of the soft breaking type}\cite{Girardello:1981wz} even when
$g(\Phi)$ contains only  renormalizable operators, unless
$g$ is a trivial constant, $g(\Phi) \sim m^3$ with $m$ some mass 
scale. 
Note that the same term is obtained
if instead of Eq.~(\ref{eq:W2non0}) we choose $W_2(z)=0$ and
$W_1(z) = m_{p\ell}^{-1} h(\zeta)$, since now one negative power
of $m_{p\ell}$ is absorbed in the gravitino mass. Other hard breaking terms appear if the hidden sector fields are excited
around their VEVs. The leading ones can be read from the soft terms in Eq.~(\ref{eq:VDSSWS}) and Eq.~(\ref{eq:2ndline}) by the replacement 
$b_i \to \zeta^i/m_{p\ell}$ in the $A$ parameter. The two types of hard breaking, the one involving only visible sector fields and
the one involving visible and hidden sector fields do not appear in Eq.~(\ref{eq:VDSSWS}) for two distinct reasons: the first
type is suppressed by two powers of $m_{p\ell}$, and the second type, because hidden sector fields are not excited in the low energy visible
sector. The situation is different as concerns the $S$ fields in the NSWS case, because on the one hand they appear in parts of the
superpotential where usually only hidden sector fields reside, and on the other hand they are excited at low energy in the visible
sector. Whence the appearance of the $A^{(S)}$ and  $A'^{(S)}_i$ terms in $V_{\text{LE}}^{\text{NSWS}}$, that are suppressed by only 
one negative power of $m_{p\ell}$.}

\gilbertzero{In order to assess whether the terms $A^{(S)}$ and $A'^{(S)}_i$ indeed induce
SUSY breaking we consider now the F-terms VEVs associated with the $S$ fields and hidden
sector fields. Assuming no contribution
to SUSY breaking from the purely visible sector, one has from Eqs.~(\ref{eq:Fterm}, \ref{eq:K-Non-SW} -- \ref{eq:hDSNSWS}):}
\gilbertzero{
\begin{eqnarray}
&\langle F_{{S^{r}}^*}\! \rangle& \! \sim  m_{p\ell}  M_{1 1} \langle  \omega_{1 1}  \rangle \mu^*_r + \frac{\langle S^r \rangle^*}{m_{p\ell}}\left(M_{1 0}^2 \langle \omega_{1 0} \rangle + M_{1 1} \langle  \omega_{1 1} \rangle \mu^*_q \langle S^q \rangle \right), \\
&\langle F_{\zeta^*} \! \rangle& \!\! \sim \! M_{1 0}^2 \langle \frac{\partial \omega_{1 0}}{\partial z} \rangle \!+ \! M_{1 1} \langle \frac{\partial  \omega_{1 1}}{\partial z} \rangle
\mu^*_q \langle S^q \rangle
\!+\!  \langle z \rangle^* \!\!\left(M_{1 0}^2 \langle \omega_{1 0}  \rangle \!+\! M_{1 1} \langle  \omega_{1 1} \rangle \mu^*_q \langle S^q \rangle \right) \!\! , 
\end{eqnarray}}
\gilbertzero{where the repeated $q$ index indicates a sum and $\zeta, z$ span all the hidden sector fields $\zeta^i, z^i$.}
\gilbertzero{One sees from these expressions that $ \langle \omega_{1 1} \rangle \neq 0$ contributes to SUSY breaking irrespective of
the values of $\langle S^q \rangle$, while $\langle \frac{\partial \omega_{1 1}}{\partial z} \rangle \neq 0$ would contribute only if at
least one $S^q$ develops a nonvanishing VEV. It follows that the $A^{(S)}$ term will always contribute hard breaking of SUSY
while the contribution of $A'^{(S)}_i$ to the hard breaking will depend on the dynamics of the $S$ fields. The magnitude of these hard
breaking terms will be discussed further in the following two subsections.}


\subsection{A simple non-Soni-Weldon solution \label{sec:simple-NSWS}}
\gilbertzero{We consider now the NSWS in the simple configuration where the $\Xi$ function in Eq.~(\ref{eq:finalformWM-10}) does
not depend on the $S$-type fields.} This configuration is even mandatory in the case of a singleton partition
of the $S$-type fields set. In this case all the $\xi(z)$ functions can be taken vanishing since $\Xi$
is $S$-independent. Equation (\ref{eq:xirelation0}) becomes trivial and non-constraining, implying that one can choose $P=k_1$, 
and $n_p=1, \; \forall p$ in Eq.~(\ref{eq:partition0}).
The superpotential of the NSWS 
simplifies to,

\beqa
W(z, S, \widetilde{\Phi}) = 
m_{p\ell}\Big[W_{1,0}(z) + 
 S^p W_{1, p} (z)\Big]
+ \   W_0(z, \widetilde{\Phi}) +  S^p W_{0, p} (z) \ ,  \label{eq:Non-SW}
\eeqa
where $W_{1,0}(z),W_{1,p}(z), W_{0,p}(z)$ and $W_{0}(z,\widetilde{\Phi})$ are arbitrary holomorphic functions of
$z^i$ and $\widetilde{\Phi}^a$, whereas $S^p$ stands for an arbitrary number $k_1$ of $S$-type fields in the visible sector, 
and we denote by $W_0$ the $S$-independent $\Xi$ function. Note also that the $k_1$ functions $W_{1, p}(z)$, corresponding 
to the partition sum appearing in Eq.~(\ref{eq:finalformWM0}), are now all independent functions. 
The K\"ahler potential is minimal and given by Eq.~(\ref{eq:K-Non-SW}). 
\gilbertzero{The simple dependence on $S^p$ in Eqs.~(\ref{eq:K-Non-SW}, \ref{eq:Non-SW}) will allow to carry out a thorough study of the vacuum structure and to determine 
conditions on the hidden sector that guarantee $\langle S \rangle \ll m_{p\ell}$. It will also show an important 
implication on the magnitude of the hard breaking terms.} 
Making use of \eqref{eq:Non-SW},
\eqref{eq:K-Non-SW}, the potential \eqref{eq:V} can be recast in the following form:
\beqa
V(z,S,\widetilde{\Phi}) & = & \exp \left( z^i z^{i *} + \frac{S^p S^{p *} + \widetilde{\Phi}^a \widetilde{\Phi}^{a *}}{m_{p\ell}^2} \right) 
\Big[ m_{p\ell}^2 V_2(z) + m_{p\ell} V_1(z) + V_0(z,S,\widetilde{\Phi})  \nonumber \\
& &  + \frac{V_{-1}(z,S,\widetilde{\Phi})}{m_{p\ell}} + \frac{V_{-2}(z,S,\widetilde{\Phi})}{m_{p\ell}^2} + \frac{V_{-3}(z,S,\widetilde{\Phi})}{m_{p\ell}^3} + \frac{V_{-4}(z,S,\widetilde{\Phi})}{m_{p\ell}^4} \Big] \ \ \label{eq:Vsimple}
\eeqa
with 
\beqa
V_2(z) & = & W_{1, p} \overline{W}_{1, p} , \\
V_1(z) & = & W_{0,p} \overline{W}_{1, p} + W_{1, p} \overline{W}_{0,  p} , \\
V_0(z,S,\widetilde{\Phi}) & = & \frac{\partial W_0}{\partial \widetilde{\Phi}^a} \frac{\partial \overline{W}_0}{\partial \widetilde{\Phi}^{a *}} + \left( |\widetilde{\rho}_{\widetilde{1}i}|^2 - 3 \right) \widetilde{W}_1 \overline{\widetilde{W}}_1 + W_{0,p} \overline{W}_{0,  p} \nonumber \\
& & \;\;\;\;\;\;\;\;\;\;\;\;\;\; + \;\; S^p W_{1, p} \overline{\widetilde{W}}_1 + S^{p *} \widetilde{W}_1 \overline{W}_{1, p} , \\
V_{-1}(z,S,\widetilde{\Phi}) & = & \Bigg\{ \left( \widetilde{\rho}_{\widetilde{1}i} \overline{\widetilde{\rho}}_{0 i} - 3 + \widetilde{\Phi}^{a *} \frac{\partial}{\partial \widetilde{\Phi}^{a *}} \right) \widetilde{W}_1 \overline{W}_0 + S^p W_{1, p}  \overline{W}_0  \nonumber \\
& & \;\;\; + \;\;  \left( \left( \widetilde{\rho}_{\widetilde{1}i} \overline{\widetilde{\rho}}_{0 pi} - 2 \right) \widetilde{W}_1 + S^q W_{1,q} \right) S^{p *} \overline{W}_{0,  p} \Bigg\} + \text{h.c.} , 
\eeqa
\beqa
V_{-2}(z,S,\widetilde{\Phi}) & = & \left[ |\widetilde{\rho}_{0i}|^2 - 3 + \widetilde{\Phi}^a \frac{\partial}{\partial \widetilde{\Phi}^a} + \widetilde{\Phi}^{a *} \frac{\partial}{\partial \widetilde{\Phi}^{a *}} \right] W_0 \overline{W}_0 + \left( \widetilde{\Phi}^a \widetilde{\Phi}^{a *} + S^p S^{p *} \right) \widetilde{W}_1 \overline{\widetilde{W}}_1 \nonumber \\
& &  + \;\; S^p S^{q *}  \left( \widetilde{\rho}_{0pi} \overline{\widetilde{\rho}}_{0 qi} - 1 \right) W_{0,p} \overline{W}_{0, q}  \\
& &+ \left\{ S^p \left[ \widetilde{\rho}_{0pi} \overline{\widetilde{\rho}}_{0 i}  -2 + \widetilde{\Phi}^{a *} \frac{\partial}{\partial \widetilde{\Phi}^{a *} } \right] W_{0,p} \overline{W}_0 + \text{h.c.} \right\} , \nn \\
V_{-3}(z,S,\widetilde{\Phi}) & = &  \left( S^p S^{p *} + \widetilde{\Phi}^a \widetilde{\Phi}^{a *} \right) \left( \overline{W}_0 + S^{q *} \overline{W}_{0, q} \right) \widetilde{W_1} + \text{h.c.}  , \\
V_{-4}(z,S,\widetilde{\Phi}) & = & \left( S^p S^{p *} + \widetilde{\Phi}^a \widetilde{\Phi}^{a *} \right) \left( W_0 + S^q W_{0, q} \right) \left( \overline{W}_0 + S^{r *} \overline{W}_{0, r} \right) , 
\eeqa
where we have defined the functions
\beqa
\label{eqs:fcts}
\begin{aligned}
& \widetilde{\rho}_{\widetilde{1} i} = \frac{\partial}{\partial z^i} \left( \frac{K}{m_{p\ell}^2} + \ln \frac{\widetilde{W}_1}{m_{p\ell}^2} \right), \\
& \widetilde{\rho}_{0i} = \frac{\partial}{\partial z^i} \left( \frac{K}{m_{p\ell}^2} + \ln \frac{W_{0}}{m_{p\ell}^3} \right), \\
& \widetilde{\rho}_{0pi} = \frac{\partial}{\partial z^i} \left( 
\frac{K}{m_{p\ell}^2} + \ln \frac{W_{0,p}}{m_{p\ell}^2} \right), 
\end{aligned}
\eeqa
with $\widetilde{W}_1 \equiv W_{1, 0} + S^p W_{1, p}$.\footnote{Unless stated otherwise, summation over repeated indices, including 
three occurrences of the same index, is understood.} 
\gilbertzero{Apart from the exponential prefactor, the potential
is quartic in the $S$-fields.} 
A first general remark is that the couplings of the $S$-fields
to the other observable sector fields $\widetilde{\Phi}$ appear only in the terms $V_{-1,-2,-3, -4}$ with negative powers of $m_{p\ell}$
in the potential, while the couplings of the $S$ fields to the (reduced) hidden sector fields $z$ occur already in the $V_0$ term. 
This does not allow to decide in advance on the relative magnitudes of these couplings since the $z$ fields contain a Planck mass 
suppression, and furthermore different physical scales lighter than $m_{p\ell}$ that are expected to exist, would affect differently 
the magnitudes of these couplings.   
To proceed we make explicit these mass scales through the mass dimensions of the various superpotential contributions defining reduced 
functions and {\sl reduced} visible sector fields $\phi$ as follows:

\beqa
\begin{aligned}
 {W}_{1,0}  &=  M_1^2 \omega_1 , \\
 {W}_{1, p}  &=  M_{2, p} \omega_{1p} , (\text{no sum on $p$}), \\
 {W}_{0, p}  &=  M_{3, p} ^2 \omega_{0p} , (\text{no sum on $p$}), \\
 {W}_0  &=  M_4^3 \omega_0(z, \phi) , \\
 \widetilde{\Phi} &=  M_4 \phi\ .
\end{aligned}
\label{eq:scales}
\eeqa
By definition $M_1, M_{2, p},M_{3, p}, M_4 \ll m_{p\ell}$, but it is also plausible that the scale $M_4$ involving the visible sector is 
much smaller
than the other scales occurring in the hidden sector.   
As seen on closer inspection of  the terms in Eq.~(\ref{eq:Vsimple}), the potential is, apart from the exponential prefactor,
a quartic polynomial in the $S$ fields. The $S$ sector can be thoroughly studied, in particular the structure of the vacuum in the 
$S$ fields directions. An explicit computation of the VEVs $\langle S^p \rangle$ allows to determine the conditions
that guarantee $\langle S^p \rangle \ll m_{p\ell}$, otherwise there would be an inconsistency with the fact that the $S$ fields are not
part of the hidden sector!  
Requiring furthermore the vacuum energy to remain tiny (at the tree-level), the relevant vacuum conditions read:
\beqa
 \langle V \rangle &=&0,  \label{eq:vac1}\\
 \left \langle \frac{\partial V}{\partial S^{p *}} \right \rangle &=& \left \langle \frac{\partial V}{\partial S^p} \right \rangle =0, 
 \label{eq:vac2} \\
 \left \langle \frac{\partial V}{\partial z^{i *}} \right \rangle &=& \left \langle \frac{\partial V}{\partial z^i}\right \rangle = 0 \ . 
 \label{eq:vac3}
\eeqa

In principle one should add to Eqs.~(\ref{eq:vac1} -- \ref{eq:vac3}) the extrema conditions on the visible sector fields
as well. However the latter are expected to hold at much lower scales than the one at which the $S$ fields develop VEVs, and are
thus unessential in determining the aforementioned consistency conditions on $\langle S^p \rangle$. Also, necessary 
positivity conditions
on the second order derivatives will be checked {\sl a posteriori} through the positivity of the squared masses of the $S$ and $z$
fields. 

\gilbertzero{
To go further in the study of the vacuum structure, 
we assume for simplicity hereafter a common scale $M$ in the hidden sector such that 
\begin{equation}
M_1 \simeq M_{2, p} \simeq M_{3, p}= \xi^{-1} M_4 = M \equiv \epsilon m_{p\ell}, \label{eq:scaleM}
\end{equation}
 with $\epsilon \ll 1$ (and possibly $\xi \ll 1$ as well),} and consider expansions in powers of the $\epsilon$ parameter. 
\gilbertzero{It turns out that keeping only the leading order in $\epsilon$ misses important physical contributions, as the consistancy requirement
$\langle S^p \rangle \ll m_{p\ell}$ enforces at this order the potential to be essentially flat in the $S$-fields directions.  Including all contributions to ${\cal O}(\epsilon^4)$
proves sufficient as it takes into account all the hidden sector and part of the observable sector.\footnote{The neglected higher powers 
of $\epsilon$, up to power $8$, all involve the observable sector.} } 

\gilbertzero{
We address now the issue of the size of $\langle S^p \rangle$. For the sake of argument, let us assume from now on 
that we have one single $S$ field (i.e.
 $p$ takes only one value), and one single
 hidden sector field $z$.
Parameterizing in general the VEV of $S$ as
$\langle S \rangle \equiv  e^{i\, \theta} \lambda m_{p\ell}$, where $\lambda$ is real-valued and positive and $\theta$ is the phase 
of $S$ at the minimum of the potential, Eq.~(\ref{eq:vac1}) leads
to a quartic equation for $\lambda$ with terms controlled by
$m_{p\ell}^2$, $M m_{p\ell}$ and $M^2$.}

\gilbertzero{
This equation can be solved for $\lambda$. 
However, since $S$ is not in the hidden sector, a consistent solution should correspond to 
$\lambda = {\cal O}(\frac{M}{m_{p\ell}}) \ll 1$. The 
requirement $\lambda \ll 1$ cannot be
satisfied in general due to dangerous terms in $m_{p\ell}^2$ and $M m_{p\ell}$ that do not vanish with vanishing $\lambda$. 
This puts a necessary constraint on the VEV of $\omega_{1p}(z)$ in the form, }
\begin{equation}
\langle \omega_{1p}(z) \rangle =  \epsilon \rho e^{i \, \alpha} \ .
\label{eq:w1p}
\end{equation}


\gilbertzero{
Focusing on the case $\langle S \rangle =0$ implies 
that Eq.~(\ref{eq:vac1}) should be satisfied
 globally rather than order by order
in the $M, m_{p\ell}$ expansion. Equations (\ref{eq:vac1}, \ref{eq:vac2})
lead then to the determination of $\rho$, }

\beqa
&&\rho_{\pm} = \Big\langle -Re [e^{-i \, \alpha} \omega_{0p}(z)] \; \pm  \nonumber \\
&&        \sqrt{Re [e^{-i \, \alpha} \omega_{0p}(z)]^2 
           + 3 |\omega_1(z)|^2 - 
          |\omega_{0p}(z)|^2 - 
            |z^\dag \omega_1(z) + \omega_1'(z)|^2 - \xi^4 |\partial_\phi
            \omega_0(z, \phi)|^2}  \Big\rangle, \nonumber \\
             \label{eq:rho}
\eeqa
and to the supplementary constraint, 
\begin{eqnarray}
\langle \omega_{1p}'(z) \rangle 
     &=&
       \left \langle
      (e^{i \, \alpha}\,\rho + \omega_{0p}(z))\,\big(\frac{2\,\overline\omega_1(z^\dag)}{z\,\overline{\omega}_1(z^\dag) + \overline{\omega}_1'(z^\dag)} 
      -z^\dag \big) -  \omega_{0p}'(z)  \right \rangle \, \epsilon + 
       {\cal O}(\xi^3 \epsilon^2 ) \nonumber \\
       \label{eq:w1pprime} \ .
\end{eqnarray}
\gilbertzero{where the primes denote derivatives with respect to $z$ (or $z^\dag$). 
One finds that the potential is bounded from below in the $S$ field direction 
 and the $S$ mass is generically nonvanishing given by:}


\begin{eqnarray}
&&m_S^2 \equiv \langle \partial_{S^{p*}} \partial_{S^p} V \rangle_{|\langle S \rangle =0} = \frac{e^{|b_i|^2} M^4}{m_{p\ell}^2} 
\frac{1}{
\langle |z^\dag \omega_1(z) + \omega_1'(z)|^2 \rangle } \times \nonumber \\
&& ~~~~~~~~~~ \langle \big (3 |\omega_1(z)|^4 + ( |z^\dag \omega_1(z) + \omega_1'(z)|^2- 3  |\omega_1(z)|^2)^2  \nonumber \\
&&   ~~~~~~~~~~ + 
       \xi^4 \, ( |z^\dag \omega_1(z) + \omega_1'(z)|^2 -4 \, |\omega_1(z)|^2 ) 
        |\partial_{\phi} \omega_0(z, \phi)|^2 \big) \rangle + 
{\cal O}(\frac{M^5}{m_{p\ell}^3}), \nonumber \\
&& ~~~~~= {\cal O}(m_{3/2}^2),  \label{eq:mS2}
\end{eqnarray}
where we neglected the visible sector VEVs as compared to
$m_{p\ell}$ and  used Eq.~(\ref{eq:gravNSWDS}) in the last line.
Note that the above expression is the same for both values of
$\rho = \rho_{\pm}$ given by Eq.~(\ref{eq:rho}). Note also
that the physical requirement $m_S^2 \geq 0$ is automatically
satisfied in the limit $\xi \ll 1$ and would otherwise imply
a further constraint on the VEVs of the involved superpotential
functions and their derivatives. 



\gilbertzero{
Another constraint resulting from Eq.~(\ref{eq:vac3})  
with $\langle S \rangle =0$, allows to express $\langle \omega_1''(z) \rangle$
in terms of $\langle \omega_1(z) \rangle, 
\langle \omega_1'(z) \rangle, \langle \omega_0(z, \phi) \rangle$
and $\langle \partial_{\phi} \omega_0(z, \phi) \rangle$.
However, in contrast with
$\langle \omega_{1p}(z) \rangle$ and  $\langle \omega_{1p}'(z) \rangle$
as given by Eqs.~(\ref{eq:w1p}, \ref{eq:w1pprime}), we find that
$\langle \omega_1''(z) \rangle$ is neither $\epsilon$ suppressed nor 
$\rho_{\pm}$ dependent.  }


\gilbertzero{
One can also obtain the hidden sector scalar field squared 
mass which, 
on the $\langle S \rangle =0$ vacuum, takes the form}

\begin{equation}
m_\zeta^2 = \frac{1}{m_{p\ell}^2} \langle \partial_z \partial_{z^\dag} V \rangle_{|\langle S \rangle =0} , 
\end{equation}
which reads at leading order in $\epsilon$,

\begin{eqnarray}
m_\zeta^2 &&= \epsilon^2 m_{p\ell}^2 \, 
 e^{ |b_i|^2} \,
   \langle \, (\,|\omega_{1p}(z)|^2  + |z^\dag \omega_{1 p}(z) + \omega_{1p}'(z)|^2
      \,) \,\rangle,  \label{eq:z-mass} \\
&&      = {\cal O}(m_{3/2}^2),
\end{eqnarray}
\gilbertzero{where in the last line we took into account the suppression due to Eqs.~(\ref{eq:w1p}, \ref{eq:w1pprime}),
otherwise  $m_\zeta^2$ would have been ${\cal O}(M^2)$. However, this suppression is not generic; masses larger than $m_{3/2}$ can be obtained
when relaxing the common scale assumption Eq.~(\ref{eq:scaleM}) as will be discussed later on.\footnote{\gilbertzero{Note also that since we are merely interested in orders of
magnitude here, we have not considered the contributions of the off-diagonal terms of the mass matrix
of the $S, \zeta$ system, as these contribute the same order of magnitude to the eigenmasses.}}
} 
Although we considered for simplicity a common scale $M$ in the above discussion, it is easy to trace
back the dependence on the various $M_i$ scales from Eq.~(\ref{eq:scales}). 
To clarify the possible physical role the $S^p$ fields could play in the NSWS, in particular in the 
simplest context of a single field studied in this section, a few comments are in order:

\begin{itemize}
\item[]{1/} the $S^p$ are considered as belonging to the visible sector
in the sense that,
like the $\widetilde{\Phi}^a$, they never occur with positive powers of $m_{p\ell}$ in the
potential $V$, as one can easily check upon injecting \eqref{eq:Non-SW}
in the master form \eqref{eq:Vgen}, even though they
occur in the first positive power of $m_{p\ell}$ in the superpotential. 
Thus, like the $\widetilde{\Phi}^a$, the $S^p$ are low energy degrees of freedom.
\item[]{2/} $W_0(z, \widetilde{\Phi}^a)$ in \eqref{eq:Non-SW} describes the usual matter fields interactions
of the supersymmetric SM and its extensions.
\item[]{3/} the interactions between the $S^p$ and the $\widetilde{\Phi}$ fields are always Planck
suppressed! This means that although in the visible sector, the $S^p$ fields are not
`observable' through SM interactions. There could, however, still be sizable couplings to a
beyond the SM observable sector, if the associated scale $M_4$ is much larger than the electroweak scale,
such as for instance in a GUT sector.
\item[]{4/} In the single $S$ field case, the $S$ field is massless before SUSY breaking and
 acquires a soft mass of order $m_{3/2}$ after SUSY breaking. 
\end{itemize}
 Given the above properties it is useful to go beyond the common scale \gilbertzero{assumption Eq.~(\ref{eq:scaleM})} by
considering hierarchical configurations among the various scales. We discuss hereafter briefly two such configurations
\gilbertzero{(postponing a more elaborate discussion of other configurations to the next subsection)}:

\begin{itemize}
\item[a)] \begin{equation} M_{2,p}, M_{3,p}, M_4 \ll M_1 \ll m_{p\ell} . \label{eq:scaleM1} \end{equation} 
In this case the leading interactions of the $S$ fields involve
the hidden sector fields with couplings scaling as $\left(M_1/m_{p\ell}\right)^4$. 
\item[b)] \begin{equation} M_1, M_{3,p}, M_4 \ll M_{2,p} \ll m_{p\ell} . \label{eq:scaleM2} \end{equation} 
Here too the leading interactions of the $S$ fields involve
the hidden sector fields, however, now the couplings scale as $\left(M_{2,p}/m_{p\ell}\right)^2$.
\end{itemize}
\gilbertzero{In configuration a) the hidden sector scalar has a 
mass of ${\cal O}(M_1^2/m_{p\ell}) \simeq {\cal O}(m_{3/2}^2)$ and in configuration b) a mass of ${\cal O}(m_{3/2}^2)$ or 
${\cal O}(M_4^2/m_{p\ell})$ (if $M_1 < M_4$)}.   If these scales are such that some of the hidden degrees of freedom are excited
at the end of an inflationary phase, and thermalized through say some gauge interactions, both a) and b) configurations would feature 
scattering or production of $S$ field quanta. Depending on the magnitudes of $M_1$ and $M_{2,p}$, configuration a) would then possibly provide 
a setting for freeze-in scenarios for the $S$ fields due to the feeble couplings \cite{Hall:2009bx}, while configuration b)
could rather fit the conventional freeze-out scenario given the larger couplings \cite{Jungman:1995df}, provided that $S$ is
initially thermalized. Note also that 
in these configurations the decay of the $S$ fields to visible sector fields is highly suppressed, and to the hidden sector
fields kinematically forbidden. 

 The above properties suggest that in specific models, the $S$ can be relatively
light, ${\cal O}(m_{3/2})$, and produced in the early universe through the hidden
sector interactions at the end of inflation more copiously than by the observable
sector degrees of freedom. {\sl It will be interesting to investigate further
the possibility that the $S$ fields provide a good dark matter candidate.} 
Of course one should consider as well
their supersymmetric fermionic partners. \\

\gilbertzero{Let us end this subsection by two related comments:}

\begin{itemize}
\item
{\sl The hidden sector and the cosmological constant:}
As stated in the discussion preceding Eq.~(\ref{eq:w1p}) one needs to fine-tune $\langle \omega_{1 p}(z) \rangle$ to a small number 
of order $\frac{M}{m_{p\ell}}$ in order to achieve a vanishing vacuum energy. This is somewhat an indication of the criticality
of the simple NSWS considered in section \ref{sec:simple-NSWS}, since $\omega_{1 p}(z)$ away from $z=\langle z \rangle$ should 
definitely be nonzero otherwise the NSWS does not exist altogether. It is worth noting that the issue is expected to be less critical 
in the case of non-canonical K\"ahler where, as noted at the end of \ref{app:non-flat}, there is more freedom in fine-tuning the 
vacuum energy to a small number by requiring conditions such as Eq.(\ref{eq:no-scale-like1}) close to the no-scale condition.
\gilbertzero{We will also see in the next section that relaxing the
assumption of common scale $M$ reduces the fine-tuning of $\langle \omega_{1 p}(z) \rangle$.}

\item \gilbertzero{{\sl Flat directions and inflation:}
The model configutation studied in the present section does not
feature F-flat directions in the parts of the  potential that involve the $S$ field.
However, as already noted in Section \ref{sec:comparison}, 
such flat directions become endemic to our 
general solutions as soon as there are two or more distinct 
$S$ fields coupled directly, through $\Xi$, to the other 
($\tilde \Phi$) fields 
of the visible sector, cf. Eqs.~(\ref{eq:finalformWM-10}, 
\ref{eq:UdefFINAL0}). The existence of these directions, together
with the fact that they are lifted
by the non-vanishing (derivatives of the) $W_{1,p}(z)$ functions in Eq.~(\ref{eq:finalformWM0}), suggest an interesting playground for inflation, since this lifting is expected to be small
when the hidden sector scalar fields are close to their 
VEV values, Eq.~(\ref{eq:w1pprime}).  (see also the discussion in the
next section.) Furthermore, since the NSWS favors particle 
physics models that are specific extensions of the NMSSM, 
it would be interesting to consider variants to the Higgs-like
inflation scenarios \cite{Einhorn:2009bh,Ferrara:2010yw} in this context.}
\end{itemize}

\subsection{The magnitude of the hard breaking \label{sec:hardbreaking}}
{It should be clear that the properties of the $S$-field
studied in the previous section are not generic to all the NSWS. In 
particular, the fact that the $S$-field has Planck suppressed couplings
not only to the hidden sector but also 
to the visible sector, is a consequence of two assumptions --the superpotential $\Xi$ is assumed to be $S$
independent which is a special case. The mass scales appearing in the hidden sector are either 
all of the same order and larger than the one in the visible sector, Eq.~(\ref{eq:scaleM}) or they follow one of the two 
other configurations
Eq.(\ref{eq:scaleM1}) or Eq.(\ref{eq:scaleM2}) of the previous section-- 
These assumptions can be relaxed. For instance, with at least two $S$-fields one can easily construct
a superpotential  $\Xi$ with non suppressed couplings  between these fields and the rest of the visible sector,
as discussed in Section \ref{sec:model-pheno-1}. 
There is however a lesson from the previous section that is
expected to hold even in the presence of a more general $S$-{\sl dependent} $\Xi$: It is always possible
to find minima of the potential consistent with the fact that $S$ is not in the hidden sector, i.e. necessarily 
such that $\langle S \rangle \ll m_{p\ell}$, but at the price of requiring  $\langle \omega_{1p}(z) \rangle$ and $\langle \omega_{1p}'(z) \rangle$ 
to be ${\cal O}(\frac{M}{m_{p\ell}})$ with $M$ some common physical scale much below $m_{p\ell}$, 
cf. Eqs.~(\ref{eq:w1p},  \ref{eq:w1pprime}). If one assumes a common scale, $ M_{1 0} = M_{1 1} \equiv M$, in the slightly different NSWS 
configuration studied in Section \ref{subsec:NSWS}, the same suppression is expected, i.e. $\langle \omega_{11}(z) \rangle \approx  
\langle \frac{ \partial}{\partial z}\omega_{1 1}(z) \rangle = {\cal O}(\frac{M}{m_{p\ell}})$, where we assumed for simplicity a single hidden sector field. 
It then follows from Eqs.~(\ref{eq:gravNSWDS}, \ref{eq:ASdef}, \ref{eq:ASprimedef})  that the hard SUSY breaking terms in 
Eqs.~(\ref{eq:2ndline}, \ref{eq:3rdline}) are ${\cal O}(\frac{M^2}{m_{p\ell}^2}) \sim {\cal O}(\frac{m_{3/2}}{m_{p\ell}})$,
while in Eq.~(\ref{eq:1stline}) the linear term  in the $S$ fields is ${\cal O}(\frac{m_{3/2}^2}{m_{p\ell}})$ and the quadratic term
is ${\cal O}(\frac{m_{3/2}^2}{m_{p\ell}^2})$. The latter induces at the one-loop level SUSY hard breaking contributions to the diagonal entries of the squared mass
matrix of the (conventional) observable sector scalar fields $\tilde \Phi^a$. These loops of virtual $S^p$ (that are quadratically 
divergent with no cancellations by loops of the $S^p$ fermionic partners) are proportional to $m_{S^p}^2$.
However, since  $m_{S^p}^2 \sim {\cal O}(m_{3/2}^2)$, cf. Eq.~(\ref{eq:mS2}), the net effect is ${\cal O}(\frac{m_{3/2}^2}{m_{p\ell}^2})$,
up to loop factors.
Furthermore, if the superpotential contains a term similar to the first term of Eq.~(\ref{eq:extNMSSM}), then SUSY hard breaking
contributions to the off-diagonal entries of the $H_u, H_d$ squared mass matrix are similarly generated at the one-loop level 
proportional to $m_{S^p}^2$.
Since  $m_{S^p}^2 \sim {\cal O}(m_{3/2}^2)$, the net off-diagonal contribution to the squared mass matrix
is ${\cal O}(\frac{m_{3/2}}{m_{p\ell}})$. 
However, despite this formal suppression, a substantial effect 
could still be obtained  by appealing to some model-dependence in the hidden sector such that

\begin{equation} \langle |z^\dag \omega_1(z) + \omega_1'(z)|^2 \rangle \ll 1 . \label{eq:verysmall}
\end{equation}
Indeed, as seen from Eq.~(\ref{eq:mS2}), this configuration can in principle allow for parametrically large $m_{S^p}$. 

The situation can be different if the common scale assumption is relaxed while keeping for simplicity $\langle S^p \rangle \simeq 0$; in this case the discussion is carried out mainly in terms of $M_1$ and $M_4$. 
Indeed, the dependence on $M_{2,p}, M_{3,p}$
drops out from the $S, \zeta$ mass matrix once Eq.~(\ref{eq:vac1}) (zero cosmological constant) and Eq.~(\ref{eq:vac2}) are imposed,
as well as from the gravitino mass because we take $\langle S^p \rangle \simeq 0$. The gravitino mass scales as
\begin{equation}
m_{3/2} \sim \left|\frac{M_1^2}{m_{p\ell}} \langle \omega_1 \rangle + \frac{M_4^3}{m_{p\ell}^2} \langle \omega_0 \rangle \right| \ ,
\label{eq:gravmass}
\end{equation}
where we include now the contribution from the visible sector that we had neglected for simplicity in section \ref{subsec:NSWS},
as stated after Eq.~(\ref{eq:gravNSWDS}). \gilbertzero{(Here $M_4$ denotes the largest physical scale in the visible sector and would obviously not contribute to the gravitino mass if it corresponded
to the electroweak scale.)} It also follows, as a consequence of relaxing the common scale assumption, that 
$\langle \omega_{1p}(z) \rangle$ and 
$\langle \omega'_{1p}(z) \rangle$ become less fine-tuned than in Eqs.~(\ref{eq:w1p}, \ref{eq:w1pprime}), and are of order
$\max\{\frac{m_{3/2}}{M_{2 p}}, \frac{M_1^2}{M_{2 p} m_{p\ell}}, \frac{M_{3 p}^2}{M_{2 p} m_{p\ell}}\}$. 
Moreover, since $\langle |\langle \omega_0 \rangle| \rangle$ and $\langle |\langle \omega_1 \rangle| \rangle$,
are expected to be typically ${\cal O}(1)$,  
 one identifies three relevant mass scale hierarchy configurations:
\begin{equation} \left(\frac{M_4^3}{m_{p\ell}}\right)^{1/2} \ll M_1 \ll m_{p\ell} , \label{eq:scaleM1prime} \end{equation} 
\begin{equation} (M_1^2 m_{p\ell})^{1/3} \ll M_4 \ll m_{p\ell} , \label{eq:scaleM4} \end{equation} 
\begin{equation} M_4 \sim {\cal O}((M_1^{2} m_{p\ell})^{1/3}) \ll m_{p\ell} . \label{eq:scaleM4prime} \end{equation} 
The first configuration, Eq.~(\ref{eq:scaleM1prime}), is compatible with Eq.~(\ref{eq:scaleM1}), while configurations (\ref{eq:scaleM4}, \ref{eq:scaleM4prime}) can be realized only if $M_1 \ll M_4$.  
Computing the $S$ and $\zeta$ eigenmasses in this more general case (but assuming for simplicity one $S$ and one $\zeta$ fields), 
one finds that in configurations (\ref{eq:scaleM1prime}) and (\ref{eq:scaleM4}) the $S$ field mass is again ${\cal O}(m_{3/2})$
and the resulting hard breaking contributions to the visible sector squared masses is again ${\cal O}(\frac{m_{3/2}^2}{m_{p\ell}^2})$ suppressed. In contrast, configuration (\ref{eq:scaleM4prime}) allows to decrease significantly $m_{3/2}$ with respect to $M_1$
if the relative complex phase $\theta$ between   $\langle \omega_1 \rangle$ and $\langle \omega_0 \rangle$ satisfies
$\frac{2 \pi}{3} \lneqq |\theta| \leq \pi$. The $S$ field mass is
now controlled by $M_1$ for $M_1 \gg \sqrt{m_{3/2}m_{p\ell}}$, 
and $m_S^2$ becomes ${\cal O}(M_1^2 m_{3/2}/m_{p\ell})$. 
Considering again the SUSY hard breaking contribution 
from Eq.~(\ref{eq:1stline}), the effect on the squared masses
of the scalar visible sector is now ${\cal O}(M_1^6/(16 \pi^2 m_{3/2}m_{p\ell}^5)) $ (where we have included  a $1/16\pi^2$ 
loop factor). As an illustration, for $m_{3/2} \sim {\cal O}(1)$~TeV and $M_1 \sim {\cal O}(10^{16})$~GeV  this effect can easily 
lead to a $\sim$20\% increase in the scalar masses. Moreover, as 
mentioned above, if a term of the form $\lambda {\cal U} H_u \cdot H_d$ is 
present in the superpotential, cf. Eq.~(\ref{eq:extNMSSM}), 
the hard SUSY breaking term in Eq.~(\ref{eq:2ndline}) would contribute to the off-diagonal squared 
mass entries in the Higgs sector, which becomes now ${\cal O}(M_1^4/(16 \pi^2 m_{3/2}m_{p\ell}^3)) $. A sizable effect that increases the squared mass of the lightest Higgs state 
can ensue. For example, a $\sim$ 5 -- 20\% loop correction to the tree-level coupling $\lambda$ obtains, for
$m_{3/2} \sim {\cal O}(1)$~TeV, when $M_1 \lesssim {\cal O}(10^{15})$~GeV and $M_4 \sim {\cal O}(10^{16})$~GeV. (Recall that $M_4$ is the largest scale in the visible sector, which of course does not preclude the presence of much smaller scales, such as the electroweak scale, that would however not play a role in the effects of hard breaking.)}  \\
\gilbertzero{The above features favor the presence of very high scales both in the visible (GUT scale) and hidden sectors. They may be welcome as they provide new sources to increase the SM-like ${\cal H}$-boson mass and
the other SUSY scalar partners. This is expected to reduce the amount of fine-tuning required to account simultaneously for the
$125$~GeV mass of the former, the increasing experimental exclusion limits
on the masses of the latter, and the $Z$-boson mass through radiative electroweak symmetry
breaking in minimal and next-to-minimal SUSY extensions. A detailed phenomenological study of these features lies outside the scope of the present paper. We note, however, that although a sizable hard SUSY breaking is likely to modify the rational about TeV scale naturalness that is usually based on soft SUSY breaking, the hierarchy problem appears in the present scenario embodied in 
the large cancellation in Eq.~(\ref{eq:gravmass}) if one requires
electroweak scale gravitino mass. Another option would be to allude to model-dependent configurations in the hidden sector that could lead to Eq.~(\ref{eq:verysmall}).} \\


{\sl The fermionic $S$-sector:}
\gilbertzero{Throughout this paper we have discussed only} the scalar components of the $S$-sector. The consistency requirement (\ref{eq:the_condition})
which has been realized through the $F$-term potential of the  scalar sector should be automatically satisfied, due to Supersymmetry, 
in the fermionic sector. 
(Note that the $D$-term contributions to the potential, Eq.~(\ref{eq:VD}) always satisfy (\ref{eq:the_condition}) as long
as the hidden sector is not charged under visible sector gauge groups.)
One can then in principle construct the full Lagrangian involving the $S$-type chiral superfields using the well-known
recipes. The 
fermion-scalar couplings are obtained from the Lagrangian term
\beq 
- \frac{1}{2} e \exp\left(K/2m_{p\ell}^2\right) \mathfrak{D}_I {\cal D}_J W \chi^I \chi^J + h.c. \ ,
\eeq
where ${\cal D}_J W$ is as defined in Eq.(\ref{eq:DI}), \newline
$\mathfrak{D}_I {\cal D}_J W \equiv W_{I J} + \frac{1}{m_{p\ell}^2} \left(K_{I J} W + K_I {\cal D}_J W + K_J {\cal D}_I W  - K_I K_J W \right)  -\Gamma^K_{I J} {\cal D}_K W$, the $\Gamma^K_{I J}$
being the Christoffel symbols associated with the K\"ahler manifold,  
and the $\chi^I$ scan the fermionic components of all hidden and visible sectors; see \cite{WessBagger199203} for further details. 
For instance, from 
Eqs.~(\ref{eq:Kf}, \ref{eq:Non-SWgen} --\ref{eq:finalformWM-10}) one obtains readily the couplings of the NSWS case. We do not
discuss further the other parts of the Supegravity Lagrangian in the present paper. 

\section{Conclusion \label{sec:conclusion}}
\gilbertzero{Guided by the familiar consistency requirement of
separation between as widely different scales as the Planck and
electroweak scales}, we have proven in this paper the existence of new possibilities, within $N=1$ Supergravity, for the construction of low energy 
supersymmetric models consistent with gravity mediation of SUSY breaking to 
the observable sector in the flat space-time limit. 
In the minimal K\"ahler case, \gilbertzero{and assuming (some of) the
hidden sector VEVs to be of the order of the Planck mass}, we provided a detailed proof for a complete classification of the physically 
consistent superpotentials, recovering the conventional forms and exhibiting a large class of new ones as well. 
The latter are endowed with a new type of fields that, although in the visible sector, feature unusual properties
regarding their couplings to the hidden sector and to the conventional observable sector. We then argued that this can open up new 
model building possibilities beyond the MSSM where the dynamics of this new type of fields, uncharged under the visible sector \gilbertzero{gauge} symmetries,  
can modify the conventional structure of the soft SUSY breaking parameters and lead \gilbertzero{as well to unusual hard breaking terms} and typically to a much lighter gravitino than
in the conventional gravity mediation scenarios. Moreover, these fields can be naturally charged under
symmetries of the hidden sector, but also possibly under other secluded or dark gauge sectors, thus in principle allowing to 
relate SUSY breaking to spontaneous breaking of the corresponding symmetries in these sectors. 
Alternatively, in the simplest new form for the superpotential the new fields acquire masses of the same order as those of the conventional 
visible sector but couple extremely weakly to the latter, suggesting dark matter candidates whose 
relic density would be rather determined by the hidden sector interactions. The most general forms feature
flat directions in the multiple new fields sectors that involve as well the hidden sector, and could possibly provide a 
new playground for inflationary model scenarios. \gilbertzero{Last but not least, we argued that the occurrence of the hard SUSY breaking terms could have interesting spin-offs for particle physics SUSY phenomenology.} These results set the stage for further detailed studies.

Finally, we provided in the non-minimal K\"ahler case two special new solutions satisfying conditions reminiscent of the no-scale one, 
and suggesting the existence of much richer structures. 
A systematic treatment of the non-minimal K\"ahler case is outside the scope of the present paper and will be the subject of a companion
paper.

\appendix

\section{Preliminaries \label{app:prelimin}}
This appendix is devoted to the proofs of Propositions that are used in the next appendix. The reader interested in the main
proof of the existence of new solutions can skip the details of the present appendix and go directly to \ref{app:proof-flat}. 

\begin{proposition}\label{prop:holo}
Let $P(z^i), Q(z^i)$ and $R(z^i)$  be three arbitrary multivariate holomorphic functions satisfying the functional identity
\beqa
\label{eq:master}
\sum \limits_{i=1}^n
\Big(\frac{\partial Q}{\partial z^i}  + z^{i*} Q \Big) 
\Big(\frac{\partial \overline{P}}{\partial z^{i*}}  + z^i \overline{P}\Big)  = \alpha Q \overline{P}  + \overline{Q} R \ , 
\eeqa
with $i$ labeling $n$ complex fields  and $\alpha$ an 
arbitrary complex number. If $P$ is not identically vanishing, then 
\beqa
Q=0 \ , \ \ \forall z^i \ ,\nn
\eeqa
except for the special case $n=1, \alpha=1$ with $R$ nonzero.
\end{proposition}

To establish the above Proposition, we proceed by 
{\sl reductio ad absurdum} proving that $Q\neq0$
 leads to a contradiction.

If $Q\neq0$ in a domain of the $(z^i)$'s, then dividing by  
$Q \overline{P}$ both sides of \eqref{eq:master} one gets,

\beqa
\label{eq:master1}
\sum \limits_{i=1}^n
\Big(\frac{\partial \ln Q}{\partial z^i}  + z^{i*}  \Big) 
\Big(\frac{\partial \ln \overline{P}}{\partial z^{i*}}  + z^i \Big)  = 
\alpha + \Big(\frac{R}{Q} \Big) \Big( \frac{\overline{Q}}{\overline{P}} \Big)
\ .
\eeqa
Then, applying on both sides of 
Eq.(\ref{eq:master1}) respectively $\displaystyle \frac{\partial^2}{\partial z^a \partial z^{b *}},  \frac{\partial^2}{\partial z^a \partial z^{b}}$ and $\displaystyle \frac{\partial^2}{\partial z^{a *} \partial z^{b *}}$, where
$a,b$ are two arbitrary indices, one finds respectively

\beqa
\label{eq:mastersecond}
\sum \limits_{i=1}^n 
\frac{\partial^2 \ln Q}{\partial z^a \partial z^i} 
\frac{\partial^2 \ln \overline{P}}{\partial z^{i*} \partial z^{b*}} +
\delta^{a b} = \frac{\partial}{\partial z^a} 
\Big(\frac{R}{Q} \Big) 
\frac{\partial}{\partial z^{b *}} \Big( \frac{\overline{Q}}{\overline{P}}\Big)
 \ ,
\eeqa

\beqa
\label{eq:master2}
\sum \limits_{i=1}^n
\frac{\partial^3 \ln Q}{\partial z^a \partial z^b \partial z^i}
\Big( \frac{\partial \ln \overline{P}}{\partial z^{i*}} + z^i\Big) +
2 \frac{\partial^2 \ln Q}{\partial z^a \partial z^b} =
\Big(\frac{\overline{Q}}{\overline{P}} \Big) \frac{\partial^2}{\partial z^a \partial z^b}\Big( \frac{R}{Q}\Big) 
\eeqa
and
\beqa
\label{eq:master22}
\sum \limits_{i=1}^n
\frac{\partial^3 \ln \overline{P}}{\partial z^{a *} \partial z^{b *} \partial z^{i *}}
\Big( \frac{\partial \ln Q}{\partial z^{i}} + z^{i *}\Big) +
2 \frac{\partial^2 \ln \overline{P}}{\partial z^{a *} \partial z^{b *}} =
\Big( \frac{R}{Q}\Big) \frac{\partial^2}{\partial z^{a *} \partial z^{b *}}\Big(\frac{\overline{Q}}{\overline{P}} \Big)  \ .
\eeqa

Moreover, we will need the following lemma which we prove first: 

\begin{lemma}
\label{lem:lemma1} 
Let $Q$ and $P$ be nonzero holomorphic functions of the 
$n$ fields $z^i$ satisfying Eq.~(\ref{eq:master}), then 
\beqa
\frac{\partial}{\partial z^{j}} \Big( \frac{Q}{P}\Big)
 \neq 0, \forall j=1,...,n \ . \label{eq:lemma1}
 \eeqa
\end{lemma}

\begin{demo}
If there exists an $a$ such that
\beqa
\frac{\partial}{\partial z^{a}} \Big( \frac{Q}{P}\Big)
 = 0, \nonumber
 \eeqa
then
\beqa
Q(z^1,...,z^a,...,z^n) = \zeta(z^1,...,z^{a-1},z^{a+1}...,z^n) 
P(z^1,...,z^a,...,z^n),
\eeqa
where $\zeta$ is an arbitrary (nonzero) function of all 
the $z^i$'s but
$z^a$. Injecting this form back into Eq.~(\ref{eq:mastersecond})
and taking $a=b$ one finds

\beqa
\sum \limits_{i=1}^n 
\frac{\partial^2 \ln P}{\partial z^a \partial z^i} 
\frac{\partial^2 \ln \overline{P}}{\partial z^{i*} \partial z^{a*}} +
1 = \sum \limits_{i=1}^n 
\Big|\frac{\partial^2 \ln P}{\partial z^a \partial z^i}\Big|^2 
+ 1 =0 \ ,
\eeqa 
which is obviously impossible to satisfy whatever the form of $P$.
\end{demo}

We now prove Proposition \ref{prop:holo}

\begin{demo} Let us first define the following $n \times n$ symmetric matrices: 
 
\beqa
\label{eq:matrixdef}
\mathbf{Q}^{a b} \equiv i \frac{\partial^2 \ln Q}{\partial z^a \partial z^b},
\;\; 
\mathbf{\overline{P}}^{a b} \equiv i \frac{\partial^2 \ln \overline{P}}{\partial 
z^{a*} \partial z^{b*}} \ .
\eeqa

Using obvious notations, \eqref{eq:mastersecond} can then 
be rewritten in matrix form as

\beqa
\label{eq:matrixmastersecond}
\mathbf{Q} \mathbf{\overline{P}} = \mathbf{1} -
 \boldsymbol{\nabla}_z (R/Q)  \otimes 
 \big[\boldsymbol{\nabla}_z (Q/P)\big]^\dag  \ .
\eeqa
Taking the determinant one finds
\beqa
\det \mathbf{Q} \mathbf{\overline{P}} &=& \det \Big( \mathbf{1} -
 \boldsymbol{\nabla}_z (R/Q)  \otimes 
 \big[\boldsymbol{\nabla}_z (Q/P)\big]^\dag \Big) \label{eq:detmaster0} \\
 &=& 1 - \mbox{Tr} \Big(\boldsymbol{\nabla}_z (R/Q)  \otimes 
 \big[\boldsymbol{\nabla}_z (Q/P)\big]^\dag \Big), \label{eq:detmaster}
\eeqa
where in the last equation we used the fact that any $n \times n$ matrix
that is given by a tensor product of two vectors has its trace as 
the only nonzero invariant.
We are thus led to considering separately
the cases\\ $\mbox{Tr} \Big(\boldsymbol{\nabla}_z (R/Q)  \otimes 
 \big[\boldsymbol{\nabla}_z (Q/P)\big]^\dag \Big) \neq 1$ or $=1$:

\begin{itemize}
\item[\sl i)] 
\begin{equation} \mbox{Tr} \Big(\boldsymbol{\nabla}_z (R/Q)  \otimes 
 \big[\boldsymbol{\nabla}_z (Q/P)\big]^\dag \Big) \neq 1 \ .
 \label{eq:Trnoone}
 \end{equation}
 \end{itemize}

This equation is expected to hold generically for a large class
of functions $Q/P,R/Q$, since there cannot be a 
constraint  among the independent fields $z^i, z^{i*}$. 
Equation \eqref{eq:detmaster} then implies 
$\det \mathbf{Q}.\mathbf{\overline{P}} \neq 0$ so that both matrices
$\mathbf{Q}$ and $\mathbf{\overline{P}}$ are invertible. It follows
from \eqref{eq:matrixmastersecond}
that

\beqa
\label{eq:matrixmaster1}
\mathbf{Q} &=& (\mathbf{\overline{P}})^{-1} -
\big( \boldsymbol{\nabla}_z (R/Q)  \otimes 
 \big[\boldsymbol{\nabla}_z (Q/P)\big]^\dag\big) (\mathbf{\overline{P}})^{-1}  \nonumber \\
 &\equiv& (\mathbf{\overline{P}})^{-1} +
i \boldsymbol{\nabla}_z (R/Q)  \otimes   \boldsymbol{\lambda}_Q^\dag , 
\eeqa
where we define the vector
\beqa
\label{eq:deflambdaQ}
  \boldsymbol{\lambda}_Q^\dag \equiv 
  i \big[\boldsymbol{\nabla}_z (Q/P)\big]^\dag 
  \big(\mathbf{\overline{P}} \big)^{-1} \ .
\eeqa
Applying $\partial/\partial z^a$ to \eqref{eq:matrixmaster1}, holomorphy leads to,
\beqa
\label{eq:matrixmaster11}
\frac{\partial}{\partial z^a}\mathbf{Q} 
&=&  i
\Big(\frac{\partial}{\partial z^a} \boldsymbol{\nabla}_z (R/Q) \Big)  \otimes  \boldsymbol{\lambda}_Q^\dag  \ , \ \ \forall z^a ,
\eeqa
and in component form, cf. \eqref{eq:matrixdef},
\beqa
\label{eq:matrixmaster3}
\frac{\partial^3 \ln Q}{\partial z^a \partial z^b \partial z^c} =  
 \frac{\partial^2 (R/Q)}{\partial z^a \partial z^b}
 \lambda^{c *}_Q , \;\;\;\;\;\;\;\; \forall \; a,b,c =1,...,n, \;\; \mbox{and} \;\; \forall z^a, z^b, z^c .
\eeqa
The  $\lambda^{c *}_Q$ are by definition $z^i$-independent.
Moreover, if there exists at least one set $(a,b)$ such that 
$\frac{\partial^2(R/Q)}{\partial z^a \partial z^b} \neq 0$  
then the above equations become holomorphically separable 
for all the  $\lambda^{c *}_Q$'s, thus implying that the latter
 must be $z^{i *}$ independent as well. A set $(a,b)$ with
 $\frac{\partial^2(R/Q)}{\partial z^a \partial z^b} \neq 0$ should
 indeed exist. [If not, then $\frac{\partial^2(R/Q)}{\partial z^a \partial z^b} = 0$  {\sl for all} $a,b$, which 
 would imply from \eqref{eq:matrixmaster3} that $\frac{\partial^3 \ln Q}{\partial z^a \partial z^b \partial z^c} = 0$ {\sl for all} $a, b, c$, 
 and finally plugging both in \eqref{eq:master2} would lead to
 $\frac{\partial^2 \ln Q}{\partial z^a \partial z^b}=0$ {\sl for all} $a,b$, that is $\mathbf{Q}=0$, contradicting the fact that
 this matrix is invertible, cf. \eqref{eq:matrixdef} - \eqref{eq:Trnoone}.]
 We thus conclude that the vector $\boldsymbol{\lambda}_Q^\dag$ 
 must be a $z^i, z^{i*}$-independent quantity. It then follows immediately from the holomorphy structure of 
 \eqref{eq:matrixmaster1} that $\mathbf{\overline{P}}$ should be constant
 as well. Moreover, the constancy of $\boldsymbol{\lambda}_Q^\dag$ and $\mathbf{\overline{P}}$ implies through 
 \eqref{eq:deflambdaQ} the constancy of $\big[\boldsymbol{\nabla}_z (Q/P)\big]^\dag$.
In components this reads,
$\frac{\partial^3 \ln \overline{P}}{\partial z^{a *} \partial z^{b *} \partial z^{c *}} = 0$ and $\frac{\partial^2}{\partial z^{a *} \partial z^{b *}}\Big(\frac{\overline{Q}}{\overline{P}} \Big)=0$ 
$\forall a, b, c$.  Plugging the latter constraints in \eqref{eq:master22}, one finds $\frac{\partial^2 \ln \overline{P}}{\partial z^{a *} \partial z^{b *}} =0$ $\forall a, b$, thus contradicting
the fact that $\mathbf{\overline{P}}$ is invertible. 

{\sl This completes the proof that \eqref{eq:master} admits no solution with a nonzero function $Q$, 
when \eqref{eq:Trnoone} is satisfied.}

\begin{itemize}
\item[\sl ii)] let us consider now the non-generic case where
\end{itemize}
\begin{equation} \mbox{Tr} \Big(\boldsymbol{\nabla}_z (R/Q)  \otimes 
 \big[\boldsymbol{\nabla}_z (Q/P)\big]^\dag \Big) = 1 ,
 \label{eq:Trone}
 \end{equation}
so that $\det \mathbf{Q} \mathbf{\overline{P}} =0$, see \eqref{eq:detmaster}, and at least one of the 
$\mathbf{Q}$ or $\mathbf{P}$ matrices
is not invertible.
Naively speaking, one expects the sufficient and 
necessary conditions for the above equation to be:
$\frac{\partial}{\partial z^j} 
\Big(\frac{R}{Q} \Big)$,  
$\frac{\partial}{\partial z^{j *}} \Big( \frac{\overline{Q}}{\overline{P}}\Big)$
nonzero constants for some $j$'s, and possibly 
$\frac{\partial}{\partial z^j} 
\Big(\frac{R}{Q} \Big)=0$ and  
$\frac{\partial}{\partial z^{j *}} \Big( \frac{\overline{Q}}{\overline{P}}\Big)$ 
arbitrary functions for some other $j$'s. [Recall however that
$\frac{\partial}{\partial z^{j *}} \Big( \frac{\overline{Q}}{\overline{P}}\Big)
=0$ is forbidden by Lemma \ref{lem:lemma1}.] 

The above naive expectation, true for $n=1$, is not the most 
general solution when $n>1$! We note first that 
there should exist at least one
$k$ such that $\frac{\partial}{\partial z^k} 
\Big(\frac{R}{Q} \Big) \neq 0$ $\forall z^k$, otherwise 
\eqref{eq:Trone} cannot be satisfied. Relabelling $k=1$,
the general solution for \eqref{eq:Trone},
treated as a linear partial differential equation 
for $\overline{Q}/\overline{P}$ takes the form
\cite{CourantHilbert198901,decuyper1972modèles}, 

\begin{equation}
\frac{\overline{Q}}{\overline{P}} = \frac{z^{1 *}}{\frac{\partial}{\partial z^1} 
\left(\frac{R}{Q} \right) } + {\cal H}\left[\xi^1 z^{2 *}  - \xi^2 z^{1 *} , ...,
\xi^1 z^{j *} - 
\xi^j z^{1 *}, ...,
\xi^1 z^{n *} - 
\xi^n z^{1 *} ;z \right] \label{eq:Trsol}
\end{equation}
with
\begin{equation}
{\xi^j} \; {\frac{\partial}{\partial {z^1}} \left(\frac{R}{Q} \right)}= {\xi^1} \; {\frac{\partial}{\partial {z^j}} \left(\frac{R}{Q} \right)} \label{eq:Trsolprime}
\end{equation}
\noindent
and  ${\cal H}[...;z]$ denotes an arbitrary function of 
$2 n -1$ entries, where all the $z^{i*}$ appear in the 
first $n-1$ entries as given by the specified linear combinations 
of $z^{i \ast}$ and $z^{1 \ast}$, and the $z^i$ correspond
to the last $n$ entries. It follows that, despite the 
generality of ${\cal H}$, the first 
term linear in $z^{1 *}$ on the right-hand side of 
\eqref{eq:Trsol} cannot be reabsorbed in ${\cal H}$, albeit
at the expense of replacing it by another $z^{j*}$ if $\frac{\partial}
{\partial z^j} (R/Q) \neq 0$. This might suggest that the 
requirement of anti-holomorphy of the sum of the two terms on the 
right-hand side (to match the anti-holomporphy of $\overline{Q}/\overline{P}$) should 
be satisfied separately for each term, and thus 
$\frac{\partial}{\partial z^1} (R/Q)$ should be a constant 
function of the $z^i$'s. Although sufficient, it is not obvious
that these requirements are necessary. In general the $\xi^j$'s
can also be functions of $z$ such that the anti-holomporphy 
condition results in a new partial differential equation
for ${\cal H}$ where non-trivial compensations between the two
contributions on the right-hand side of 
\eqref{eq:Trsol} cannot be a priori excluded. Rather than studying
such a differential equation on ${\cal H}$ it will prove more
efficient to write down the general solution for $\ln \overline{P}$ 
as dictated by Eq.~(\ref{eq:master1}) assuming now that the 
functions $\frac{\overline{Q}}{\overline{P}}$ and $\frac{R}{Q}$ take the general 
forms dictated by \eqref{eq:Trsol} and \eqref{eq:Trsolprime}. The general solution of Eq.~(\ref{eq:master1})
then reads\footnote{recall that the 
general solution 
to a multi-variable function $f(x_1, x_2,...x_n)$ satisfying a first order linear partial differential equation of the form
\beqa
\sum_{i=1}^n (a_i x_i + b_i) \frac{\partial f}{\partial x_i} = B(x_1, x_2, ..., x_n),
\nonumber
\eeqa
where all the $a_i$ are nonzero constants, 
is given by 
\beqa
f(x_1, x_2,...) &=& \frac{1}{a_1}\int_{x_0}^{x_1} \frac{dx}{x + c_1} B\Big(x, \cdots,  \Big(\frac{x + c_1}{x_1 + c_1}\Big)^{a_i/a_1}
( x_i + c_i) - c_i,   \nonumber \\
&& ~~~~~~~~~~~~~~~~~~~~~~~~~~~\cdots, \Big(\frac{ x + c_1}{ x_1 + c_1}\Big)^{a_n/a_1}(x_n + c_n) - c_n\Big) \nonumber \\
&&+\;C\Big( \frac{x_2 + c_2}{(x_1 + c_1)^{a_2/a_1}}, \cdots,  \frac{x_i + c_i}{(x_1 + c_1)^{a_i/a_1}}, \cdots, 
 \frac{x_n + c_n}{(x_1 + c_1)^{a_n/a_1}}\Big) \nonumber \ ,
\eeqa
where $C$ is an arbitrary function of $n-1$ variables, $x_0$ an arbitrary reference constant and $c_i \equiv \frac{b_i}{a_i}$. 
}
\beqa
\ln \overline{P} &=& \int_{(z^{1 *})_{in}}^{z^{1 *}} \frac{d\tau^\ast}{\tau^\ast + \partial_{z^1} \ln Q} \Big \{ \alpha + \frac{R}{Q}[z] \frac{\overline{Q}}{\overline{P}}\big [\tau^\ast, \cdots,  \nonumber \\
&&(\tau^\ast + \partial_{z^1} \ln Q) \, r_{i 1} - \partial_{z^i} \ln Q, \cdots, 
(\tau^\ast + \partial_{z^1} \ln Q) \, r_{n 1} - \partial_{z^n} \ln Q\big ]\Big\} \nonumber \\
&& - (z^{1 *}-(z^{1 *})_{in})\sum_{i \geq 1}^n r_{i 1} \, z^i
+ C\big[ r_{21}, \cdots, r_{i1}, \cdots, r_{n1}; z\big] \, , \label{eq:lnPstar}
\eeqa
where $(z^{1 *})_{in}$ is an arbitrary initial condition reference points,  
$C$ an arbitrary function of the 
$r_{i 1}$'s and the $z$ fields [the occurence of non-indexed $z$ denotes dependence on all $z^i$ fields], 
and we have defined
\beqa
r_{i 1} \equiv \frac{\partial_{z^i} \ln Q + z^{i *}}{ \partial_{z^1} \ln Q + z^{1 *}} \, . \label{eq:rdef}
\eeqa
The requirement of anti-holomorphy of $\ln \overline{P}$ in all $z^i$'s puts strong constraints on the right-hand side of 
Eq.~(\ref{eq:lnPstar}).
Taken at the special point $z^{1 *} = (z^{1 *})_{in}$, the anti-holomorphy of $\ln \overline{P}$ implies that 
$C\big[ r_{21}, \cdots, r_{i1}, \cdots, r_{n1}; z\big]$ should be anti-holomorphic as well in all $z^i$ fields including
$z^1$ since $(z^{1 *})_{in}$ can be chosen arbitrarily. This constraint is readily satisfied since the function $C$ is
arbitrary. Then from Eq.~(\ref{eq:lnPstar}) with $z^{1 *} \neq (z^{1 *})_{in}$
follows that the sum of the first two terms on the right-hand side
should be separately anti-holomorphic in all $z^i$'s. In particular, the derivative of this sum with respect to $z^{1 *}$ should 
also be anti-holomorphic. One then obtains the compact condition
\beqa
 \alpha + \frac{R}{Q}[z^1,\cdots] \frac{\overline{Q}}{\overline{P}}\big [z^{1 *},\cdots, z^{i *},\cdots, z^{n *}
 ]  -\sum_{i \geq 1}^n (z^{i *} + \partial_{z^i} \ln Q) \, z^i  \nonumber \\
 \equiv (z^{1 *} + \partial_{z^1} \ln Q) \; {\cal F}_1[z^\ast],  \label{eq:lnPstar1}
\eeqa
where we define
\beqa
{\cal F}_1[z^\dag] = \partial_{z^{1 *}} \ln \overline{P} - \partial_{z^{1 *}} C\, \label{eq:Fdef}
\eeqa
which is by assumption anti-holomorphic in all $z^{i *}$'s. Note that in Eq.~(\ref{eq:lnPstar1}) all dependence on the 
$z^i$'s has dropped out from $\frac{\overline{Q}}{\overline{P}}$ due to the replacement $\tau^\star=z^{1*}$. It is intuitively clear 
why this 
equation cannot hold if $n > 1$, whatever the poduct $\frac{R}{Q} \times \frac{\overline{Q}}{\overline{P}}$ of holomorphic and anti-holomorphic functions may 
be, due to the presence of the term $\sum_{i \geq 1}^n z^{i *} z^i$. To proceed with the proof, it is convenient to 
treat separately the multi-field and single-field cases:\\

\noindent
$\bullet$ {$ n \geq 2$:}
applying $\partial^2/\partial {z^{k *}}^2$ for a given $k\neq 1$ 
to Eq.~(\ref{eq:lnPstar1}), one obtains

\beqa
\frac{R}{Q}[z^1,\cdots] \frac{\partial^2}{\partial {z^{k *}}^2} \frac{\overline{Q}}{\overline{P}}\big [z^{1 *},\cdots] = (z^{1 *} + \partial_{z^1} \ln Q)
\frac{\partial^2}{\partial {z^{k *}}^2}{\cal F}_1[z^{1 *}, \cdots] \label{eq:lnPstar2}
\eeqa
which by holomorphy implies either 
\beqa
\frac{\partial^2}{\partial {z^{k *}}^2} \frac{\overline{Q}}{\overline{P}} =0 \label{eq:lnPstar3} \ ,
\eeqa 
or 
\beqa
\frac{R}{Q} = \text{constant} \;\; \text{and} \;\; \partial_{z^1} \ln Q = \text{constant} \label{eq:lnPstar4} \ .
\eeqa
Indeed if Eq.~(\ref{eq:lnPstar3}) is not satisfied for at least one $k(\neq 1$) then Eq.~(\ref{eq:lnPstar2}) takes the
form
\beqa
\frac{f(z)}{z^{1 *} + g(z)} =h(z^\dag)
\eeqa
which, from $\partial_{z^k} h(z^\ast) =0$ for all $k$ and the fact that $z^{1 *}$ cannot be a function of the $z$ fields, 
leads to $f$ and $g$ constant in $z$, whence Eq.~(\ref{eq:lnPstar4}). Now the latter equation cannot be satisfied since
 $\frac{R}{Q} = \text{constant}$ in all $z$ fields contradicts Eq.~(\ref{eq:Trone}). It follows that 
 Eq.~(\ref{eq:lnPstar3}) should be satisfied {\sl for all} $k\neq 1$. Note that having chosen $z^{1 *}$ as a reference field when
 writing Eq.~(\ref{eq:lnPstar}) entailed only the assumption that $z^{1 *} + \partial_{z^1} \ln Q$ is not a vanishing function
 of $z^{1 *}$. But this is true for any other $z^{k *} + \partial_{z^k} \ln Q$. We could thus rewrite Eq.~(\ref{eq:lnPstar})
 in terms of any other reference field $z^{k *}$ and permute the roles of the $1$
 and $k$ indices in the thread of equations (\ref{eq:lnPstar} -- \ref{eq:lnPstar3}) thus arriving at 
\beqa
\partial^2_{z^{1 *}} \frac{\overline{Q}}{\overline{P}} =0  \ .
\eeqa 
{\sl Condition (\ref{eq:lnPstar3}) should thus be satisfied for all} $k$. Furthermore, this condition implies 
from Eq.~(\ref{eq:lnPstar2}) that
\beqa
\partial^2_{z^{k *}} \; {\cal F}_1[z^{1 *}, \cdots] =0 , \; \; \forall k, \label{eq:lnPstar3prime}
\eeqa 
since $z^{1 *} + \partial_{z^1} \ln Q$ cannot be a vanishing function. The most general allowed forms for $\frac{\overline{Q}}{\overline{P}}$
and  ${\cal F}_1$ are then,

\beqa
\frac{\overline{Q}}{\overline{P}}[z^\dag] = \alpha_0 + \sum_{i \geq 1}^n \alpha_i z^{i *} + \sum_{j > i \geq 1}^n \alpha_{ i j} z^{i *} z^{j *} ,\\
{\cal F}_1[z^\dag] = \beta_0 + \sum_{i \geq 1}^n \beta_i z^{i *} + \sum_{j > i \geq 1}^n \beta_{ i j} z^{i *} z^{j *},
\eeqa
where the $\alpha$'s, $\beta$'s are ($z$-independent) constants and the double sums hold only for $i \neq j$. One can now plug 
these forms back into Eq.~(\ref{eq:lnPstar1}) and identify the various monomials. One then immediatly sees that monomials of
the form $z^{1 *} z^{i *} z^{j *}$ with $i\neq j$ appear only once, on the right-hand side of Eq.~(\ref{eq:lnPstar1}), implying
\beqa
\beta_{ i j} = 0 \;\; \forall i \neq j \ .\label{eq:betaij}
\eeqa
Note also that the monomial $(z^{1 *})^2$ appears only once, thus
\beqa
\beta_1 = 0 \ . \label{eq:beta}
\eeqa
Then collecting monomials of the form $z^{1 *} z^{i *}$ with $i\neq 1$ implies
\beqa
\alpha_{ i 1} \times \frac{R}{Q}[z]  - \beta_i = 0 , \;\; \forall i \neq 1 \ .
\eeqa
Note that the only other function of $z$, namely $\partial_{z^1} \ln Q$ does not appear in the above equation
as a consequence of Eq.~(\ref{eq:betaij}). 
Since $\frac{R}{Q}[z]$ cannot be a constant, as noted previously, it follows that
\beqa
\alpha_{ i 1} = \beta_i = 0 , \;\; \forall i \neq 1 \label{eq:alphabeta} \ .
\eeqa
Collecting now all the contributions to the monomials $z^{1 *}$ and  $z^{i *}$ with $i\neq1$ one gets respectively
\beqa
\alpha_{1} \times \frac{R}{Q}[z] - z^1 - \beta_0 = 0 \;\; \text{and} \; \; \alpha_{i} \times \frac{R}{Q}[z] - z^i=0, \;\;  \forall i \neq 1 \ ,
\eeqa
where we have used Eqs.~(\ref{eq:beta}, \ref{eq:alphabeta}) and again $\partial_{z^1} \ln Q$ does not contribute.
Since the above constraints should hold functionally in $z$, then $\alpha_1$ and $\alpha_i$ cannot be vanishing.
It follows that,
\beqa
\frac{R}{Q}[z] = \frac{z^1 + \beta_0}{\alpha_{1}} = \frac{z^i}{\alpha_{i}}, \;\;  \forall i \neq 1,
\eeqa
which is obviously a contradictory requirement. Thus $n \geq 2$ does not allow for a consistent solution.\\

\noindent
$\bullet$ {$ n = 1$:} in this case of single field the arbitrary function $C$ becomes $z^\ast$ independent, cf. Eqs.~(\ref{eq:lnPstar},
\ref{eq:rdef}), and Eq.~(\ref{eq:Fdef}) becomes ${\cal F}_1[z^\ast] = \partial_{z^{1 *}} (\ln P\ast)$. In this case 
Eq.~(\ref{eq:lnPstar1}) reduces exactly to Eq.~(\ref{eq:master1}) with $n=1$ and we can use directly the latter or its
derivatives. Moreover, Eq.~(\ref{eq:Trone}) simplifies to
\beqa
\Big(\frac{R}{Q} \Big)'  \times 
\Big( \frac{\overline{Q}}{\overline{P}}\Big)' = 1,
\eeqa
where the primes denote here and in the sequel 
derivatives with respect to either $z^1$ or $z^{1 *}$ consistently. The latter equation and holomorphy imply that $\frac{R}{Q}$ and
$\frac{\overline{Q}}{\overline{P}}$ should be strictly linear in $z^1$ and $z^{1 *}$ respectively:
\beqa
\frac{\overline{Q}}{\overline{P}}[z^{1 *}] &=& \alpha_0 + \alpha_1 z^{1 *}, \label{eq:Trsol2} \\
\frac{R}{Q}[z^1] &=& \gamma_0 + \gamma_1 z^1 \label{eq:Trsol1} \ .
\eeqa
where the $\alpha$'s and $\gamma$'s are complex-valued constants with $\alpha_1, \gamma_1$ necessarily nonzero 
and satisfying $\alpha_1 \gamma_1 = 1$. 
When combined with Eq.~(\ref{eq:mastersecond}) with $n=1$ it also implies, 
\beqa
(\ln Q)'' (\ln \overline{P})'' = 0  \ .
\eeqa
Moreover, one sees from \eqref{eq:Trsol2} that
$(\ln Q)''$ and $(\ln \overline{P})''$ cannot be simultaneously 
vanishing since $\alpha_1$ cannot vanish. Choosing without loss of generality $(\ln \overline{P})''=0$ 
and using equations \eqref{eq:Trsol2}, \eqref{eq:Trsol1} in \eqref{eq:master2}, leads then immediately to 
$(\ln P)'=\alpha_0/\alpha_1$. Also equation \eqref{eq:Trsol2} yields $(\ln Q)'=
\alpha_1^\ast/(\alpha_1^\ast z^1 + \alpha_0^\ast) + \alpha_0/
\alpha_1$. Injecting these expressions in \eqref{eq:master1}
one finds
\beqa
\Big(\frac{\alpha_1^\ast}{\alpha_1^\ast z^1 + \alpha_0^\ast}  + \frac{\alpha_0}{
\alpha_1} + z^{1 *} \Big) \Big(\frac{\alpha_0^\ast}{
\alpha_1^\ast} + z^1 \Big)= \alpha + (\gamma_1 z^1 + \gamma_0) (\alpha_1 z^{1 *} + \alpha_0)  \ .
\eeqa
Close inspection shows that this equation is valid 
{\sl if and only if}:
$\gamma_1= (\alpha_1)^{-1}, \gamma_0 = \alpha_0^\ast |\alpha_1^\ast|^{-2}$, and  $\alpha =1$. 
This identifies the special case referred to at the end
of Proposition \ref{prop:holo} showing the existence of a nonzero $Q$ in this special case. We do not write
explicitly this solution here since the relevant value of $\alpha$
in our study is $\neq 1$. This ends the proof  
for the case where \eqref{eq:Trone} is valid.\end{demo}

 {\sl We have thus completed the proof of Proposition \ref
 {prop:holo},  namely that when $P$ is not identically vanishing, 
equation \eqref{eq:master} implies in general the unique solution 
$Q=0$.}\\

In the next section we will encounter cases where Proposition \ref
 {prop:holo} applies directly, but also cases that have either more
 restricted or more general forms than this Proposition. We give here the
 two corresponding corollaries.

 \begin{corollary}\label{corol:holo1}
Let $P$ and $R$  be two  arbitrary holomorphic functions satisfying the following functional identity
\beqa
\label{eq:corolmaster}
\sum \limits_{i=1}^n
\Big|\frac{\partial P}{\partial z^i}  + z^{i*} P \Big|^2 
  = \alpha |P|^2  + \overline{P} R \ ,
\eeqa
with $i$ labeling $n$ complex fields  and $\alpha$ an 
arbitrary complex number. Then the only solution is 
\beqa
P=0 \ , \ \ \forall z^i \ .\nn
\eeqa
\end{corollary} 
\begin{demo}
Clearly, Eq.~(\ref{eq:corolmaster}) is a special case of (\ref{eq:master}) where $P=Q$. Then $P=0$ follows immediatly
from Lemma \ref{lem:lemma1}, since if $P\neq0$ then Eq.~(\ref{eq:lemma1}) cannot hold when $P=Q$.
\end{demo}

\begin{corollary}\label{corol:genholo}
Let $P_r(z^i), Q_{l q}(z^i) $ and $R_{r l q}(z^i)$  be three sets of  arbitrary multivariate holomorphic functions satisfying the functional identity
\beqa
\label{eq:genmaster}
\sum \limits_{i=1}^n
\Big(\frac{\partial Q_{lq}}{\partial z^i}  + z^{i*} Q_{lq} \Big) 
\Big(\frac{\partial \overline{P}_r}{\partial z^{i*}}  + z^i \overline{P}_r\Big)  = \alpha Q_{lq} \overline{P}_r  + \sum_{p=1}^k 
\overline{Q}_{r p} R_{p l q} \ , 
\eeqa
with $i$ labeling $n$ complex fields, $r, l, p, q=1,\dots, k$  and $\alpha$ an 
arbitrary complex number. If $P_r$ is not identically vanishing for at least one $r$, then generically
\beqa
Q_{lq}=0 \ , \ \ \forall z^i, l,q \ .\nn
\eeqa
\end{corollary}
The proof is mostly a mere generalization of the proof of Proposition \ref{prop:holo}  
albeit with some technical complications. \gilbertzero{For the sake of conciseness we refrain from giving this somewhat lengthy 
proof here and refer the reader to appendix A of the arXiv version $1$ (arXiv:1611.10327v1, \cite{Moultaka:2016frs}).}

\section{Proof of the existence of new solutions}
\subsection{Minimal K\"ahler \label{app:proof-flat}}
We give here the main steps of the proof that in the case of  canonical K\"ahler potential Eq.~(\ref{eq:Kf}), equations  
\eqref{eq:SW} and \eqref{eq:Non-SWgen} of Section \ref{sec:NSW-flat}
constitute together the {\sl sufficient} and {\sl necessary} 
general forms of the superpotential that lead to a 
consistent low energy Lagrangian. 


Collecting in \eqref{eq:Vgen00} all coefficients of the same 
power $m_{p\ell}^c$ with $c \geq 0$ 
one finds for $V_{M,c} [z, z^\dag, \Phi, \Phi^\dag]$:

\beqa
 V_{M,c} [z, z^\dag, \Phi, \Phi^\dag] &=& 
 \sum_{n_{-}^{(0)} \leq n \leq n_{+}^{(0)}}\frac{\partial W_n}{\partial \Phi^a} 
\frac{\partial\overline W_{c-n}}{\partial \Phi^{a *}}    \nonumber \\
 && + \sum_{n_{-}^{(2)} \leq n \leq n_{+}^{(2)}}\left( \Big(\frac{\partial W_n}{\partial z^i} + z^{i *} W_n\Big) 
\Big(\frac{\partial \overline W_{c-n+2}}{\partial z^{i *}} + z^i \overline W_{c-n+2}\Big) \right. \nonumber \\
&& \left. \;\;\;\; + \;\; \Phi^a \frac{\partial W_n}{\partial \Phi^a} \overline W_{c-n+2}  + \Phi^{a *} \frac{ \partial\overline W_n}{\partial \Phi^{a *}}
 W_{c-n+2} - 3 W_n \overline W_{c-n+2}\right)  \nonumber \\
 && + \;\; \sum_{\!\!\!\!\! n_{-}^{(4)} \leq n \leq n_{+}^{(4)} } W_n \overline W_{c-n+4} \Phi^{a *} \Phi^a \label{eq:Vgen}
\eeqa
where we define
\beqa
&& n_{+}^{(s)} = \min[M, c+s], \\
&& n_{-}^{(s)} = \max[0, c - M + s] .
\eeqa
Recall that  $W_r$ denotes the coefficient of $m_{p\ell}^r$ 
in the superpotential expansion Eq.~(\ref{eq:Wexpansion}) with
$r$ an integer verifying 
$0 \leq r \leq M$ by definition. 
The latter constraint is expressed in Eq.~(\ref{eq:Vgen}) through 
the summation bounds $n_{-}^{(s)} \leq n \leq n_{+}^{(s)}$, with $s=0,2,4
$, corresponding to the three different sums occurring in each
$m_{p\ell}^c$ coefficient. We will refer to these three sums as, 
$0$-sum, $2$-sum and $4$-sum. Note also that the 
implicit constraints
$n_{-}^{(s)} \leq n_{+}^{(s)}$ yield the conditions under which the
corresponding $s$-sum contributes to a given power of $m_{p\ell}$. 
They can be stated as follows: 
\begin{equation}
\text{{\sl a given $s$-sum contributes to
a given $m_{p\ell}^c$ term if and only if $c \leq 2 M -s$}.} \label{eq:text}  
\end{equation}
Recall that requiring the fields $\Phi^a$  not to belong to the 
hidden sector, \gilbertzero{that is not to appear in parts of $V_F$ that diverge in the limit $m_{p\ell} \to \infty$,} cf. (\ref{eq:the_condition}), 
is equivalent to requiring that each $V_{M, c}[z, z^\dag, \Phi, \Phi^\dag]$  with $c>0$ in Eq.~(\ref{eq:Vgen}), 
should be functionally independent of $\Phi, \Phi^\dag$, cf. Eq.~(\ref{eq:masterequation}). 
We write this as
\beqa
V_{M, c}[z, z^\dag, \Phi, \Phi^\dag] \p 0
\eeqa
and take throughout the proof the notation $E \sim_X 0$
to mean that $E$ has no functional dependence on $X$ or
$X^\dag$. 
We now study the above master equation for each value of $M$ and 
each strictly positive value of $c$.

Note first that if $M=0$ then $c=0$ and there are obviously no conditions to be imposed, 
\beqa
\label{eq:SW0}
 W(z,\Phi) = W_0(z,\Phi) \ , 
\eeqa
with $W_0(z,\Phi)$ an arbitrary function.\\

We assume now $M \geq 1$ and study all $c$ values in decreasing order.

\noindent
$\bullet \; \mathbf{c=2M :}$
relevant when $M>0$, the highest power of $m_{p\ell}$   gets 
according to property \ref{eq:text} contributions only from
the $0$-sum in the master equation \eqref{eq:Vgen}. The 
corresonding constraint reads, 
\beqa
\sum_a \Big|\frac{\partial W_M}{\partial \Phi^a}\Big|^2  \p0 \ . 
\label{eq:WM}
\eeqa
Since the lefthand side is a sum of positive numbers there can
be no cancellation among the $\Phi^a$ dependent terms to get
a $\p0$ result. Thus one must have
\beqa
\Big|\frac{\partial W_M}{\partial \Phi^a}\Big|^2  \p0 \ , \nn
\eeqa
separately for each $\Phi^a$. 
It follows that $\frac{\partial W_M}{\partial \Phi^a}$ should be a $\Phi^a$
independent function for all $\Phi^a$. The general form for
$W_M$ is thus
\beqa
W_M(z,\Phi)= W_{M,0}(z) +  \Phi^a W_{M,a}(z) \ , \label{eq:firstform}
\eeqa
where $W_{M,0}(z)$ and $W_{M,a}(z)$ denote arbitrary holomorphic 
functions of the $z^i$'s, and the repeated $a$ index is understood
as a sum ($a=1,...,k$).\\

\noindent
$\bullet \; \mathbf{c=2M-1 :}$
again, due to \ref{eq:text} only the $0$-sum
contributes leading to the constraint

\beqa 
 \frac{\partial \overline{W}_M}{\partial \Phi^{a *}} \frac{\partial W_{M-1}}{\partial \Phi^a}  + {\rm h.c.} \p 0 \ , \label{eq:WM-1}
\eeqa
with a summation on $a$ understood. 
Using Eq.~(\ref{eq:firstform}) this constraint reads,
\beqa 
\overline{W}_{M,a}(z^\dag) \frac{\partial  W_{M-1}}{\partial \Phi^a}  +
       {\rm h.c.} \p 0 \label{eq:secondformEq}\ ,
\eeqa
which is obviously equivalent to requiring
\beqa 
\overline{W}_{M,a}(z^\dag) \frac{\partial  W_{M-1}}{\partial \Phi^a}  \p 0 \label{eq:secondformEq1}\ ,
\eeqa
[as can be seen for instance by taking the derivative of 
Eq.~(\ref{eq:secondformEq}) with respect
to $\Phi^b$ for any $b$ where the
h.c. part drops out in this operation due to holomorphy.]
We recast Eq.~(\ref{eq:secondformEq1}) in the form
\beqa 
\overline{W}_{M,a}(z^\dag) \frac{\partial  W_{M-1}}{\partial \Phi^a}  = f(z, z^\dag)\label{eq:secondformEq2}\ ,
\eeqa
where $f$ denotes an arbitrary function depending only on 
the hidden sector fields. 
 
We generically assume $f$ to be a nonzero function. By this we mean that there exists at least 
one connected domain (not just isolated points) in the $z, z^\dag$ field space where $f$ 
does not vanish. In such a domain Eq.~(\ref{eq:secondformEq2}) can be rewritten as
\beqa 
g_{a}(z, z^\dag) \frac{\partial  W_{M-1}}{\partial \Phi^a}  = 1\label{eq:secondformEq3}\ ,
\eeqa
with $g_{a}(z, z^\dag) \equiv \frac{\overline{W}_{M,a}(z^\dag)}{f(z, z^\dag)}$. Equation~(\ref{eq:secondformEq3}) is a linear partial 
differential equation for $W_{M-1}$ as a function
of the $\Phi^a$'s with coefficients independent of these variables. We have already encountered a similar equation in the previous
appendix, Eq.~(\ref{eq:Trone}). The most general solution of Eq.~(\ref{eq:secondformEq3}) will have a form similar
to Eq.~(\ref{eq:Trsol}). Indeed, in order for Eq.~(\ref{eq:secondformEq3}) to be satisfied at least one $g_a$ 
function (or equivalently one $\overline{W}_{M,a}$) should be nonzero, call it $\overline{W}_{M,1}$. The general solution for 
$W_{M-1}$ takes then the form (see also \cite{CourantHilbert198901,decuyper1972modèles}),
\beqa
 W_{M-1}(z, \Phi^a) = \frac{f(z, z^\dag) }{\overline{W}_{M,1}(z^\dag)} \Phi^{1} + { \Omega_1}({\cal U}^{1 2}_{\Phi} , \dots,
{\cal U}^{1 a}_{\Phi}, \dots,
{\cal U}^{1 k}_{\Phi} ;z, z^\dag),
 \label{eq:WM-1sol}
\eeqa
where we have defined
\beqa
{\cal U}^{1 a}_{\Phi}= \xi^1_a \Phi^{a} - 
\xi^{a, 1} \, \Phi^{1}, \; {\rm for}\, a=2,\dots,k\ ,
\label{eq:Udef0}
\eeqa
with $\xi^1_a$ and $\xi^{a, 1}$ satisfying
\beqa
{\xi^1_a}  \; \overline{W}_{M,a}(z^\dag) = {\xi^{a, 1}} \; \overline{W}_{M,1}(z^\dag)   \label{eq:xirelation} \ , \ \text{(no summation over $a$)}, 
\eeqa
and where $\Omega_1$ is an {\sl arbitrary} function of $k-1$ entries with the dependence on combinations of the  $\Phi^a$'s as shown 
in Eq.~(\ref{eq:Udef0}), and $2\ell$ entries in the $z^i$'s and $z^{i *}$'s. 
(Although we assumed $f$ to be nonzero, in fact 
the case where $f$ is the zero function holds as a special case, putting simply $f=0$ in the general solution Eq.~(\ref{eq:WM-1sol}).)
To keep the discussion general, we assume that the two sets of parameters $\xi^{a, 1}$ and $\xi^1_a$ can be any 
functions of $z, z^\dag$ satisfying Eq.~(\ref{eq:xirelation}).
Note that we write $\xi^1_a$ rather than simply $\xi^1$ to keep track of the fact that $\Phi^1$ has been singled out among
all the $\Phi^a$ in writing the general solution Eq.~(\ref{eq:WM-1sol}) and to stress as well 
that $\xi^1_a$ in Eq.~(\ref{eq:xirelation}) is in general different for different values of the $a$ index.
Similarly we write $\xi^{a, 1}$, rather than just $\xi^{a}$, to indicate that if we chose another reference field than
$\Phi^1$, say $\Phi^2$, we do not necessarily have $\xi^{a, 1} = \xi^{a, 2}$.
Indeed to keep the discussion generic one should consider that
there are more than one nonzero $\overline{W}_{M,a}$ apart from
$\overline{W}_{M,1}$. For instance if $\overline{W}_{M,2} \neq 0$ then one can chose $\Phi^2$ to play the role of $\Phi^1$,
and a general solution similar to Eq.~(\ref{eq:WM-1sol}) reads,

\beqa
 W_{M-1}(z, \Phi^a) = \frac{f(z, z^\dag) }{\overline{W}_{M,2}(z^\dag)} \Phi^{2} + { \Omega_2}({\cal U}^{2 1}_{\Phi}, \dots,
{\cal U}^{2 a}_{\Phi}, \dots,
{\cal U}^{2 k}_{\Phi} ;z, z^\dag)\ .
 \label{eq:WM-1solalternative}
\eeqa
Here the ${\cal U}^{2 a}_{\Phi}$ are defined as in 
Eq.(\ref{eq:Udef0}) with $1$ replaced by $2$, and as just noted we assume in general $\xi^1_a \neq \xi^2_a$ and 
$\xi^{a, 1} \neq \xi^{a, 2}$. Since $\Omega_1$ and $\Omega_2$ are arbitrary functions of their respective entries,
it is legitimate to ask whether each of them can span all possible solutions, so that we can rely on just one of the two
equations (\ref{eq:WM-1sol}) or (\ref{eq:WM-1solalternative}), or else one would need to combine them
to get more general solutions. In the former case (\ref{eq:WM-1sol}) and (\ref{eq:WM-1solalternative}) should be equivalent,
implying:
\beqa
{ \Omega_2}({\cal U}^{2 1}_{\Phi} , \dots,
{\cal U}^{2 a}_{\Phi}, \dots,
{\cal U}^{2 k}_{\Phi} ;z, z^\dag) = { \Omega_1}({\cal U}^{1 2}_{\Phi} , \dots,
{\cal U}^{1 a}_{\Phi}, \dots,
{\cal U}^{1 k}_{\Phi} ;z, z^\dag) && \nonumber \\
- \;\frac{f(z, z^\dag) }{\xi^{2,1} \overline{W}_{M,1}(z^\dag)} {\cal U}^{1 2}_{\Phi},
~~~~~~~~~~~~~&&\label{eq:difference}
\eeqa
where we used Eq.~(\ref{eq:xirelation}) with $a=2$ when writing the $f$-dependent term. This means that for each $\Omega_1$ there
should exist an $\Omega_2$ satisfying the above equation. The two functions should then be constrained
in two ways:
\begin{itemize}
\item[(i)] their respective forms should be such that the entries
${\cal U}^{2 a}_{\Phi}$ can be transformed into the entries
${\cal U}^{1 a}_{\Phi}$.
\item[(ii)] $\Omega_1$ and/or $\Omega_2$ should have linear contributions in order to account for
the linear term on the right-hand side of 
Eq.~(\ref{eq:difference}).
\end{itemize}
In fact requirement (i) is trivially satisfied due to the
identities
\beqa
{\cal U}^{2 1}_{\Phi} &=& - \frac{\xi^2_1}{\xi^{2,1}} {\cal U}^{1 2}_{\Phi} = - \frac{\xi^{1,2}}{\xi^1_2} {\cal U}^{1 2}_{\Phi}, 
\label{eq:identity0} \\
{\cal U}^{2 a}_{\Phi} &=& \frac{\xi^2_a}{\xi^1_a} {\cal U}^{1 a}_{\Phi}
- \frac{\xi^{a, 2}}{\xi^1_2} {\cal U}^{1 2}_{\Phi},
\label{eq:identity}
\eeqa
meaning that any function of the ${\cal U}^{2 a}_{\Phi}$ entries
is automatically a function of the ${\cal U}^{1 a}_{\Phi}$ entries
and vice-versa.\footnote{\label{footnote:4} To establish  (\ref{eq:identity}) we used the identity $\xi^{2,1}/\xi^1_2 = \xi^2_a \xi^{a,1}/(\xi^1_a \xi^{a,2})$
which follows from Eq.~(\ref{eq:xirelation}) and its equivalent where $1$ is replaced by $2$. Also (\ref{eq:identity0}) is a consequence
of (\ref{eq:identity}) with $a=1$ and upon use of the obvious identities $\xi^{1,1}=\xi^1_1$ and ${\cal U}^{1 1}=0$.} In this case,
requirement (ii) can be arranged for by simply adding a linear term in ${\cal U}^{1 a}_{\Phi}$ or ${\cal U}^{2 a}_{\Phi}$. One is thus
tempted to conclude that either of the two forms, (\ref{eq:WM-1sol}) or (\ref{eq:WM-1solalternative}), is sufficient to span
all possible solutions, as well as any other similar form taking a different $\Phi^a$, for which $\overline{W}_{M,a}$ is nonzero,
as a reference field. 
However, as we will see below the requirement of holomorphy of $W_{M-1}(z, \Phi^a)$ leads in some instances to obstructions such
that (i) and (ii) become insufficient or their generalisation not trivially satisfied. 

We note first that the holomorphy in $z$ of $W_{M-1}(z, \Phi^a)$ requires
that the last $\ell$ entries in $\Omega_1$ should be left out
since they cannot be canceled by a $z^\dag$ dependence in 
$\xi$. This is easily seen by considering the holomorphy of
 $W_{M-1}$ in $z$ at the field space point $\Phi^a=0, \forall a=1,\dots, k$, implying the holomorphy of 
 $\Omega_1(0,...,0,...;z, z^\dag)$ and thus that the last $\ell$ entries should be absent. Taking
 this into account one finds that the $\xi^1_a$'s with $a\neq 1$ should also be holomorphic in $z$. 
Indeed, the holomorphy of $W_{M-1}(z, \Phi^a)$
at the field space points defined by $\Phi^a=0, \forall a=1,\dots, b-1, b+1,k$ and $\Phi^b \neq 0$
for any given $b\neq1$, implies the holomorphy of 
$\Omega_1(0,...,0,\xi^1_b \Phi^b,0,...0;z)$ and thus the holomorphy
of $\xi^1_b$, $\forall b \neq1$. By following the same line of reasoning one finds similarly 
that $\Omega_2$ should not have the last $z^\dag$ entries and that the $\xi^2_a$'s should be holomorphic
in $z$. This obviously holds as well for any function $\Omega_r$ and parameters $\xi^r_a$'s, when
$\Phi^r$ plays the role of the reference field. 

In contrast, there is no reason at this point to require the $\xi^{a, 1}, \xi^{a, 2}, ...\xi^{a, r}$ to be holomorphic, 
as cancellations are in principle possible within the entries of each $\Omega$ function or
between $\Omega$ and the $f(z, z^\dag)$ dependent terms present 
in Eqs.~(\ref{eq:WM-1sol}), (\ref{eq:WM-1solalternative}), etc. However, assuming the $\xi^{a, 1}, \xi^{a, 2}, ...\xi^{a, r}$
to be non-holomorphic leads to one of the obstructions alluded to after Eq.~(\ref{eq:identity}). Let us illustrate the case with 
indices $1,2$. The coefficient of ${\cal U}^{1 2}_{\Phi}$
in (\ref{eq:identity}) and possibly in (\ref{eq:identity0}) becomes non-holomorphic as soon as $\xi^{a, 2}$ is non-holomorphic for
a subset of the $a$ indices, since as proven above all the $\xi^1_a$'s (including $\xi^1_2$) must be holomorphic. 
Now to show that $\Omega_2$ spans all possible solutions, one uses (\ref{eq:identity0}) and (\ref{eq:identity}) to express
$\Omega_2$ as a function of the ${\cal U}^{1 a}_{\Phi}$'s. However this involves redefining $\Omega_2$ by reabsorbing the
coefficients of the ${\cal U}^{1 a}_{\Phi}$'s and in particular $\xi^{a, 2}$, in the last entries of $\Omega_2$ containing the explicit 
dependence on the $z$ fields.
But this is impossible in the present case since we showed that holomorphy enforces the last $\ell$ entries of the $\Omega$'s not to
contain $z^\dag$ while $\xi^{a, 2}$ does by assumption, whence the obstruction.  
This means that in general $\Omega_2$, and similarly 
$\Omega_1$ and $\Omega_i$, cannot span all possible solutions for $W_{M-1}(z, \Phi^a)$ 
if they depended only on the  ${\cal U}^{2 a}_{\Phi}$'s, ${\cal U}^{1 a}_{\Phi}$'s and ${\cal U}^{i a}_{\Phi}$'s respectively.
To obtain the general solution one should include {\sl all} the ${\cal U}^{i a}_{\Phi}$'s as explicit entries. The
most general solution takes then the form,
\beqa
 W_{M-1}(z, \Phi^a) = \frac{f(z, z^\dag) }{\overline{W}_{M,1}(z^\dag)} \Phi^{1} + { \Omega}( \dots,
{\cal U}^{r a}_{\Phi}(z, z^\dag), \dots \; ;z),
 \label{eq:WM-1solprime}
\eeqa
where $r=1,2,...$ label all the fields $\Phi^r$ that can be used as reference fields, i.e. for which
$\overline{W}_{M, i}(z^\dag)$ is nonzero, and
\beqa
{\cal U}^{r a}_{\Phi} (z, z^\dag)= \xi^r_a(z) \Phi^{a} - 
\xi^{a,r}(z, z^\dag) \Phi^{r}, \; {\rm for}\, a=1,\dots,k\ ,
\label{eq:Udefr}
\eeqa 
with
\beqa
{\xi^r_a}(z)  \; \overline{W}_{M,a}(z^\dag) = {\xi^{a, r}}(z, z^\dag) \; \overline{W}_{M,r}(z^\dag) \ ,  \ \text{(no summation over $a$)}.
 \label{eq:xirelationr}
\eeqa
The first term on the right-hand side of Eq.~(\ref{eq:WM-1solprime}) is chosen to be the same as in 
Eq.~(\ref{eq:WM-1sol}). This is not a loss of generality. It can be traded for by any other term of the same form, 
with $1$ replaced by $r$, or even an averaged sum of such terms. 
The reason is that this linear term does not suffer from the aforementioned holomorphy obstruction
since the coefficient of the linear term in ${\cal U}^{1 2}_{\Phi}$ in 
Eq.~(\ref{eq:difference}) is in general non-holomorphic.
One can thus always pull out of $\Omega$ a linear term, where the non-holomorphy is embedded in the $\xi^{a,r}$, 
to match the choice made in Eq.~(\ref{eq:WM-1solprime}). Note that there can be subsets of $r$ and $a$ for which
$\xi^{a,r}$ is holomorphic in $z$. These subsets can be related to each other through (\ref{eq:identity0}), (\ref{eq:identity})
without any holomorphy obstruction. Given our notations in Eq.~(\ref{eq:WM-1solprime}) this leads to redundant entries in $\Omega$, 
but this is harmless and we stick for simplicity to these notations.




Although the holomorphy of each of
the $\xi^{a,r}$'s  and a proper choice of $f(z, z^\dag)$ would be obviously sufficient 
to ensure the holomorphy of $W_{M-1}$, the reverse is far from
straightforward to prove, as cancellations are in principle possible. 
We address this question in detail below.

But first a crucial remark: as easily seen from Eqs.~(\ref{eq:WM-1solprime} -- \ref{eq:xirelationr}),  
$\Omega$ becomes a totally arbitrary function of the $\Phi^a$'s for the $a$ 
labels that correspond to vanishing functions $\overline{W}_{M,a}(z^\dag)$, as this implies $\xi^{a,r}=0$ and 
${\cal U}_\Phi^{r a} = \xi^r_a \Phi^a$, while it 
remains arbitrary, 
albeit a function of the field combinations given by
${\cal U}^{r a}_{\Phi}$, for the $a$ labels that correspond to nonzero functions 
$\overline{W}_{M,a}(z^\dag)$. {\it This mathematical difference 
leads to an important physical difference which is at the heart
of the existence of the NSWS discussed in Section \ref{sec:NSW-flat}.}
It is thus convenient for the ongoing
discussion to label distinctly the $\Phi^a$ fields corresponding
to the two classes of dependence. We denote by
$\widetilde{\Phi}^a$ with $a=1,\dots, k_0$ the $k_0$ fields that
correspond to $\overline{W}_{M,a}(z^\dag)=0$ and denote by $S^p$ with
$p=1, \dots, k_1$ the $k_1$ fields $\Phi^r$ that correspond
to $\overline{W}_{M,r}(z^\dag)\neq0$, with $k=k_0+k_1$. We will refer
to these two distinct classes as $\widetilde{\Phi}$-type and $S$-type respectively.  
In terms of this relabeling Eq.~(\ref{eq:WM-1solprime}) takes the form
\beqa
 W_{M-1}(z, S, \widetilde{\Phi}) = \frac{f(z, z^\dag) }{\overline{W}_{M,1}(z^\dag)} S^{1} + { \Omega}(\dots, {\cal U}^{r p}_{S}, \dots,
; \widetilde{\Phi}^1, \dots, \widetilde{\Phi}^{k_0} ;z),
 \label{eq:WM-1solsecond}
\eeqa
where now 
\beqa
{\cal U}^{r p}_{S} (z, z^\dag)= \xi^r_p(z) S^{p} - 
\xi^{p,r}(z, z^\dag) S^{r}, \; {\rm with}\, r,p=1,\dots,k_1\ .
\label{eq:Udef}
\eeqa
Note that when there is only one single $S$-type field, say $S^1$, then 
${\cal U}_S^{1p}=0$ and $W_{M-1}$ becomes a sum of a linear function in 
$S^1$ and an arbitrary $S$-independent function of $\widetilde{\Phi}^a$
and $z$. Also the case with no $S$-type fields is implicitly
included in Eq.~(\ref{eq:WM-1solsecond}) by dropping all dependence
on $S$. 
  
 We come now to the requirement of the holomorphy of $W_{M-1}$ in $z^i$. 
 This is expressed as,
\beqa
\frac{\partial}{\partial z^{i *}} W_{M-1}(z, S,  \widetilde{\Phi}) =0, \; \forall i \ .
\label{eq:Eqholomorphy}
\eeqa
The full exploitation of this constraint turned out to be technically very involved. We present it here in the simplified
case where $\Omega$ is a function of a single set of ${\cal U}^{r p}_{S}$ variables, say ${\cal U}^{1 p}_{S}$ with
$p=2, \dots, k_1$. In fact this simplified assumption will prove sufficient to obtain the most general solution.
Upon use of (\ref{eq:WM-1solsecond}) equation (\ref{eq:Eqholomorphy}) implies 
\beqa
\frac{\partial}{\partial z^{i *}}\frac{f(z, z^\dag) }{\overline{W}_{M,1}(z^\dag)}-\sum_{p \geq 2}^{k_1} \frac{\partial \xi^p}{\partial z^{i *}} \frac{\partial}{\partial {\cal U}_S^{1 p}} {\Omega}({\cal U}^{1 2}_{S} , \dots,
{\cal U}^{1 p}_{S}, \dots,
{\cal U}^{1 k_1}_{S}; \widetilde{\Phi}^1, \dots, \widetilde{\Phi}^{k_0} ;z) =0, \nonumber \\
 \;\; \forall i=1,...,\ell, ~~~~\label{eq:EqOmegahat}
\eeqa 
where we have used the fact that 
$\xi^1_p$ is holomorphic in $z$, denoted the $\xi^{p, 1}(z, z^\dag)$ by $\xi^{p}$ for simplicity, 
and dropped out an overall common factor $S^1$ when writing the above equation. 
This equation can be again treated as a linear partial differential 
equation for $\Omega$ for each given $i$. If 
$\frac{\partial}{\partial z^{i *}}\frac{f(z, z^\dag) }{\overline{W}_{M,1}(z^\dag)}\neq 0$ there should exist at 
least one $p\geq 2$, call it $p=2_b$, such that $\frac{\partial 
\xi^{2_b}}{\partial z^{i *}} \neq 0$ in order for Eq.~(\ref
{eq:EqOmegahat}) to be satisfied. 
The general solution of Eq.~(\ref{eq:EqOmegahat}) for the unknown function $\Omega$  is similar to that 
of Eq.(\ref{eq:secondformEq2}) for $W_{M-1}$, the ${\cal U}^{1 p}_{S}$ playing the role
of the $\Phi^a$, the $\frac{\partial \xi^p}{\partial z^{i *}}$
the role of the $\overline{W}_{M, a}$  of Eqs.(\ref{eq:WM-1sol},
\ref{eq:xirelation}), $\frac{\partial}{\partial z^{i *}}\frac{f(z, z^\dag) }{\overline{W}_{M,1}(z^\dag)}$ the role of $f(z, z^\dag)$
and the index $2_b$ the role of the index $1$. It reads, 

\beqa
\Omega=\left(\frac{\partial \xi^{2_b}}{\partial z^{i *}}\right)^{-1} \!\!\!\! \frac{\partial}{\partial z^{i *}}\frac{f(z, z^\dag) }{\overline{W}_{M,1}(z^\dag)} {\cal U}^{1 2_b}_{S} +  
\Xi_i({\cal V}^{2_b 2}_{S,i}, \dots,{\cal V}^{2_b p}_{S,i}, \dots, {\cal V}^{2_b k_1}_{S,i}; \dots,\widetilde{\Phi}^a; z), \nonumber \\
\label{eq:Omegahat0}
\eeqa
 compare with Eq.~(\ref{eq:WM-1sol}), 
 where $\Xi_i$ is an arbitrary function of its entries and
 we have defined
 \beqa
 {\cal V}^{2_b p}_{S, i} \equiv \gamma_i^{2_b} {\cal U}^{1 p}_{S} - 
 \gamma^p_i {\cal U}^{1 2_b}_{S}\ ,
 \label{eq:Vdef}
 \eeqa
 (compare with Eq.~(\ref{eq:Udef})), with the $\gamma$'s satisfying
 \beqa
 \gamma^p_i \frac{\partial \xi^{2_b}}{\partial z^{i *}}=\gamma_i^{2_b} \frac{\partial \xi^p}{\partial z^{i *}}
 \label{eq:gammarelation}, \;\; \forall p=2, \dots,k_1\; , \; 
 \eeqa
 (compare with Eq.~(\ref{eq:xirelation}); note that ${\cal V}^{2_b 2_b}_{S,i} =0 $.) The $\Xi_i$ functions do not have explicit
 dependence on $z^\dag$ since $\Omega$ does not. This is a direct consequence of the holomorphy of $W_{M-1}$ in $z$ at the
 field space point $\widetilde{\Phi}^a = S^p =0, \forall a,p$.  
 Similarly we note also that $\gamma_i^{2_b}$ should be holomorphic in $z$ as a consequence of the holomorphy of  
$W_{M-1}$ at the field space point $\widetilde{\Phi}^a = S^p =0, \forall a,p = 1,...,r-1, r+1,...$
and $S^r \neq 0$ where $r\neq 1, 2_b$. Indeed the only term left in 
$W_{M-1}$ in this case, cf. Eqs.~(\ref{eq:WM-1solsecond}, \ref{eq:Omegahat0}), is
$\Xi_i(0,...,0,\gamma_i^{2_b} \xi^1_r S^r,0,...0;z)$. The holomorphy 
of $\gamma_i^{2_b}$ in $z$ then follows from the holomorphy of $
\xi^1_r$, and of that of $W_{M-1}$ in $z$ at this field space point.

 For the sake of generality we have assumed that there is at least
 one nonholomorphic function $\xi^{2_b}$, i.e.
 $\frac{\partial \xi^{2_b}}{\partial z^{i *}} \neq 0$, but 
 we allow
 as well a subset $\xi^q$, with $2_b < q_{0b} \leq q \leq k_1$ 
 of the $\xi$ functions to be holomorphic. It follows 
 from Eq.~(\ref{eq:gammarelation}) that 
 $\gamma^q_i =0$ for this subset, and the associated entries
 ${\cal V}^{2_b q}_{S,i}$ reduce to ${\cal U}^{1 q}_S$ where the
 common $\gamma_i^{2_b}$ factors are absorbed in a re-definition
 of $\Xi_i$. Equation (\ref{eq:Omegahat0}) takes then the form
 \beqa
\Omega=\left(\frac{\partial \xi^{2_b}}{\partial z^{i *}}\right)^{-1}\!\!\! \frac{\partial}{\partial z^{i *}}\frac{f(z, z^\dag) }{\overline{W}_{M,1}(z^\dag)} {\cal U}^{1 2_b}_{S} + \Xi_i({\cal V}^{2_b 2}_{S,i}, \dots,{\cal V}^{2_b (q_{0b}-1)}_{S,i},~~~~~~~~&& \nonumber \\
{\cal U}^{1 q_{0b}}_S \dots, {\cal U}^{1 k_1}_S; \dots,\widetilde{\Phi}^a; z).&&
\label{eq:Omegahat}
\eeqa

In the following it will prove necessary  to distinguish the two generic cases:

\begin{itemize}
\item[]{\sl single nonholomorphy}: $\frac{\partial}{\partial z^{i *}}\frac{f(z, z^\dag) }{\overline{W}_{M,1}(z^\dag)} \neq 0$
only for one given $i$, 
\item[]{\sl multiple nonholomorphy}: $\frac{\partial}{\partial z^{i *}}\frac{f(z, z^\dag) }{\overline{W}_{M,1}(z^\dag)} \neq 0$
for several $i$'s.
\end{itemize}
[The special case where $\frac{\partial}{\partial z^{i *}}\frac{f(z, z^\dag) }{\overline{W}_{M,1}(z^\dag)} = 0$ for all $i$ will
be addressed later on.]
For both single and mutiple nonholomorphies  there should exist at 
least one $p\geq 2$, call it $p=2_b$, such that $\frac{\partial 
\xi^{2_b}}{\partial z^{i *}} \neq 0$ in order for Eq.~(\ref
{eq:EqOmegahat}) to be satisfied. 
In both cases there can
exist other indices $2_c$ of the same type, i.e. satisfying 
$\frac{\partial \xi^{2_c}}{\partial z^{i *}} \neq 0 $; 
moreover, in 
the {\sl multiple nonholomorphy} case one should also consider indices satisfying {\sl simultaneously}
$\frac{\partial \xi^{2_d}}{\partial z^{i *}} \neq 0$ and 
$\frac{\partial \xi^{2_d}}{\partial z^{j *}} \neq 0$, with $i\neq j$ and $d$ being possibly equal or not to $b$ or $c$,... 
These distinctions actually
lead to different requirements on the allowed general forms
for $W_{M-1}$. 

In the {\sl single nonholomorphy} case this form of $\Omega$ 
is necessary  and sufficient 
to insure the holomorphy of $W_{M-1}$ and the latter is finally given by,
\beqa
 W_{M-1}(z, S,  \widetilde{\Phi}) &=& \frac{f(z, z^\dag) }{\overline{W}_{M,1}(z^\dag)} S^{1} +\Big(\frac{\partial \xi^{2_b}}{\partial z^{i *}}\Big)^{-1} \frac{\partial}{\partial z^{i *}}\frac{f(z, z^\dag) }{\overline{W}_{M,1}(z^\dag)} {\cal U}^{1 2_b}_{S} \; + \nonumber \\
 && \Xi_i({\cal V}^{2_b 2}_{S,i}, \dots,{\cal V}^{2_b (q_{0b}-1)}_{S,i},
{\cal U}^{1 q_{0b}}_S \dots, {\cal U}^{1 k_1}_S; \dots,\widetilde{\Phi}^a; z). \nonumber \\
 \label{eq:WM-1solfinal1}
 \eeqa
 Note that due to the single nonholomorphy corresponding to one specific index $i$, 
 Eq.~(\ref{eq:EqOmegahat}) is by definition trivially satisfied
 for all hidden sector indices other than $i$.
 Moreover, if there exists at least one other index of
 the $2_b$ type, say $2_c$, one can readily recast Eq.~(\ref{eq:WM-1solfinal1}) in terms of $2_c$ similarly to the discussion
 following   Eq.~(\ref{eq:difference}), due to the identities
 \beqa
{\cal V}^{2_b 2_c}_{S,i} &=& - {\cal V}^{2_c 2_b}_{S,i},
\label{eq:identity0V} \\
{\cal V}^{2_c p}_{S,i} &=& \frac{\gamma^{2_c}}{\gamma^{2_b}} {\cal V}^{2_b p}_{S,i}
- \frac{\gamma^p}{\gamma^{2_b}} {\cal V}^{2_b 2_c}_{S,i}.
\label{eq:identityV}
\eeqa

In the {\sl multiple nonholomorphy} case the issue becomes more
involved. For instance considering the nonholomorphy with respect
to two distinct fiels $z^i$ and $z^j$, $\Omega$ can now be
written in two different forms. The requirement that they should
match brings extra constraints. Consider two indices $2_b$
and $2_c$ satisfying $\frac{\partial \xi^{2_b}}{\partial z^{i *}} \neq 0$ and $\frac{\partial \xi^{2_c}}{\partial z^{j *}} \neq 0$.
 Writing Eq.~(\ref{eq:Omegahat}) for the two sets $(2_b, i)$
 and $(2_c,j)$ leads to the requirement,
 \beqa
 \Xi_i({\cal V}^{2_b 2}_{S,i} \,\!\!\dots,{\cal V}^{2_b 2_c}_{S,i} \!\!\!\!\dots,{\cal V}^{2_b (q_{0b}-1)}_{S,i},
{\cal U}^{1 q_{0b}}_S \!\!\!\!\dots, {\cal U}^{1 k_1}_S, \dots,\widetilde{\Phi}^a; z) =~~~~~~~~~~~~~~~~~~~~~~\nonumber \\
\Xi_j({\cal V}^{2_c 2}_{S,j} \,\!\!\dots,{\cal V}^{2_c 2_b}_{S,j} \!\!\!\!\dots,{\cal V}^{2_c (q_{0c}-1)}_{S,j},
{\cal U}^{1 q_{0c}}_S \!\!\!\!\dots, {\cal U}^{1 k_1}_S, \dots,\widetilde{\Phi}^a; z) \, + ~~~~~~\nonumber \\ 
~~~~~~~~~~~~~~~~~~~~\Big(\frac{\partial \xi^{2_c}}{\partial z^{j *}}\Big)^{-1} \frac{\partial}{\partial z^{j *}}\frac{f(z, z^\dag) }{\overline{W}_{M,1}(z^\dag)} {\cal U}^{1 2_c}_{S} -
\Big(\frac{\partial \xi^{2_b}}{\partial z^{i *}}\Big)^{-1} \frac{\partial}{\partial z^{i *}}\frac{f(z, z^\dag) }{\overline{W}_{M,1}(z^\dag)} {\cal U}^{1 2_b}_{S}, \nonumber \\
\label{eq:constraint}
\eeqa
for given $i\neq j$. This equation is similar to Eq.~(\ref
{eq:difference}) where now the $\Xi$'s, ${\cal V}_S$'s, ${\cal U}_S$'s and $2_b$, $2_c$ 
play respectively the role of the $\Omega$'s, ${\cal U}_\Phi$'s, 
$\Phi$'s and $1$, $2$ of that equation. We should stress however
that contrary to Eq.~(\ref{eq:difference}) which is allowed not to hold when $\Omega_1$ and $\Omega_2$
span different sets of solutions, here the constraint (\ref{eq:constraint}) is mandatory since
the {\sl same} function $\Omega$ should satisfy Eq.~(\ref{eq:EqOmegahat}) for {\sl all} $i$.
 One has thus to deal   
correspondingly with requirements similar to those of (i) and (ii) 
(see the discussion following Eq.(\ref{eq:difference})):
\begin{itemize}
\item[(i)'] $\Xi_i$ and $\Xi_j$ should be such that the entries
${\cal V}^{2_b p}_{S,i}$ can be transformed into the entries
${\cal V}^{2_c p}_{S,j}$.
\item[(ii)'] $\Xi_i$ and/or $\Xi_j$ should have linear contributions in order to account for
the linear terms in ${\cal U}_S$ on the right-hand side of 
Eq.~(\ref{eq:constraint}).
\end{itemize} 
Contrary to the case of {\sl single nonholomorphy} here we
do not have identities similar to Eqs.~(\ref{eq:identity0V}, \ref{eq:identityV}), 
and (i)', (ii)' are not trivially satisfied. For
one thing the linear terms on the right-hand side of Eq.~(\ref
{eq:constraint}) cannot be recast as a linear term in ${\cal V}^
{2_b 2_c}_{S,i}$ or ${\cal V}^{2_c 2_b}_{S,j}$ without assuming additional
constraints involving $f(z, z^\dag)$. For another, unlike Eq.~(\ref
{eq:identity0V}), 
one cannot relate ${\cal V}^{2_b 2_c}_{S,i}$ to $-
{\cal V}^{2_c 2_b}_{S,j}$ as can be seen from Eq.~(\ref
{eq:Vdef}) even with the rescaling freedom in the 
definition of the $\gamma$'s, unless a constraint is imposed
involving $\gamma_i^{2_b}, \gamma_j^{2_c}, \gamma_i^{2_c}$ and $\gamma_j^{2_b}$. As a consequence one cannot find a relation similar
to Eq.~(\ref{eq:identity}) that would express the ${\cal V}^{2_c a}_{S,j}$'s as functions of the  ${\cal V}^{2_b a}_{S,i}$'s unless further
constraints are imposed on the $\gamma$'s. We come back later on to the constraints required by (i)' and (ii)'.

To be more specific, we
assume without loss of generality that $q_{0b} \geq q_{0 c}$
so that $\Xi_j$ has more ${\cal U}_S$ entries and less
${\cal V}_S$ entries than $\Xi_i$. As can be seen from the
definitions (\ref{eq:Udef}, \ref{eq:Vdef}), a ${\cal U}^{1 q}_S$
entry of $\Xi_i$ cannot be obtained from linear combinations 
involving ${\cal U}_S$ and ${\cal V}_S$ entries of $\Xi_j$.
It can thus only correspond to the same entry ${\cal U}^{1 q}_S$
of $\Xi_j$. It follows that $\Xi_j$ should be chosen
constant in the extraneous ${\cal U}^{1 q}_S$ entries that do not 
appear in $\Xi_i$. Similarly one sees from  Eqs.~(\ref{eq:Udef}, 
\ref{eq:Vdef}) that each ${\cal V}^{2_b p}_{S,i}$ 
variable of $\Xi_i$ can, at best, be obtained only from a 
linear combination
of the corresponding variable  ${\cal V}^{2_c p}_{S,j}$
and ${\cal V}^{2_c 2_b}_{S,j}$ in $\Xi_j$. It follows that $\Xi_i$
should also be taken as constant in the extraneous variables ${\cal V}^{2_b p}_{S,i}$ for which the corresponding variables
${\cal V}^{2_c p}_{S,j}$ do not appear in $\Xi_i$. 

For the remaining entries
of $\Xi_i$ and $\Xi_j$, 
the consistency of Eq.~(\ref{eq:constraint}) requires the 
existence  of functions $\alpha, \beta, \kappa, \alpha_p$ and $
\kappa_p$ holomorphic in $z$, such that the following 
linear combinations are satisfied:
\beqa
\sum_{p> 1} \kappa_p {\cal V}^{2_b p}_{S,i} 
+ \Big(\frac{\partial \xi^{2_b}}{\partial z^{i *}}\Big)^{-1} \frac{\partial}{\partial z^{i *}}\frac{f(z, z^\dag) }{\overline{W}_{M,1}(z^\dag)} {\cal U}^{1 2_b}_{S} = ~~~~~~~~~~~~~~~~~~~~~~~~~~~~~~~\nonumber \\
\sum_{p> 1} \alpha_p {\cal V}^{2_c p}_{S,j} +
\Big(\frac{\partial \xi^{2_c}}{\partial z^{j *}}\Big)^{-1} \frac{\partial}{\partial z^{j *}}\frac{f(z, z^\dag) }{\overline{W}_{M,1}(z^\dag)} {\cal U}^{1 2_c}_{S},  \label{eq:linear2}
\eeqa
and 
\beqa
{\cal V}^{2_b p}_{S,i} + \kappa {\cal V}^{2_b 2_c}_{S,i}= \alpha {\cal V}^{2_c p}_{S,j} + \beta {\cal V}^{2_c 2_b}_{S,j} \ , \;\; 
\forall p = 2, \cdots, q_{0b}-1 \ . \label{eq:linear1}
\eeqa
 
The holomorphy of $\alpha, \beta, \kappa, \alpha_p$ and $\kappa_p$  is a necessary requirement. It can be proved analogously to the case of $\gamma_i^{2_b}$ discussed after Eq.~(\ref{eq:Omegahat}). 

Equation (\ref{eq:linear2}) expresses in the most general way 
requirement (ii)'. [Note, though, that we
assumed without loss of generality $\frac{\partial \xi^{2_c}}{\partial z^{i *}} \neq 0$
and $\frac{\partial \xi^{2_b}}{\partial z^{j *}} \neq 0$,
so as not to have ${\cal U}^{1 2_c}_{S}$ and ${\cal U}^{1 2_b}_{S}$ entries respectively in $\Xi_i$ and $\Xi_j$ that would lead
to a special form replacing Eq.~(\ref{eq:linear1}).] 
\gilbertzero{Equating the coefficients of the independent fields in 
Eq.~(\ref{eq:linear2}), 
one finds}

\beqa
f(z,z^\dag) &=& \sum_{p \geq 1} \overline{W}_{M,p}(z^\dag) W_{M-1,p}(z),
\label{eq:f}
\eeqa
 where we defined 
\beqa
W_{M-1,p}(z) \equiv \xi^1_p \gamma_j^{2_c} \alpha_p \ , \forall p\neq1 \ ,
\label{eq:WM-1-def}
\eeqa
 \gilbertzero{and $W_{M-1,1}(z)$ an arbitrary holomorphic function.
 \gilbertzero{(For details on the intermediate steps see appendix B.1 of the arXiv version $1$ arXiv:1611.10327v1, \cite{Moultaka:2016frs}.)} 
 Thus $f(z,z^\dag)$ has exactly the functional form that would have been 
 expected from Eq.(\ref{eq:secondformEq2}).}

\gilbertzero{Equation (\ref{eq:linear1}) expresses requirement (i)', but only in the case where 
the $\Xi_i$'s have further nonlinear dependences on the ${\cal V}_S$ entires. Indeed if the $\Xi_i$'s were linear then
Eq.~(\ref{eq:linear2}) would satisfy by itself both requirements
(i)' and (ii)', and Eq.~(\ref{eq:linear1}) would be irrelevant.
Barring this simple case, general consistency requirements that arise when 
equating the coefficients of the independent fields 
 ${\cal U}_S^{1 2_b}$, ${\cal U}_S^{1 2_c}$ and ${\cal U}_S^{1 p}$,
 in Eq.~(\ref{eq:linear1}), cf. Eqs.~(\ref{eq:Udef}, 
 \ref{eq:Vdef}), lead to the following proportionality relations:}
\beqa
\xi^p(\xi^{2_b}(z, z^\dag), z) = 
\nu_p(z) \xi^{2_b}(z, z^\dag),
\eeqa
with $\nu_p(z)$ an arbitrary holomorphic function in $z$, 
\beqa
\frac{\overline{W}_{M, p}(z^\dag)}{\overline{W}_{M, 2_b}(z^\dag)} = \frac{\xi^1_{2_b}}{\xi^1_p} \,
\frac{\xi^p(\xi^{2_b}(z, z^\dag), z)}{\xi^{2_b}(z, z^\dag)} = \frac{\xi^1_{2_b}}{\xi^1_p} \, \nu_p(z) \equiv \mu_p, 
\label{eq:relation}
\eeqa
with $\mu$ a constant in $z, z^\dag$, and
\beqa
\frac{\gamma^p_i}{\gamma_i^{2_b} }= \frac{\gamma^p_j}{\gamma_j^{2_b}}
=\nu_p(z).
\label{eq:relationprime}
\eeqa
\gilbertzero{The above relations result from a rather involved proof 
that we
do not reproduce here. The reader is referred to  appendix B.1 of 
the arXiv version $1$ (arXiv:1611.103271v1, \cite{Moultaka:2016frs}) for
full details.}

It follows from Eq.~(\ref{eq:relationprime}) that all the $\gamma^p_i, \gamma^p_j$ are holomorphic in
$z$ since it was shown that $\gamma_i^{2_b}$ and $\gamma_j^{2_b}$
should be holomorphic. In contrast $\xi^{2_b}$ and $\xi^p$ (with $p\neq 1$) 
are not forced to be holomorphic. However, one has from Eqs.~(\ref{eq:relation}, \ref{eq:relationprime})
that
\beqa 
\gamma^p_i \xi^{2_b} - \gamma_i^{2_b} \xi^p = 0 \ ,
\label{eq:gaxirelation}
\eeqa 
which implies
\beqa
{\cal V}_{S, i}^{2_b p} &=& \xi^1_p(z) \gamma_i^{2_b}(z) S^p - \xi^1_{2_b}(z) \gamma_i^p(z) S^{2_b} =
\gamma_i^{2_b}(z) \big( \xi^1_p(z) S^p - \xi^1_{2_b}(z) \nu_p(z) S^{2_b} \big) \nonumber \\
&=& \gamma_i^{2_b}(z) \, {\cal U}_{S}^{2_b p}  \ .
\label{eq:VrelU}
\eeqa
The two $S^1$ terms, containing the nonholomorphic dependence of ${\cal V}_{S, i}^{2_b p}$ have cancelled out, leading to a holomorphic
quantity which, moreover, is proportional to a ${\cal U}_{S}^{2_b p}$ form where ${\cal U}_{S}^{2_b p} \equiv \widetilde{\xi}^{2_b}_p S^p -
\widetilde{\xi}^{p, 2_b} S^{2_b}$, cf. Eq.~(\ref{eq:Udef}), with the 
definitions $\widetilde{\xi}^{2_b}_p \equiv \xi^1_p(z)$ and $\widetilde{\xi}^{p, 2_b} \equiv \xi^1_{2_b}(z) \nu_p(z)$. 
The consistency of the 
latter definitions, see Eq.~(\ref{eq:xirelationr}), requiring the relation  
\beqa
\frac{\widetilde{\xi}^{p, 2_b}}{\widetilde{\xi}^{2_b}_p} = \frac{\overline{W}_{M, p}(z^\dag)}{\overline{W}_{M, 2_b}(z^\dag)} \ ,
\eeqa 
follows immediately from (\ref{eq:relation}).
This leads to the important conclusion that the function $\Xi_i$ in Eq.~(\ref{eq:Omegahat}) is in fact 
reduced to a function of only ${\cal U}_S$'s with no dependence on $z^\dag$ and where the $\gamma_i^{2_b}(z)$ factor
in Eq.~(\ref{eq:VrelU}) is absorbed in a redefinition of $\Xi_i$. Moreover, the ${\cal U}_S$'s have to correspond at 
least to two disjoint sets of $S$-type fields since the relation (\ref{eq:VrelU}) is valid only for $p=2, \cdots, q_{0b} -1$, 
while the initial ${\cal U}_S^{1 p}$ entries in  Eq.~(\ref{eq:Omegahat}) correspond by construction
to $p= q_{0b}, \cdots, k_1$. One can easily check that the above properties established for $2_b$ are also valid for $2_c$
or for that matter any supplementary index $2_d$ of the same type. {\sl The general form should thus contain several disjoint $S$-field 
sets corresponding to disjoint sets of ${\cal U}_S$ entries.}

Before summarizing the results we note that the linear contribution
Eq.~(\ref{eq:linear2}) can now be recast in the form
\beqa
\sum_{p> 1}^{k_1} \alpha_p {\cal V}^{2_c p}_{S,j} +
\Big(\frac{\partial \xi^{2_c}}{\partial z^{j *}}\Big)^{-1} \frac{\partial}{\partial z^{j *}}\frac{f(z, z^\dag) }{\overline{W}_{M,1}(z^\dag)} {\cal U}^{1 2_c}_{S}
= \sum_{p \geq 1}^{k_1} W_{M-1, p}(z) S^p - \frac{f(z, z^\dag)}{\overline{W}_{M, 1}(z\dag)} S^1, \nonumber \\ \label{eq:linearterm}
\eeqa
where we have used Eqs.~(\ref{eq:WM-1-def}, \ref{eq:f}, 
\ref{eq:xirelation}), the holomorphy of $\xi^1_p$ in $z$ and 
Eq.~(\ref{eq:gammarelation}). Note that the right-hand side of
the above equation is independent of the $j$ index, in accordance 
with the requirement of Eq.~(\ref{eq:linear2}).

Putting all the pieces together we can now write down the general solutions for
$W_M$ and $W_{M-1}$. This entails the choice of an arbitrary partition of the set of $k_1$ $S$-type fields into $P$ 
disjoint subsets. We thus have
\beqa
k_1 = \sum_{p\geq 1}^P n_p \ , \label{eq:partition}
\eeqa
where $p=1,...,P$ label the subsets and $n_p$ denotes the number of $S$-type fields
forming the $p^{th}$ subset.  We also denote by $S^{p_s}$, with $s=1,...,n_p$, the $S$-type fields of the $p^{th}$ subset.
 With these notations the general solutions are summarized as follows:  

\beqa
W_M(z, S, \widetilde{\Phi})&=& W_{M,0}(z) +  \sum_{p \geq 1}^P W_{M,p}(z) \sum_{s \geq 1}^{n_p}  \; \mu_{p_s}^*\,S^{p_s} \ , \label{eq:finalformWM} \\
W_{M-1}(z, S, \widetilde{\Phi})&=& \sum_{q \geq 1}^{k_1} W_{M-1,q}(z) 
\; S^{q}  + \Xi(...,
{\cal U}^{p p_s}_S...;...,\widetilde{\Phi}^a,...; ...,z^i,...), \nonumber \\
\label{eq:finalformWM-1}
\eeqa
where $\Xi$ is an arbitrary function of all its entries, 
\beqa
{\cal U}^{p p_s}_{S} = \xi_{p_s}(z) S^{p_s} - 
\xi^{p_s}(z) S^{p_1}, \; {\rm with}\; p=1,\dots,P, \; {\rm and} \; s=1, \dots, n_p, 
\label{eq:UdefFINAL}
\eeqa
and
\begin{itemize}
\item[-] $\xi_{p_s}(z)$ and $\xi^{p_s}(z)$ should satisfy:
              \begin{equation}
                \mu_{p_s}  \; \xi_{p_s}(z) = \mu_{p_1}  \; \xi^{p_s}(z), \label{eq:ximureplation}
              \end{equation}
          normalized to   $\xi_{p_1}(z)=\xi^{p_1}(z)$,
(note the non-ambiguous simplified notation,  $\xi^p_{p_s}\to \xi_{p_s}$, $\xi^{p_s}_p\to \xi^{p_s}$),
\item[-] $W_{M, 0}(z), W_{M, p}(z), W_{M-1, q}(z)$ and $\xi_{p_s}(z)$ (or $\xi^{p_s}(z)$) are arbitrary holomophic functions of the 
$z^i$ fields, 
\item[-]  $\mu_{p_s}$ are arbitrary complex-valued constants,
\item[-] $P$, $n_p$, and $k_1$ are arbitrary positive integers satisfying Eq.~(\ref{eq:partition}), 
\item[-] $a=1, \dots, k_0$ with $k_0$ an arbitrary positive integer, labels the $\widetilde{\Phi}$-type fields.
\item[-] $q=1, \dots, k_1$ with $k_1$ an arbitrary positive integer, labels the $S$-type fields.
\item[-] $i=1, \dots, \ell$ with $\ell$  an arbitrary positive integer,  labels the $z$ fields.
\end{itemize}
We stress that while the $\mu_{p_s}$'s enter Eq.~(\ref{eq:ximureplation}), it is their complex-conjugate that 
enter Eq.~(\ref{eq:finalformWM}).
Note finally that the sum in Eq.~(\ref{eq:finalformWM-1}) is a shorthand notation for
\beqa
\sum_{q \geq 1}^{k_1} W_{M-1,q}(z) 
\; S^{q}  \equiv 
\sum_{p \geq 1}^P  \sum_{s \geq 1}^{n_p} W_{M-1,p_s}(z) \; \,S^{p_s} . \label{eq:notation}
\eeqa

We close this discussion by a few remarks---the result we arrived at implies that all the relevant $\xi$'s are holomorphic
in $z$. This is not to mean that it would have been equivalent to choosing from the start {\sl all} the $\xi$'s holomorphic 
in Eq.~(\ref{eq:WM-1solprime}). In the latter case all the ${\cal U}^{r,a}_S$ sets would be related to each other
thanks to identities similar to Eq.~(\ref{eq:identity}). The holomorphy structure in Eq.~(\ref{eq:xirelationr}) would then imply
that the $\overline{W}_{M, a}(z^\dag)$ are {\sl all} expressed in terms of one single function of $z^\dag$ up to constant factors.
This is obviously less general than what we found through the above detailed analysis and corresponds simply to the
special choice $P=1$ and $n_1=k_1$ in (\ref{eq:partition}), (\ref{eq:finalformWM}), (\ref{eq:finalformWM-1}), that is the trivial partition 
of the $S$-type fields set---the important point in our general result is that if some of the $\xi$'s are not holomorphic then
cancellations should take place separately in each entry of $\Xi$ in such a way that generates partitions of disjoint subsets of
$S$-type fields---as stated after Eq.~(\ref{eq:Eqholomorphy}) the proof leading to Eqs.~(\ref{eq:finalformWM}, 
\ref{eq:finalformWM-1}) has been carried out
in a simplified version of Eq.~(\ref{eq:WM-1solsecond}) where just one set of ${\cal U}^{r p}_S$ entries was assumed for $\Omega$ 
with fixed $r=1$. It is legitimate to ask whether in the absence of this simplifying assumption one would get more general solutions. 
But we see no memory of this assumption in the final form of our solutions. In particular the $\Xi$ function
contains all the $S$-type fields albeit in the form of partitions into disjoint subsets. \\


\noindent
$\bullet$ {\bf a non SW solution:}
If we truncate the expansion in Eq.~(\ref{eq:Wexpansion}) at $M=1$,
putting $W_M=0$ for all $M\geq 2$, then the above solutions give 
us the most general form of the superpotential,

\beqa
W(z, S, \widetilde{\Phi})&=& m_{p\ell} W_1(z, S, \widetilde{\Phi}) +
W_0(z, S, \widetilde{\Phi})  , 
\eeqa
where $W_1$ and $W_0$ are given respectively by
Eq.~(\ref{eq:finalformWM}) and Eq.~(\ref{eq:finalformWM-1})
with $M=1$. Note that the SW solution Eq.~(\ref{eq:SW})
meets this one only in the case where the $S$-type fields
sector is not present and $W_2(z):=0$, thus confirming that the SW form
is not the most general.

When $M>1$ additional constraints must be studied originating
from lower strictly positive $c$ powers of $m_{p\ell}$ in \eqref
{eq:Vgen} that we now consider.\\

\noindent
$\bullet \; \mathbf{c=2M-2:}$ 
relevant when $M\geq 2$, the term with $c=2M-2$  in the 
master equation \eqref{eq:Vgen} 
gets contributions from the $0$-sum and the $2$-sum terms
and leads to the constraint,
\beqa
&&\Big(\frac{\partial W_{M-2}}{\partial \Phi^a}\frac{\partial \overline W_{M}}{\partial \Phi^{a *}} + \text{h.c.} \Big) +
\frac{\partial W_{M-1}}{\partial \Phi^a}\frac{\partial \overline W_{M-1}}{\partial \Phi^{a *}} +
\big({\cal D}_i W_M\big)\big(\overline {\cal D}_i \overline W_M\big) \nonumber \\
&& ~~~~~~~~~~~+
 \Big( \Phi^a  \frac{\partial W_{M}}{\partial \Phi^a} 
\overline W_{M} + \text{h.c.} \Big)
-3 W_M \overline W_M\sim_\Phi 0, \nonumber \\
\label{eq:WM-2}
\eeqa
where we use from now on the shorthand notation,
\beqa
{\cal D}_i W \equiv \frac{\partial W}{\partial z^i} + z^{i *} W , \\
\overline {\cal D}_i \overline W \equiv \frac{\partial \overline W}{\partial z^{i *}} + z^i \overline W .
\eeqa
Recall that summation over repeated indices is understood. 
Applying $\displaystyle \frac{\partial}{\partial \widetilde{\Phi}^b}
\frac{\partial}{\partial \widetilde{\Phi}^{b *}}$ to the above 
equation and using Eqs.~(\ref{eq:finalformWM} \ref{eq:finalformWM-1}) one obtains
\beqa
\Big|\frac{\partial^2 W_{M-1}}{\partial \Phi^a \partial \widetilde{\Phi}^b} \Big|^2 = \Big|\frac{\partial^2 \Xi }{\partial \Phi^a \partial \widetilde{\Phi}^b} \Big|^2 =0 \label{eq:Xiconstraint} \ .
\eeqa
Since this constraint is a sum over positive definite terms, it is valid separately for fixed $a,b$.
Applying it to the case $\Phi^a \equiv S^p$ implies,
\beqa
\frac{\partial^2 \Xi }{\partial S^p \partial \widetilde{\Phi}^b} = 0, 
\eeqa
that is
$\Xi$ should be a direct sum of exclusively $\widetilde{\Phi}$
and $S$ dependent functions
\beqa
\Xi = \widetilde{\Xi}(\dots, \widetilde{\Phi}^b, \dots; z) + 
\Xi_S(\dots, S^p, \dots;z ) \label{eq:Xiform}\ .
\eeqa 
Applying further Eq.~(\ref{eq:Xiconstraint}) in the case
$\Phi^a \equiv \widetilde{\Phi^a}$ implies that $\widetilde{\Xi}$
should be linear in $\widetilde{\Phi}$ which we write as
\beqa
\widetilde{\Xi} =  \widetilde{W}_{M-1, a}(z) \;\widetilde{\Phi}^a + 
 \widetilde{W}_{M-1, 0}(z) \label{eq:Xiform1} \ .
\eeqa
Back to Eq.~(\ref{eq:WM-2}) to which we apply now
$\displaystyle \frac{\partial}{\partial {S}^p}
\frac{\partial}{\partial {S}^{p *}}$, one finds
\beqa
\Big|\frac{\partial^2 W_{M-1}}{\partial \Phi^a \partial {S}^p} \Big|^2   \sim_\Phi 0 \ .\label{eq:Xiconstraint1}
\eeqa
Given Eqs.~(\ref{eq:finalformWM-1}, \ref{eq:Xiform},
\ref{eq:Xiform1}), the above constraint is trivially satisfied
in the case $\Phi^a \equiv \widetilde{\Phi}^a$, while the case
$\Phi^a \equiv {S}^q$ leads to

\beqa
\Big|\frac{\partial^2 \Xi_S }{\partial S^p \partial {S}^q} \Big|^2
\sim_\Phi 0 \ .
\eeqa
The function $\Xi_S$ is thus at most quadratic in the $S$ fields. This is 
sufficient but necessary as well as can be seen by applying 
again
$\displaystyle \frac{\partial}{\partial {S}^r}
\frac{\partial}{\partial {S}^{r *}}$ to the above equation to
obtain
\beqa
\Big|\frac{\partial^3 \Xi_S }{\partial S^p \partial {S}^q \partial {S}^r} \Big|^2 = 0 \ .
\eeqa
Again, since this is a sum of positive definite terms then each
term of the sum should be seperately zero and the above equation
should hold also for any {\sl fixed} set of $p,q, r$. 
Thus $\Xi_S$ should  be necessarily of the form
\beqa
\Xi_S = \frac12 \sum_{p, q}  W_{M-1, p q}(z) \; S^p S^q+ 
\sum_{p}  W^{(S)}_{M-1, p}(z) \; S^p + W_{M-1, 0}(z) \ ,
\eeqa
where the functions $W^{(S)}_{M-1, p}(z)$ and $W_{M-1, p q}(z)$, 
the latter defined to be symmetric in $p,q$, 
are arbitrary albeit related among each other in such a way
that $\Xi_S$ is a function of the combinations ${\cal U}_S$ or
${\cal V}$ as dictated by Eq.~(\ref{eq:finalformWM-1}). Thus when $M \geq 2$ the 
function $W_{M-1}$ as given in Eq.~(\ref{eq:finalformWM-1}) is 
constrained further to a polynomial form,
\beqa
W_{M-1}(z, S,  	\widetilde{\Phi})&=&
\frac12 \sum_{p, q}  W_{M-1, p q}(z) \, S^p S^q + 
\sum_{p}  (W_{M-1, p}(z)+W^{(S)}_{M-1, p}(z)) \, S^p \nonumber \\
&&+
\widetilde{W}_{M-1, a}(z) \;\widetilde{\Phi}^a + 
  W_{M-1, 0}(z) \ , 
\label{eq:finalformWM-1prime}
\eeqa
where we kept distinct the two contributions to the linear part
in $S^p$ since $W^{(S)}$ must have a specific form, while
$W_{M-1, 0}(z) + \widetilde{W}_{M-1, 0}(z)$ is denoted by
$W_{M-1, 0}(z)$ without loss of generality. Although necessary,
the above form is not yet sufficient to fulfill Eq.~(\ref{eq:WM-2}). Injecting it back
into Eq.~(\ref{eq:WM-2}) and using Eq.~(\ref{eq:finalformWM})
one finds,
\beqa
&\sum_{p, q}  A_{p q} (z,z^\dag) \;S^p S^{q *} + 
\Big [\sum_p B_q(z, z^\dag) \; S^q + \overline{W}_{M, p}(z^\dag) 
\frac{\partial}{\partial S^p} W_{M-2}(z, S^r, \widetilde{\Phi}^a) 
 + \text{h.c.} \Big] \nonumber \\
 &~~~~~~~~~\sim_\Phi 0 \label{eq:WM-2prime}
\eeqa
with
\beqa
&&A_{p q}(z,z^\dag) 
= \sum_r W_{M-1, r p} \overline{W}_{M-1, r q} + \nonumber \\
&&\Big( \frac{\partial W_{M, p}}{\partial z^i} + z^{i *} W_{M, p}\Big) \Big( \frac{\partial \overline{W}_{M, q}}{\partial z^{i *}} + z^i \overline{W}_{M, q}\Big) - W_{M, p} \overline{W}_{M, q} \ , \label{eq:Apq}
\eeqa
\beqa
&& B_q(z, z^\dag)= \sum_p W_{M-1, p q} \big(\overline{W}_{M-1,p} + \overline{W}_{M-1,p}^{(S)}\big) + \nonumber \\
&&\Big( \frac{\partial W_{M, q}}{\partial z^i} + z^{i *} W_{M, q}\Big) \Big( \frac{\partial \overline{W}_{M, 0}}{\partial z^{i *}} + z^i \overline{W}_{M, 0}\Big) - 2 W_{M, q} \overline{W}_{M, 0} \ ,
\eeqa
and we have absorbed the remaining $\widetilde{\Phi}$- and 
$S$-independent
contributions in the right-hand side of (\ref{eq:WM-2prime}).
The latter equation is a partial differential equation
in the still unknown function $W_{M-2}$ (and $\overline{W}_{M-2}$)
with respect to the $S$ (and $S^\dag$)
fields. There is thus {\sl a priori} no reason to zero the
$B_q(z, z^\dag)$ and $A_{p q}(z,z^\dag)$ coefficients of the $S$
polynomial in order to match Eq.~(\ref{eq:WM-2prime}).  
However since the term $\overline{W}_{M, p}(z^\dag) 
\frac{\partial}{\partial S^p} W_{M-2}(z, S^r, \widetilde{\Phi}^a)$
and its hermitian conjugate present in Eq.(\ref{eq:WM-2prime})
are respectively holomorphic and anti-holomporphic in $S$,
one has necessarily
\beqa
A_{l r}(z,z^\dag) = 0 \; \forall l, r,  \label{eq:Apqzero}
\eeqa   
as can be easily seen by acting with the operator
$\displaystyle \frac{\partial}{\partial {S}^l}
\frac{\partial}{\partial {S}^{r *}}$ on Eq.(\ref{eq:WM-2prime}).
This equation reduces then to
\beqa
\sum_p B_q(z, z^\dag) S^q + \overline{W}_{M, p}(z^\dag) 
\frac{\partial}{\partial S^p} W_{M-2}(z, S^r, \widetilde{\Phi}^a)  
 = g(z, z^\dag) \sim_\Phi 0 \ ,\label{eq:WM-2prime1}
\eeqa
where $g$ denotes an arbitrary $\Phi$ independent function.\footnote{That the
h.c. in Eq.~(\ref{eq:WM-2prime}) can be dropped in (\ref{eq:WM-2prime1}) is intuitively clear, but can be seen by taking the derivative with respect to $S^r$ or with respect to 
$\widetilde{\Phi}^a$, showing respectively that the left-hand side of 
(\ref{eq:WM-2prime1}) should be $\sim_S 0$ and $\sim_{\widetilde{\Phi}} 0$.} This
differential equation is similar to Eq.~(\ref{eq:secondformEq2})
except that now we have a dependence on the variables $S^r$
in the nonhomogeneous part. Using well-known techniques in
the theory of quasi-linear differential equations of first order
 we find the
general solution\footnote{We refrain from detailing 
here the steps leading to Eq.~(\ref{eq:WM-2sol}). Suffice it to
recall that the general solution should be a function of an independent set of integrals of the characteristic ordinary 
differential equations system associated with Eq.~(\ref{eq:WM-2prime1}); for more details see e.g. \cite{CourantHilbert198901,decuyper1972modèles}. The upshot is that the general solution
to a multi-variable function $f(x_1, x_2,...x_n)$ satisfying a differential equation of the form
\beqa
\sum_{i=1}^n a_i \frac{\partial f}{\partial x_i} = b(x_1, x_2, ...),
\nonumber
\eeqa
where all the $a_i$ are $x$ independent and at least $a_1$
nonvanishing if $b$ is nonvanishing,
is
\beqa
f(x_1, x_2,...) = \frac{1}{a_1} \int_{x_0}^{x_1} dx \; 
b(x, \dots, x_i - \frac{\xi^i}{\xi^1} (x_1 - x), \dots) +
{\cal F}\left(\xi^1 x_2 - \xi^2 x_1, \dots, \right.\nonumber \\ 
\left. \xi^1 x_i - \xi^i x_1,
\dots, \xi^1 x_n - \xi^n x_1\right)
\nonumber
\eeqa
where $\xi^1 a_i = \xi^i a_1$, ${\cal F}$ an arbitrary
function of $n-1$ variables, and $x_0$ an arbitrary reference
constant.}
\beqa
W_{M-2} (z,  S^r, \widetilde{\Phi}^a) &=& \Big( \frac{g(z, z^\dag)}{\overline
{W}_{M, 1}}  -  \sum_{q \geq 2} B_q S^q \Big) S^1 + \frac12
\Big( \sum_{q \geq 2} \frac{\xi^q}{\xi^1} B_q  -  B_1 \Big) (S^1)^2  \nonumber \\
& & +  \; \Gamma(\dots, {\cal U}_S^{1 q}, \dots, \widetilde{\Phi}^a; z) \label{eq:WM-2sol}.
\eeqa 
Here ${\cal U}_S^{1 q}$ is as defined in Eq.~(\ref{eq:Udef})
and $\Gamma$ an arbitrary function. As 
in the case of Eq.~(\ref{eq:WM-1sol}), we assumed when writing the 
above 
solution  at least one nonzero $W_{M, p}$ to ensure that
the set of fields of the $S$-type is not empty. However,
as already mentioned after Eq.~(\ref{eq:WM-1solsecond}), the case
of no $S$-type fields is implicitly accounted for in the above
equation by dropping out from it all the $S$ dependence. Note
also that in the case of a single $S$-type field the summations
in Eq.~(\ref{eq:WM-2sol}) are trivially absent and $\Gamma$
becomes $S$-independent. Finally, one should in principle study further the constraint that the holomorphy of $W_{M-2}$ 
puts on Eq.(\ref{eq:WM-2sol}). However, it turns out that this is unnecessary to reach a definite conclusion, as the constraints originating
from lower values of $c$ to be studied below are much more restrictive.  
\\

\noindent
$\bullet \; \mathbf{c=2M-3:}$
relevant when $M \geq 2$, this term gets contributions from the $0$-sum and the $2$-sum terms. When $M \geq 3$
it leads to the constraint,
\beqa
&\displaystyle \Bigg[\frac{\partial W_{M}}{\partial \Phi^a}\frac{\partial \overline W_{M-3}}{\partial \Phi^{a *}} + \frac{\partial W_{M-2}}{\partial \Phi^a}\frac{\partial \overline W_{M-1}}{\partial \Phi^{a *}}
+\Big(\frac{\partial W_M}{\partial z^i} + z^{i *} W_M\Big)\Big(\frac{\partial \overline{W}_{M-1}}{\partial z^{i *}} + z^i \overline{W}_{M-1}\Big) &\nn\\
& + \Phi^a  \frac{\partial W_{M}}{\partial \Phi^a} 
\overline W_{M-1} + \Phi^a  \frac{\partial W_{M-1}}{\partial \Phi^a} 
\overline W_{M} 
-3 W_M \overline W_{M-1} \Bigg]+ \text{h.c.} & \nonumber \\
 &\sim_\Phi 0 .&
\label{eq:WM-3}
\eeqa
Specializing to $M=2$, the $W_{M-3}$ dependent term is absent by definition. Moreover, 
anticipating that $\frac{\partial W_{M}}{\partial \Phi^a}=0$ for $M \geq 3$, a result we will prove
independently below (see Eq.~(\ref{eq:emptyS}) in the $c=2 M - 4$ case), then again the $W_{M-3}$ dependent term is absent
in Eq.~(\ref{eq:WM-3}). For the sake of conciseness we can thus safely replace in the subsequent discussion Eq.~(\ref{eq:WM-3}) by
\beqa
&\displaystyle \Bigg[\frac{\partial W_{M-2}}{\partial \Phi^a}\frac{\partial \overline W_{M-1}}{\partial \Phi^{a *}} +
\Big(\frac{\partial W_M}{\partial z^i} + z^{i *} W_M\Big)\Big(\frac{\partial \overline{W}_{M-1}}{\partial z^{i *}} + 
z^i \overline{W}_{M-1}\Big) & \nonumber \\
&+  \Phi^a  \frac{\partial W_{M}}{\partial \Phi^a} 
\overline W_{M-1} + \Phi^a  \frac{\partial W_{M-1}}{\partial \Phi^a} 
\overline W_{M} -3 W_M \overline W_{M-1} \Bigg] + \text{h.c.} &\nonumber \\
& \;\;\;\;\;\;\;\;\;\;\;\;\;   \sim_\Phi 0 \ , & \label{eq:WM-3prime}
\eeqa
and use it for all $M\geq2$. Note, though, that we kept concistently a $\frac{\partial W_{M}}{\partial \Phi^a}$ term 
(harmless when $M \geq3$) since at this level it is not vanishing in the case $M=2$. Taking into account
$W_M$ and $W_{M-1}$ given by Eqs.~(\ref{eq:finalformWM}), (\ref{eq:finalformWM-1prime}), one rewrites (\ref{eq:WM-3prime})
as
\beqa
&\displaystyle \Bigg[\frac{\partial W_{M-2}}{\partial \Phi^a}\frac{\partial \overline W_{M-1}}{\partial \Phi^{a *}} + 
\frac{1}{2} K_{3, p q r} S^p S^q S^{r *} +
K_{2, p r} S^p S^{r *} + \widetilde{K}_{2, b r} \widetilde{\Phi}^b S^{r *} & \nonumber \\
& + \frac{1}{2} K'_{2, p q} S^p S^q + K_{1 r} S^{r *} + K'_{1, p} S^p + \widetilde{K}_{1, b} \widetilde{\Phi}^b \Bigg] 
+\; \text{h.c.} & \nonumber \\
& \;\;\;\;\;\;\;\;\;\;\;\;\;  \sim_\Phi 0, & \label{eq:WM-3prime1}
\eeqa
where the various $K$ factors are well-defined $S$-, $\widetilde{\Phi}$-independent functions of $z, z^\dag$. 
Here we list only the ones relevant for the
remainder of the discussion, 
\beqa
K_{3, p q r} &=& (\mathcal{D}_i W_{M-1, p q})(\overline{\mathcal{D}}_i \overline{W}_{M, r}), \label{eq:K3}\\
\widetilde{K}_{2, b  l} &=& (\mathcal{D}_i \widetilde{W}_{M-1, b})(\overline{\mathcal{D}}_i \overline{W}_{M, l}) - \widetilde{W}_{M-1, b} \overline{W}_{M, l} \label{eq:tildeK2}, \\
\widetilde{K}_{1, b} &=& (\mathcal{D}_i \widetilde{W}_{M-1, b})(\overline{\mathcal{D}}_i \overline{W}_{M, 0}) -2 \; 
\widetilde{W}_{M-1, b} \overline{W}_{M, 0}.
\label{eq:tildeK1}
\eeqa
Recall that the $a$ index in Eq.~(\ref{eq:WM-3prime1}) runs over all $S$ and $\widetilde{\Phi}$ fields. The holomorphy structure
of this equation allows to write down through an appropriate choice of derivatives simple necessary constraints:
\begin{itemize}
\item[1)] acting with the operator $\displaystyle \frac{\partial^2}{\partial S^{l *} \partial \widetilde{\Phi}^b}$ on Eq.~(\ref{eq:WM-3prime1}) 
with $b,l$ arbitrary fixed indices, one finds
\beqa
\overline{W}_{M-1,l p} \; \frac{\partial^2 W_{M-2}}{\partial S^p \partial \widetilde{\Phi}^b} + \widetilde{K}_{2, b  l} =0 \ .
\eeqa
\item[2)] acting with the operator $\displaystyle \frac{\partial^2}{\partial S^l \partial S^m \partial S^{n *}}$ on Eq.~(\ref{eq:WM-3prime1})
with $l,m,n$ arbitrary fixed indices one finds
\beqa
\overline{W}_{M-1,n q} \; \frac{\partial^3 W_{M-2}}{\partial S^q \partial S^l \partial S^m} + K_{3, n l m} =0 \ .
\eeqa
\end{itemize}
Combined with Eqs.~(\ref{eq:K3}, \ref{eq:tildeK2}), these holomorphy constraints take the following form:
\beqa
\overline{W}_{M-1,l p}(z^\dag) \; \alpha_{p b}(z)=\big(\mathcal{D}_i \widetilde{W}_{M-1, b}(z)\big)\;\big( \overline{\mathcal{D}}_i \overline{W}_{M, l}(z^\dag)\big) - \widetilde{W}_{M-1, b}(z) \; \overline{W}_{M, l}(z^\dag),&&\nonumber \\
\; \forall b, l,~~~~~~~~~ &&\label{eq:hol1}
\eeqa
and
\beqa
\overline{W}_{M-1,r p}(z^\dag) \; \alpha_{p q l}(z) = \big(\mathcal{D}_i W_{M-1, l q}(z)\big) \; 
\big(\overline{\mathcal{D}}_i \overline{W}_{M, r}(z^\dag)\big),  \;\; \forall l,q,r. \label{eq:hol2} 
\eeqa
The last equation has exactly the form of Proposition \ref{corol:genholo}. As a result, the set of functions
$W_{M-1, l q}(z)$ should all be vanishing if there exists at least one $r$ for which $W_{M,r}$ is nonzero. 
Injecting $W_{M-1, l q}(z)=0$ in (\ref{eq:Apq}) leads through (\ref{eq:Apqzero})
to 
\beqa
\big(\mathcal{D}_i W_{M, p}(z)\big) \big(\overline{\mathcal{D}}_i \overline{W}_{M, q}(z^\dag)\big) = 
W_{M, p}(z) \overline{W}_{M, q}(z^\dag),   \;\; \forall p, q,
\eeqa
which is of the form of Proposition \ref{prop:holo} with $\alpha =0$, $R=0$. However, in the special cases $p=q$
the above equation takes the form of  Proposition \ref{prop:holo} with $P=Q\equiv W_{M,p}$. This implies through  \ref{corol:holo1} that all $W_{M,p}$ should be 
vanishing, 
in contradiction with our starting assumption that at least one of them is not. This shows that the only consistent
result is
\beqa
\frac{\partial W_{M}}{\partial S^p }= W_{M,p}(z) = 0 , \; \forall p=1,...,k_1.  \label{eq:emptySfor2}
\eeqa
Recall that the proof of this result is strictly speaking valid here only for $M=2$. 
\begin{equation}
\text{ {\sl It implies that there are no $S$-type fields when $M=2$}}. \label{eq:emptyStext2}
\end{equation}
 As we anticipated earlier, to prove its validity for $M>2$ one needs to consider the $c=2M-4$
 case to which we turn now. 

\noindent
$\bullet \; \mathbf{c=2M-4 :}$ relevant when $M \geq 3$, this term gets contributions from the $0
$-sum, $2$-sum and $4$-sum terms. When $M \geq 4$
it leads to the constraint,
\beqa
\Bigg[\frac{\partial W_{M}}{\partial \Phi^a}\frac{\partial \overline W_{M-4}}{\partial \Phi^{a *}} + \frac{\partial W_{M-1}}{\partial \Phi^a}\frac{\partial \overline W_{M-3}}{\partial \Phi^{a *}} +
\Big(\frac{\partial W_M}{\partial z^i} + z^{i *} W_M\Big)\Big(\frac{\partial \overline{W}_{M-2}}{\partial z^{i *}} + z^i \overline{W}_{M-2}\Big) && \nonumber \\
~~~+  \Phi^a  \frac{\partial W_{M}}{\partial \Phi^a} 
\overline W_{M-2}
  + \Phi^a  \frac{\partial W_{M-1}}{\partial \Phi^a} 
\overline W_{M-1} + \Phi^a  \frac{\partial W_{M-2}}{\partial \Phi^a} 
\overline W_{M} 
-3 W_M \overline W_{M-2} \Bigg]+ \text{h.c.}&& \nonumber \\
+ \left|\frac{\partial W_{M-2}}{\partial \Phi^a}\right|^2
+\left|\frac{\partial W_{M-1}}{\partial z^i} + z^{i *} W_{M-1}\right|^2
-3 \left|W_{M-1}\right|^2  + \left|W_{M}\right|^2 \left|\Phi^a \right|^2 \;\sim_\Phi 0,~~~~~~~~~&& \nonumber \\
 ~~~~~~~~~~~~~~~~~~~~~~~~~~~~~~~~~~~~~~~~~~~~~&& \label{eq:WM-4}
\eeqa
and the special case $M=3$ is obtained by dropping in the above
equation the contribution of $W_{M-4}$.

Acting with the operator $\displaystyle \frac{\partial}{\partial \widetilde{\Phi}^b}
\frac{\partial}{\partial \widetilde{\Phi}^{b *}}$ where summation over 
the $b$ index is understood, on Eq.~(\ref{eq:WM-4}) and using
Eq.~(\ref{eq:finalformWM-1prime}) we obtain
\beqa
\Big|\frac{\partial^2 W_{M-2}}{\partial \Phi^a \partial \widetilde{\Phi}^b}\Big|^2 + 
\Big| \frac{\partial \widetilde{W}_{M-1,b}}{\partial z^i} + 
z^{i *} \widetilde{W}_{M-1,b}  \Big|^2 
- 2  \Big| \widetilde{W}_{M-1,b} \Big|^2 +  \Big| W_M \Big|^2 =0.~~~ \label{eq:constraintM}
\eeqa
Applying to the above equation the operator $\displaystyle \frac{\partial}{\partial S^q}
\frac{\partial}{\partial S^{q *}}$ where summation over 
the $q$ index is understood, recalling that the $\widetilde{W}_{M-1,b}$ are 
$\widetilde{\Phi}$ and $S$ independent [cf. Eq.~(\ref{eq:finalformWM-1prime}) and using
Eq.~(\ref{eq:firstform}) where by definition the $\Phi^a$ are only the $S$-type fields, 
or equivalently using Eq.~(\ref{eq:finalformWM}) for any chosen partition $P$, one finds
a sum of squares satisfying
\beqa
\Big|\frac{\partial^3 W_{M-2}}{\partial S^p \partial \Phi^a \partial \widetilde{\Phi}^b}\Big|^2 +
\Big| W_{M,q} \Big|^2 =0, \label{eq:dSSM-2}
\eeqa
that implies
\beqa
W_{M,q} = 0, \; \forall q=1,...,k_1.  \label{eq:emptyS}
\eeqa
Here $W_{M,q}$ denotes generically all the $W_{M,p}(z) \; \mu_{p_s}$ 
factors appearing in Eq.~(\ref{eq:finalformWM}).
The $S$-type set is thus empty when $M \geq 3$ and, given 
(\ref{eq:emptyStext2}), we conlude that:
\begin{equation}
\text{{\sl there are no $S$-type fields when $M \geq 2$}} . \label{eq:emptyStext3}
\end{equation}
The $W_{M-1}, W_{M-2}$ functions determined previously have thus no
dependence on $S$-type fields. In particular the first sum on the 
left-hand side of Eq.~(\ref{eq:dSSM-2}) vanishes trivially and
there are no other constraints from this equation. 
Equations (\ref{eq:finalformWM}, \ref{eq:finalformWM-1prime}, \ref{eq:WM-2sol}) and (\ref{eq:WM-3prime1}) simplify now to
\beqa
W_M(z) &=&  W_{M,0}(z) \ , \label{eq:M}\\
W_{M-1}(z) &=& \widetilde{W}_{M-1, a}(z) \; \widetilde{\Phi}^a + W_{M-1, 0}(z) \ , \label{eq:M-1}\\
W_{M-2}(z) &=& \Gamma(\dots, \widetilde{\Phi}^a, \dots; z) \ ,\label{eq:M-2}
\eeqa
and
\beqa
\displaystyle \Bigg[\frac{\partial W_{M-2}}{\partial \widetilde{\Phi}^a}\frac{\partial \overline W_{M-1}}{\partial \widetilde{\Phi}^{a *}} 
+  \widetilde{K}_{1, a} \widetilde{\Phi}^a \Bigg]  + \text{h.c.}  \sim_\Phi 0. \label{eq:WM-3prime2}
\eeqa
Operating  $\displaystyle \frac{\partial }{\partial \widetilde{\Phi}^b}$ on this last equation yields,
\beqa
(\mathcal{D}_i \widetilde{W}_{M-1, b})(\overline{\mathcal{D}}_i \overline{W}_{M, 0})   = 
2 \; \widetilde{W}_{M-1, b} \overline{W}_{M, 0} \; - \; \displaystyle \overline{\widetilde{W}}_{M-1, a} 
\, \frac{\partial^2 \Gamma}{\partial \widetilde{\Phi}^a \partial\widetilde{\Phi}^b}, \label{eq:WM-3prime3}
\eeqa
where we have used (\ref{eq:M-1}, \ref{eq:M-2}) and (\ref{eq:tildeK1}). Since ${W}_M \neq 0$ by assumption of the expansion 
(\ref{eq:Wexpansion}), that is 
$\overline{W}_{M, 0} \neq 0$ cf. (\ref{eq:M}), then Eq.(\ref{eq:WM-3prime3}) which is of the form of Proposition 
\ref{corol:genholo} implies 
\beqa
\widetilde{W}_{M-1, b} = 0, \; \forall b,  \label{eq:tildeWM-1}
\eeqa
meaning that for all $M \geq 2$ only $W_{M-2}$ can depend on $\widetilde{\Phi}$ fields. In the case $M=2$ there are no further
constraints. However, when $M \geq 3$ the constraint (\ref{eq:constraintM}) becomes relevant and, combined with (\ref{eq:tildeWM-1}),
leads to a sum of squares implying $W_M= 0$, a contradiction ! Thus the expansion  (\ref{eq:Wexpansion}) should be truncated at most
at $M=2$.\\ 

\noindent
{\bf Summary}:
\begin{itemize}
\item $M=1$: the most general solution for the superpotential is 
$$W = m_{p\ell} W_1(z, S, \widetilde{\Phi}) + W_0(z, S, \widetilde{\Phi}),$$ where
$W_1$ and $W_0$ are given by Eqs.~(\ref{eq:finalformWM}) and (\ref{eq:finalformWM-1}) where all the involved functions are arbitrary,
\item $M=2$: the most general solution for the superpotential is $W = m_{p\ell}^2 W_2(z) + m_{p\ell} W_1(z)  + W_0(z, \widetilde{\Phi})$, 
where $W_2, W_1$ and $W_0$ are given by Eqs.~(\ref{eq:M}), (\ref{eq:M-1}) with $\widetilde{W}_{M-1, a} = 0$, and
(\ref{eq:M-2}), where all the involved functions are arbitrary,
\item $M \geq 3$: no solution.
\end{itemize}
This list exhausts all possibilities in the case of the minimal K\"ahler potential.

\subsection{General K\"ahler \label{app:non-flat}}
The situation when the K\"ahler potential is non-canonical is much more involved. In this appendix rather than studying
completely the non-canonical case, we point out  new difficulties  which appear in this case and
two peculiar solutions are be given. The resolution
of the analogous of equation \eqref{eq:Vgen0} for a general K\"ahler potential is much more difficult.
Assuming that the K\"ahler potential and superpotential have the expansions given in Eqs.~(\ref{eq:Kexpansion}, \ref{eq:Wexpansion}),
the actual computation of the $F$-term potential of Eq.~(\ref{eq:V}) entails three main new difficulties in comparison with
the canonical K\"ahler case: 

\begin{enumerate}
\item The inversion of the K\"ahler metric is performed  perturbatively as an expansion in negative powers of $m_{p\ell}$; the actual form of the 
expansion depends on whether the observable sector enters explicitly $K_n$ or not.
\item Since $K^{ia^*}$ and $K^{ai^*}$ are in general different from zero, the coupling between the observable and the hidden sectors
renders the constraints fulfilling the consistency requirement (\ref{eq:the_condition}) much more involved.
\item As the K\"ahler potential is a polynomial of degree $N$ in the Planck mass, when $N>2$  Taylor expanding the exponential factor
gives rise to new contributions.
 \end{enumerate}

Points $1$ and $3$ could lead to some conspiracy. Indeed, the expansion of the exponential factor
leads to an unbounded from above series in $m_{p\ell}$ while the inverse of the K\"ahler metric leads to
an unbounded from above series in $1/m_{p\ell}$. This means that we could imagine some situations where these two contributions 
cancel each other.  However, in general there is no complete cancellation between these two contributions
and dangerous terms are generated. 
In our first analysis, in order to forbid dangerous terms,
we consider the case where $N=2$. Three configurations must then be considered characterized by different explicit dependence of the 
functions 
$K_n$, $n=0,1,2$, on the observable sector as follows:
\beqa
K(Z,Z^\dag)=m_{p\ell}^2 K_2 (z,z^\dag,\Phi,\Phi^\dag) +  m_{p\ell} K_1 (z,z^\dag,\Phi,\Phi^\dag) + K_0 (z,z^\dag,\Phi,\Phi^\dag),
~~~~~\label{eq:1} \\
K(Z,Z^\dag)= m_{p\ell}^2 K_2 (z,z^\dag) +  m_{p\ell} K_1 (z,z^\dag,\Phi,\Phi^\dag) +   K_0 (z,z^\dag,\Phi,\Phi^\dag), ~~~~~~~~~~~~~\label{eq:2} 
\\
K(Z,Z^\dag)= m_{p\ell}^2 K_2 (z,z^\dag) +  m_{p\ell} K_1 (z,z^\dag) +   K_0 (z,z^\dag,\Phi,\Phi^\dag) .~~~~~~~~~~~~~~~~~~~~~~\label{eq:3}
\eeqa
Distinguishing these three cases is necessary  in particular because they lead to different results for the
 perturbative inversion of the K\"ahler metric.  
Considering each case separately, with the superpotential given by \eqref{eq:Wexpansion},
three master equations, analogous to the master equation in the flat case \eqref{eq:Vgen}, are obtained.
In each situation new solutions are obtained.

Let us go into more details for the configuration given by \eqref{eq:1}. In this case the inverse K\"ahler metric takes
the form
\beqa
K^{IJ^*}&=& \phantom{+}\frac1{m_{p\ell}^2} K_2^{IJ^*} - \frac1 {m_{p\ell}^3} K_2^{IK^*} K_1{}_{K^*L} K_2^{LJ^*} \nn \\
&&+ \frac 1{m_{p\ell}^4}
\Big(- K_2^{IK^*} K_0{}_{K^*L} K_2^{LJ^*} +  K_2^{IK^*} K_1{}_{K^*L} K_2^{LM^*}K_1{}_{M^*N} K_2^{NJ^*}\Big) \nonumber \\
&& + {\cal O}(1/m_{p\ell}^5) \nn\\
&=&  \phantom{+}\frac1{m_{p\ell}^2} K_2^{IJ^*} + \frac1 {m_{p\ell}^3} \widetilde{K}_{-3}^{IJ^*}  + \frac1 {m_{p\ell}^4 }  \widetilde{K}_{-4}^{IJ^*}
    +  {\cal O}(1/m_{p\ell}^5)  \ ,
\eeqa
where
\beqa
K_2^{IJ^*} = \Big(\frac{\partial^2 K_2}{\partial Z^I \partial Z^{J^*}} \Big)^{-1} \ .
\eeqa
Particularizing to the case where $M=2$,     
the $F$-term contribution to the scalar potential takes the form
\beqa
&&V_F=e^{K_2 + \frac1{m_{p\ell}}K_1 + \frac1 {m_{p\ell}^2} K_0} \times \nonumber \\
&&\Big(m_{p\ell}^2\Big[{\cal D}_I W_2 K_2^{IJ^*} {\cal D}_{J^*}\overline{W}_2 - 3 |W_2|^2\Big]  
+ m_{p\ell}\Big[{\cal D}_I W_2 \widetilde{K}_{-3}^{IJ^*} {\cal D}_{J^*}\overline{W}_2 \nonumber \\
&& ~~+\Big({\cal D}_I W_2  K_2^{IJ^*}({\cal D}_{J^*} \overline{W}_1 + \overline{W}_2 \frac{\partial K_1}{\partial Z^{J^*}})  
  + \text{h.c.}\Big) -3 (W_2 \overline{W}_1 + W_1 \overline{W}_2) \Big] \nn\\
&& ~~+{\cal D}_I W_2 \widetilde{K}_{-4}^{IJ^*} {\cal D}_{J^*}\overline{W}_2  +
    \Big( {\cal D}_I W_2 \widetilde{K}_{-3}^{IJ^*}({\cal D}_{J^*} \overline{W}_1 + \overline{W}_2 \frac{\partial K_1}{\partial Z^{J^*}} )
    +\text{h.c.} \Big) \nonumber\\
&& ~~+  ({\cal D}_{I}  W_1 +  W_2 \frac{\partial K_1}{\partial Z^{I}})  K_2^{IJ^*}({\cal D}_{J^*} \overline{W}_1 + \overline{W}_2 \frac{\partial K_1}{\partial Z^{J^*}}) \nonumber \\
&& ~~~~~~~~~~~~~~~~~~-3(W_2 \overline{W}_0 + W_1 \overline{W}_1 + W_0 \overline{W}_2) + {\cal O}(\frac 1{m_{p\ell}})
\Big) \ , 
\eeqa
where we have denoted
\beqa
{\cal D}_I W_2 &=& \frac{\partial W_2}{\partial Z^I} + W_2 \frac{\partial K_2}{\partial Z^I} \ , \nn\\
 {\cal D}_I W_1 &=& \frac{\partial W_1}{\partial Z^I} + W_1 \frac{\partial K_2}{\partial Z^I} \ .
 \eeqa
 As in the flat case we must now impose that the terms in $m_{p\ell}^2$ and $m_{p\ell}$ do  not depend on the observable sector.
Rather than solving these complicated equations we now exhibit two specific solutions.

If we assume that
\beqa
K_1&=&0 \ ,\nn\\
W_1&=&0\ ,
\eeqa
then $\widetilde{K}_{-3}=0$ and the term proportional to $m_{p\ell}$ drops out from $V_F$, leaving us with the dangerous
term proportional to $m_{p\ell}^2$.  
To meet the requirement (\ref{eq:the_condition}), it is then sufficient to ask that $W_2$ and $K_2$ satisfy the condition
\beqa
{\cal D}_I W_2 K_2^{IJ^*} {\cal D}_{J^*}\overline{W}_2 - 3 |W_2|^2 \p 0 \ , \label{eq:no-scale-like}
\eeqa
leaving the functions $K_0$ and $W_0$ totally free. 
the  potential  then reduces to
\beqa
V=e^{K_2 +\frac1 {m_{p\ell}^2} K_0}\Big(m_{p\ell}^2 f(z, z^\dag) + {\cal D}_I W_2 \widetilde{K}_{-4}^{IJ^*} {\cal D}_{J^*}\overline{W}_2 -
3(W_2 \overline{W}_0 +  W_0 \overline{W}_2) \nonumber \\
+ \, {\cal O}(\frac 1{m_{p\ell}})\Big), ~~~
\eeqa
where $f$ denotes an arbitrary function of the hidden sector fields only. 
Having no dangerous terms this solution is an acceptable one, not found in \cite{SW}. Equation (\ref{eq:no-scale-like}) can be in 
principle solved as a partial differential equation for $W_2$ along the same lines as the analysis done in \ref{app:proof-flat}.
Note in particular that if we choose $f(z, z^\dag) \equiv 0$ then Eq.~(\ref{eq:no-scale-like}) reduces to a no-scale-like condition
\cite{cfkn,ekn,sm-no-scale,ln} 
involving the leading terms of the K\"ahler potential and superpotential. 

Another special solution can be found as follows:
suppose that
\beqa
W_2=0 \ ,
\eeqa
and let the functions $K_0,K_1,K_2$ and $W_1,W_0$ be totally free.
Then the  potential simplifies significantly to, 
\beqa
V=e^{K_2 + \frac1{m_{p\ell}}K_1 + \frac1 {m_{p\ell}^2} K_0} \Big(
{\cal D}_{I}  W_1  K_2^{IJ^*}{\cal D}_{J^*} \overline{W}_1 -3|W_1|^2 + {\cal O}(\frac 1{m_{p\ell}}) \Big) \ ,
\eeqa
and no constraint are needed in order to obtain an acceptable solution. This second solution was not found in \cite{SW} either.
If in addition we impose a no-scale-like condition \cite{cfkn,ekn,sm-no-scale,ln},
\beqa
{\cal D}_I W_1 K_2^{IJ^*} {\cal D}_{J^*}\overline{W}_1 - 3 |W_1|^2= 0 \ , \label{eq:no-scale-like1}
\eeqa
then the potential is naturally at the order $1/m_{p\ell}$. Thus when Supergravity is broken the cosmological
constant has a $1/m_{p\ell}$ suppression.

The two new solutions above were obtained  by simply  putting to zero all dangerous terms. Proceeding as
in the canonical case by solving the tower of differential equations, we anticipate the occurrence of 
other new solutions.
A detailed study goes beyond the scope of this paper.

\bibliographystyle{utphys}
\bibliography{biblio}

\end{document}